\documentclass[12pt]{article}

\usepackage{epsfig,latexsym,amsfonts,amsmath,amsthm,amssymb,amsbsy,multirow,slashed,wasysym,textcomp,dsfont,comment,mathtools,cancel,cite,diagbox,datetime,appendix,BOONDOX-calo}
\usepackage{tocloft}
\usepackage{bbold}
\usepackage{sidecap}
\usepackage{adjustbox}
\usepackage{pgfplots}
\usepackage{oplotsymbl}
\usepackage{tikzpagenodes}
\usepackage{adforn}
\usepackage{tikz}
\usetikzlibrary{backgrounds}
\usetikzlibrary{patterns}
\usetikzlibrary{arrows.meta}
\usetikzlibrary{shapes}
\tikzstyle{bag} = [align=center]
\usetikzlibrary{decorations.pathmorphing}
\usepackage[normalem]{ulem}
\usepackage[mathscr]{eucal}

\usepackage{hyperref}
\usepackage{subcaption}

 \newcommand{\badat}{\begin{alignedat}}
 \newcommand{\eadat}{\end{alignedat}}
 \newcommand\scalemath[2]{\scalebox{#1}{\mbox{\ensuremath{\displaystyle #2}}}}
 \def\be{\begin{equation}}
\def\ee{\end{equation}}

\def\p{\partial}

\usepackage{color}

\newcommand{\pink}[1]{\textcolor{\pink}{#1}}

\definecolor{dblue}{rgb}{0.2,0.50,0.80}

\newcommand{\tikzcircle}[2][black,fill=black]{\tikz[baseline=-0.5ex]\draw[#1,radius=#2] (0,0) circle ;}%

\def\n{k}

\def\A{\mathcal{A}}

\def\I{\mathcal{I}}

\def\M{\mathcal{M}}

\def\O{\mathcal{O}}

\def\S{\mathcal{S}}
\def\T{\mathcal{T}}

\def\bh{{\bar h}}

\def\bw{{\bar w}}

\def\ba{{\bar a}}
\def\bb{{\bar b}}
\def\bc{{\bar c}}
\def\bd{{\bar d}}

\def\tA{\widetilde A}

\def\th{\widetilde h}

\def\th{\widetilde h}

\def\tpsi{\widetilde \psi}
\def\tchi{\widetilde{\chi}}

\def\tvarphi{\widetilde{\varphi}}

\def\tPhi{\widetilde{\Phi}}

\def\wb{\bar{w}}

\def\bh{{\bar h}}

\def\bw{{\bar w}}

\def\Memo{{\rm M}}
\def\Gold{{\rm G}}
\def\CS{{\rm CS}}
\def\memo{{\rm m}}
\def\gold{{\rm g}}

\numberwithin{equation}{section} 
 \pgfplotsset{compat=1.17} 
\begin{document}

 \begin{titlepage}
  \thispagestyle{empty}
  \begin{flushright}
  CPHT-RR015.032021 
  \end{flushright}
  \bigskip
  \begin{center}

                  \baselineskip=13pt {\LARGE \scshape{Celestial Diamonds:\\[5mm] 
          {\Large Conformal Multiplets in Celestial CFT}
         }}

      \vskip1cm 

   \centerline{ 
   {Sabrina Pasterski}${}^\diamondsuit{}$,
   {Andrea Puhm}${}^\blacklozenge{}$
   {and Emilio Trevisani}${}^\blacklozenge{}$
   }

\bigskip\bigskip
 
 \centerline{\em${}^\diamondsuit$ Princeton Center for Theoretical Science, Princeton, NJ 08544, USA}
 
\bigskip
 
\centerline{\em${}^\blacklozenge$   CPHT, CNRS, Ecole Polytechnique, IP Paris, F-91128 Palaiseau, France}

\smallskip

\bigskip\bigskip

\end{center}

\begin{abstract}
  \noindent

We examine the structure of global conformal multiplets in 2D celestial CFT. For a 4D bulk theory containing massless particles of spin $s=\{0,\frac{1}{2},1,\frac{3}{2},2\}$ we classify and construct all SL(2,$\mathbb{C}$) primary descendants which are organized into `celestial diamonds'. This explicit construction is achieved using a wavefunction-based approach that allows us to map 4D scattering amplitudes to celestial CFT correlators of operators with SL(2,$\mathbb{C}$) conformal dimension~$\Delta$ and spin~$J$. Radiative conformal primary wavefunctions have $J=\pm s$ and give rise to conformally soft theorems for special values of $\Delta \in \frac{1}{2}\mathbb{Z}$. They are located either at the top of celestial diamonds, where they descend to trivial null primaries, or at the left and right corners, where they descend both to and from generalized conformal primary wavefunctions which have $|J|\leq s$. Celestial diamonds  naturally incorporate degeneracies of opposite helicity particles via the 2D shadow transform relating radiative primaries and account for the global and asymptotic symmetries in gauge theory and gravity.

\end{abstract}

\end{titlepage}

\setcounter{tocdepth}{2}
{\small \tableofcontents}

\section{Introduction}

Celestial holography asserts that the quantum gravity $\S$-matrix is dual to a codimension two conformal field theory (CFT) living on the celestial sphere. This paradigm puts the symmetries of the theory front and center. 
Recognizing~\cite{Strominger:2013lka,Strominger:2013jfa} that the asymptotic symmetry group of the theory~\cite{Bondi:1962px,Sachs:1962wk}
is hidden in IR divergences and factorization theorems~\cite{Weinberg:1965nx} in typical presentations of scattering amplitudes has led to a fascinating convergence of tools from the relativity~\cite{Barnich:2010eb,Barnich:2011ct,Barnich:2011mi} and amplitudes~\cite{Cachazo:2014fwa,Stieberger:2018edy,Stieberger:2018onx,Fan:2019emx,Nandan:2019jas,Fotopoulos:2019tpe,Fotopoulos:2019vac,Fotopoulos:2020bqj,Fan:2021isc,Arkani-Hamed:2020gyp,Gonzalez:2020tpi} communities.\footnote{For reviews see~\cite{Strominger:2017zoo,Compere:2018aar,Pasterski:2019ceq}.} While the $\S$-matrix can be used to diagnose the physical relevance of a proposed symmetry extension, geometric tools have the power to provide general statements that would be difficult to extract from perturbation theory. 
This is particularly true when it comes to studying the spontaneous symmetry breaking dynamics of these asymptotic symmetries. From a geometric viewpoint, this amounts to classifying the representations of the proposed symmetries on the asymptotic data.  Within amplitudes, this amounts to understanding the so called `conformally soft sector'. 

To describe this sector, we must explain the nature of the holographic map~\cite{Pasterski:2016qvg,Pasterski:2017kqt,Pasterski:2017ylz}.  
The subleading soft graviton theorem~\cite{Cachazo:2014fwa} is isomorphic to a stress tensor Ward identity~\cite{Kapec:2014opa,Kapec:2016jld} if the asymptotic particles are in boost, as opposed to momentum, eigenstates.  For massless particles, this change of basis can be done with a Mellin transform of the energies~\cite{deBoer:2003vf,Cheung:2016iub,m4pt,Pasterski:2017kqt,Pasterski:2017ylz}. This exchanges the energy $\omega$ of each external particle for a conformal dimension $\Delta$ under the SL(2,$\mathbb{C}$) Lorentz group. Encouragingly, this change of basis does not spoil the existence of universal factorization properties originally found in the energetically soft limit.
Rather than appearing as different orders in a Laurent expansion around $\omega\rightarrow0$, these manifest themselves as poles at special values of $\Delta\in \frac{1}{2}\mathbb{Z}$ and are referred to as conformally soft theorems~\cite{Nandan:2019jas,Pate:2019mfs,Adamo:2019ipt,Puhm:2019zbl,Guevara:2019ypd,Fan:2019emx}. The operators dual to these conformally soft modes are currents of the celestial CFT.  A natural direction is to continue to build up the holographic dictionary and apply CFT methods to celestial amplitudes with the hopes of learning something new about scattering.

As with any adventure into unfamiliar territory the cartographer's role is essential. With this mindset, we examine the structure of global conformal multiplets in 2D celestial CFT and perform a classification of all SL(2,$\mathbb{C}$) primary descendants corresponding to massless particles of spin $s=\{0,\frac{1}{2},1,\frac{3}{2},2\}$.  This follows the spirit of~\cite{Penedones:2015aga} from the bootstrap literature, and is inspired by recent work on celestial null states~\cite{Banerjee:2018gce,Banerjee:2018fgd,Banerjee:2019aoy,Banerjee:2019tam,Banerjee:2020kaa,Banerjee:2020zlg,Banerjee:2020vnt,Guevara:2021abz}.
We will see that even the simple case of global conformal multiplets reveals the power of symmetry to organize conformally soft behavior, tying together questions about the vacuum structure of asymptotically flat spacetimes, constraints on celestial amplitudes, and intrinsically 2D descriptions of celestial CFT~\cite{PPT2}.

We find a nested structure of primary descendants which we refer to as the {\it `celestial diamond'}. There are three variants that appear.
The first corresponds to finite dimensional SL(2,$\mathbb{C}$) modules that descend from radiative primaries of 2D spin $J=\pm s$ and $\Delta=1-s-n$ for $n\in\mathbb{Z}_>$. These correspond to poles in celestial OPEs and have recently been argued to generate an infinite tower of symmetries which, however, yields no new constraints on the $\S$-matrix~\cite{Guevara:2021abz}. 

The second class of primary descendants corresponds to the most subleading conformally soft theorems. They appear at $J=\pm s$ and $\Delta=1-s$ for $s\in \frac{1}{2}\mathbb{Z}_{>}$. They descend to their own conformal shadows and correspond to degenerate zero-area celestial diamonds.
These primaries have no obvious asymptotic symmetry interpretation~\cite{Dumitrescu:2015fej,Campiglia:2016jdj,Campiglia:2016efb,Campiglia:2016hvg,Campiglia:2017dpg} but are responsible for recursion relations sufficient to determine certain OPE coefficients~\cite{Pate:2019lpp}.

The third variant arises for any half integer $s\ge1$ and involves a finite number of non-degenerate celestial diamonds that correspond to universal conformally soft theorems with associated conformal Goldstone and memory modes. The corresponding radiative conformal primary wavefunctions appear at the left and right corners of these diamonds and descend both to and from generalized conformal primary wavefunctions with $|J|\leq s$. This shows that the now-familiar conformally soft theorems are only part of the picture. They get completed into modules which we will show in~\cite{PPT2} include conformally soft dressing modes and soft charge operators. 

Our classification addresses a number of interesting puzzles. 
The first is a two-fold question about the spectrum: Is the scattering basis augmented 1) by modes with conformal dimension off the principal series $\Delta \in 1+i\mathbb{R}$ and 2) by non-radiative modes which have $J \neq \pm s$? Question~1) was addressed in~\cite{Donnay:2020guq} where we showed that conformally soft radiative primaries with $\Delta \notin 1+i\mathbb{R}$ (see~\cite{Donnay:2018neh}) can be expressed as contour integrals of the radiative data on the principal series basis of~\cite{Pasterski:2017kqt}. Question~2) is motivated by the observation~\cite{Pasterski:2020pdk} that there are interesting non-radiative conformal primary wavefunctions, both off-shell as well as exact bulk solutions to Einstein's equations. Here we find a home for generalized conformal primaries as parents and descendants of conformally soft radiative primaries in celestial diamonds.

The second puzzle concerns the role of the 2D shadow transform in the conformal basis. The generator of conformal symmetries requires taking a shadow of the $\Delta=0$ conformally soft graviton~\cite{Donnay:2018neh}.   This begs the existential question of when to shadow or not to shadow~\cite{ss,BP}. Here we show that celestial diamonds are closed under shadows: the radiative wavefunctions at the left and right corner of finite-area celestial diamonds as well as the two corners of zero-area celestial diamonds are related by a shadow transform. By adding a generalized conformal primary at the top of finite-area diamonds, its descendants include both the radiative primary and its shadow, so that one is never forced to smear the 2D operators.
This gives new insight into a third puzzle: the soft theorems for each helicity are not independent but rather are related to one another by the shadow transform~\cite{ss}.  We now see that this is simply a particular case of the second puzzle: both modes descend from the same generalized primary.

Celestial diamonds thus offer a natural language for describing the conformally soft sector of celestial CFT and resolve various puzzles surrounding the conformal basis, shadows and helicity degeneracies.  They unify the nested primary descendants associated to soft charges found in~\cite{Banerjee:2018gce,Banerjee:2018fgd,Banerjee:2019aoy,Banerjee:2019tam,Banerjee:2020kaa,Banerjee:2020zlg,Banerjee:2020vnt} with the finite dimensional modules of~\cite{Guevara:2021abz} and demonstrate how these relations extend to arbitrary spin. Our wavefunction-based approach provides a bulk picture of what the states correspond to, a mechanical way to get results guaranteed by representation theory, and lets us decouple what comes from dynamics versus kinematics. 

The paper is organized as follows.  In section~\ref{sec:CPWs} we set up our conventions and introduce radiative and generalized conformal primary wavefunctions.  In section~\ref{sec:global_primary_desc} we perform a general classification of SL(2,$\mathbb{C}$) primary descendant operators in 2D CFTs and discuss quantization and conformal shadows. We then classify all SL(2,$\mathbb{C}$) primary descendants corresponding to massless fields in celestial CFTs in section~\ref{sec:prim_desc}, introducing the celestial diamonds which capture this nested submodule structure. Various appendices supplement our discussion and include: a complementary construction of primary descendants via representation theory (appendix~\ref{App:rep_theory}), details of the relation between SL(2,$\mathbb{C}$) Lorentz generators and celestial derivatives (appendix~\ref{sec:CelestialSL2C}) allowing for a wavefunction-based identification of primary descendants (appendix~\ref{app:modules}), and a review of the shadow transform of bulk wavefunctions in the embedding space formalism (appendix~\ref{app:embeddingspace}).

\section{Bulk Wavefunctions for Primary Operators}
\label{sec:CPWs}

The 4D Lorentz group SO(1,3)$\simeq$ SL(2,$\mathbb{C}$) acts as the global conformal group on the 2D celestial sphere at null infinity. This allows us to recast scattering amplitudes of massless spin-$s$ particles labeled by their energy $\omega$ and a point $(w,\bw)$ on the celestial sphere as celestial CFT correlators of operators $\O_{\Delta,J}$ labeled by the SL(2,$\mathbb{C}$) conformal dimension~$\Delta$ and spin~$J$~\cite{Pasterski:2016qvg,Pasterski:2017kqt,Pasterski:2017ylz}
\begin{equation}\label{eq:pspacetocelestial}
    \A_n(\omega_i,\pm s_i; w_i,\bw_i)\stackrel{\M}{\mapsto}
    \left\langle \prod_{i=1}^n\O^{s,\pm}_{\Delta_i,J_i}(w_i,\bw_i) \right\rangle\,.
\end{equation}
This can be achieved by performing a Mellin transform
\begin{equation}\label{eq:transform}
    \M(.)=\int_0^\infty \frac{d\omega}{\omega}\omega^\Delta (\cdot)
\end{equation}
 in the energy of each external particle in the amplitude and amounts to a change of basis for the external wavepackets being scattered. The amplitude on the left hand side of~\eqref{eq:pspacetocelestial} describes probabilities for scattering momentum-eigenstates. The transform~\eqref{eq:transform} is gauge equivalent to preparing wavepackets for spin-$s$ particles that transform with definite $(\Delta,J)$ under an SL$(2,\mathbb{C})$ transformation
 \begin{equation}\label{eq:sl2c}
    X^\mu\mapsto \Lambda^\mu_{~\nu}X^\nu\,, \quad w\mapsto \frac{a w+b}{cw+d}\,,~~~\bw\mapsto \frac{{\bar a} \bw+{\bar b}}{{\bar c}\bw+{\bar d}}\,
\end{equation}
acting on the bulk point $X^\mu=(X^0,X^1,X^2,X^3)$ and boundary point $(w,\bw)$ where $ad-bc=1$ and $\Lambda^{\mu}_{~\nu}$ is the corresponding vector representation. These definite $(\Delta,J)$ wavepackets known as conformal primary wavefunctions $\Phi^s_{\Delta,J}$ are being scattered in the celestial amplitude on the right hand side of~\eqref{eq:pspacetocelestial} and will be the protagonists in this work. Given a 4D operator $O^{s}(X^\mu)$ of spin-$s$ in the Heisenberg picture, we can define a 2D operator in the celestial CFT by~\cite{Donnay:2020guq}
 \be\label{qdelta}
\O^{s,\pm}_{\Delta,J}(w,\bw)\equiv i(O^{s}(X^\mu),\Phi^s_{\Delta^*,-J}(X_\mp^\mu;w,\bw))\,
\ee
using standard inner products $(.\,,.)$ computed on a Cauchy slice in the bulk.\footnote{A standard choice for $s=0$ is the Klein-Gordon inner product.  The $s=1,2$ inner products were computed in~\cite{Donnay:2020guq} where the operator $\O^{s,\pm}_{\Delta,J}$ was denoted by $Q_{\Delta,J}$.} The $\pm$ on the operator indicates whether it corresponds to an {\it in} or an {\it out} state, and this selection is achieved via a prescription for analytically continuing the wavefunction as $ X^\mu_\pm=X^\mu\pm i\varepsilon\{-1,0,0,0\}$.
 
 The goal of this section is to show how to construct these conformal primary wavefunctions. They will appear in two forms: {\it radiative} conformal primary wavefunctions which have $J=\pm s$ and {\it generalized} conformal primary wavefunctions which have $|J|\leq s$; we will discuss them in turn in sections~\ref{sec:CPW} and~\ref{sec:gencpw}. It will be convenient to summarize some notation here. We will embed the celestial sphere into the $\mathbb{R}^{1,3}$ lightcone via the reference direction
\be\label{qmu}
q^\mu=(1+w\bw,w+\bw,i(\bw-w),1-w\bw)\,,
\ee
from which we obtain two natural polarization vectors $\sqrt{2}\epsilon_w^\mu=\p_w q^\mu$ and $\sqrt{2}\epsilon_\bw^\mu=\p_\bw q^\mu$. 
From the three null vectors $\{q^\mu,\epsilon^\mu_w,\epsilon^\mu_\bw\}$ and the spacetime vector $X^\mu$ we can construct a null tetrad for Minkowski space~\cite{Pasterski:2020pdk}
\be\label{tetrad}
l^\mu=\frac{q^\mu}{-q\cdot X}\,, ~~~n^\mu=X^\mu+\frac{X^2}{2}l^\mu\,, ~~~m^\mu=\epsilon^\mu_w+(\epsilon_w\cdot X) l^\mu\,, ~~~\bar{m}^\mu=\epsilon^\mu_\bw +(\epsilon_\bw\cdot X) l^\mu\,,
\ee
which satisfies the standard normalization conditions $l\cdot n=-1$, $m\cdot\bar{m}=1$ while all other inner products vanish.  These will serve as natural building blocks for the wavefunctions $\Phi^s_{\Delta,J}$ for $s\in \mathbb{Z}$. For $s\in \frac{1}{2}\mathbb{Z}$ we further decompose the tetrad into a spin frame
\be\label{spinframe}
l_{\alpha{\dot \beta}}=o_\alpha\bar{o}_{\dot \beta}\,,~~n_{\alpha{\dot \beta}}=\iota_\alpha\bar{\iota}_{\dot \beta}\,,~~m_{\alpha{\dot \beta}}=o_\alpha{\bar\iota}_{\dot \beta}\,,~~\bar{m}_{\alpha{\dot \beta}}=\iota_\alpha{\bar o}_{\dot \beta}\,
\ee
where
\be\label{spinframespinors}
o_\alpha=\sqrt{\frac{2}{q\cdot X}}\left(\begin{array}{c} \bw \\-1 \end{array}\right)\,,~~~\iota_\alpha=\sqrt{\frac{1}{q\cdot X}}\left(\begin{array}{c}X^0-X^3-w(X^1-iX^2) \\-X^1-iX^2+w(X^0+X^3) \end{array}\right),
\ee
and we fix the overall phase ambiguity by setting ${\bar o}_{\dot \alpha}=(o_\alpha)^*$ and ${\bar \iota}_{\dot \alpha}=(\iota_\alpha)^*$ in the region where $q\cdot X>0$ and analytically continuing from there.  The SL(2,$\mathbb{C}$) quantum numbers of these null tetrad and spin frame are summarized in table~\ref{table:tetradspinframe}.
\vspace{1em}

\begin{table}[ht!]
    \centering
    \begin{tabular}{c|c|c|c|c|c|c|c|c}
   & $l^\mu$ & $n^\mu$ & $m^\mu$ & $\bar{m}^\mu$ & $o_\alpha$ & $\bar{o}_{\dot{\alpha}}$ & $\iota_\alpha$ & $\bar{\iota}_{\dot{\alpha}}$   \\ \hline
   $\Delta$ & $0$&$0$ &$0$&$0$&$0$&$0$&$0$&$0$\\
   $J$& $0$&$0$ &$+1$&$-1$&$+\frac{1}{2}$&$-\frac{1}{2}$&$-\frac{1}{2}$&$+\frac{1}{2}$\\
    \end{tabular}
    \caption{SL(2,$\mathbb{C}$) quantum numbers for the tetrad and spin frame.}
    \label{table:tetradspinframe}
\end{table}

\subsection{Radiative Conformal Primaries}
\label{sec:CPW}
In this section we review spin $s=\{0,\frac{1}{2},1,\frac{3}{2},2\}$ conformal primary wavefunctions and their shadow transforms as well as their conformally soft limits. 
We will omit vector and spinor indices for notational convenience unless required. \vspace{1em}

\noindent{\bf Definition:} A {\it radiative conformal primary wavefunction} is a wavefunction on $\mathbb{R}^{1,3}$ which satisfies the linearized equations of motion for a massless spin-$s$ particle in vacuum and transforms under SL(2,$\mathbb{C}$) as a 4D (spinor-)tensor field of spin-$s$ and as a 2D conformal primary of conformal dimension $\Delta$ and spin $J=\pm s$, namely
\be\label{deltaj20}
\Phi_{\Delta,J}\Big(\Lambda^{\mu}_{~\nu} X^\nu;\frac{a w+b}{cw+d},\frac{{\bar a} \bw+{\bar b}}{{\bar c}\bw+{\bar d}}\Big)=(cw+d)^{\Delta+J}({\bar c}\bw+{\bar d})^{\Delta-J}D(\Lambda)\Phi_{\Delta,J}(X^\mu;w,\bw)\,,
\ee
where $D(\Lambda)$ is the 3+1D spin-$s$ representation of the Lorentz algebra.

\subsubsection*{Conformal Primary Wavefunctions}

The conformal primary wavefunction with spin $J=0$ and conformal dimension $\Delta$ is given by
\begin{equation}\label{varphi2d}
    \varphi^\Delta=\frac{1}{(- q\cdot X)^\Delta}\,.
\end{equation}
Using~\eqref{varphi2d} together with the null tetrad~\eqref{tetrad} and spin frame~\eqref{spinframespinors} we can express all other conformal primaries with spin as~\cite{Pasterski:2017ylz,Pasterski:2017kqt,Pasterski:2020pdk}
\be\begin{array}{ll}\label{CPWs}
    \psi_{\Delta,J=+\frac{1}{2}}=o\varphi^\Delta\,,&~~~{\bar\psi}_{\Delta,J=-\frac{1}{2}}={\bar o}\varphi^\Delta\,,\\
     A_{\Delta,J=+ 1}=m \varphi^{\Delta} \,,&~~~ A_{\Delta,J=- 1}=\bar{m} \varphi^{\Delta}\,, \\
      \chi_{\Delta,J=+\frac{3}{2}}=om \varphi^\Delta\,,&~~~{\bar\chi}_{\Delta,J=-\frac{3}{2}}={\bar o}{\bar m}\varphi^\Delta\,,\\
    h_{\Delta,J=+2}=m m \varphi^{\Delta} \,,&~~~  h_{\Delta,J=-2}=\bar{m} \bar{m} \varphi^{\Delta}\,.
\end{array}
\ee
Note that the SL(2,$\mathbb{C}$) spin $J$ is identified with the 4D helicity $\pm s$ for these wavefunctions which form a basis for radiative states when $\Delta\in1+i\mathbb{R}$, also known as the principal series~\cite{Pasterski:2017kqt}. 
From there one can analytically continue to the complex $\Delta$ plane~\cite{Donnay:2020guq}.

For special {\it conformally soft} values of $\Delta\in \frac{1}{2}\mathbb{Z}$ the conformal primaries~\eqref{CPWs} correspond to Goldstone modes of spontaneously broken asymptotic symmetries~\cite{Cheung:2016iub,Pasterski:2017kqt,Donnay:2018neh} whose Ward identities are equivalent to (conformally) soft theorems~\cite{Cheung:2016iub,Nandan:2019jas,Pate:2019mfs,Adamo:2019ipt,Puhm:2019zbl,Guevara:2019ypd,Fan:2019emx}.  
These Goldstone modes are summarized in table~\ref{table:Goldstone}.
Furthermore, there exist soft theorems arising from primaries which do not correspond to Goldstone modes in the conformal basis but which are, nevertheless, related via the classical double copy or supersymmetry~\cite{Pasterski:2020pdk} to primaries which do have an asymptotic symmetry interpretation. They are summarized in table~\ref{table:gold}.
We refer to~\cite{Donnay:2018neh,Donnay:2020guq,Pasterski:2020pdk,PPP,upcoming2,upcoming3} for more detailed analyses of various aspects of these primaries and their associated soft charges.

\begin{table}[ht!]
\renewcommand*{\arraystretch}{1.3}
\centering
\begin{tabular}{|c|c|c|cc|}
\hline
  & \multicolumn{1}{c|}{${A}_{\Delta,J=\pm1}$} & \multicolumn{1}{c|}{$\chi_{\Delta,J=+\frac{3}{2}}$}  & \multicolumn{2}{c|}{${h}_{\Delta,J=\pm2}$}\\
  \hline
 $\Delta$ &1 &$\frac{1}{2}$ &  1&  0 \\
 symmetry & large U(1) & large SUSY  & supertranslation & shadow superrotation\\ 
\hline
\end{tabular}
\caption{Goldstone modes of spontaneously broken asymptotic symmetries for $1\le s\le2$.  
}
 \label{table:Goldstone}

\vspace{1em}
\renewcommand*{\arraystretch}{1.3}
\centering
\begin{tabular}{|c|c|c|c|c|}
\hline
  & ${\psi}_{\Delta,J=+\frac{1}{2}}$ &   ${A}_{\Delta,J=\pm1}$ & $\chi_{\Delta,J=+\frac{3}{2}}$  & ${h}_{\Delta,J=\pm2}$\\
  \hline
 $\Delta$ &$\frac{1}{2}$ &0 &$-\frac{1}{2}$ &  -1 \\
\hline
\end{tabular}
\caption{Soft theorems without conformal Goldstones for $\frac{1}{2}\le s\le2$.
}
\label{table:gold}
\end{table}

\subsubsection*{Conformal Shadow Primary Wavefunctions}

A shadow transformation takes the conformal primary wavefunctions~\eqref{varphi2d}-\eqref{CPWs} to wavefunctions of flipped and shifted conformal dimension and flipped SL(2,$\mathbb{C}$) spin
\begin{equation}\label{ShadowPhi}
\widetilde{\Phi}_{\Delta,J}=\widetilde{{\Phi}_{2-\Delta,-J}}\,.
\end{equation}
The spin-0 conformal shadow primary of conformal dimension $\Delta$ is given by~\cite{Pasterski:2017kqt}
\begin{equation}\label{SHvarphi2d}
    \widetilde{\varphi}^{\Delta}=(-X^2)^{\Delta-1}\varphi^{\Delta}\,, 
\end{equation}
while the conformal primaries with spin get mapped to~\cite{Pasterski:2017kqt,Pasterski:2020pdk,PPP}\footnote{The shadow transform of 4D conformal primary wavefunctions $\Phi_{\Delta,J}$ can be conveniently computed using the embedding space formalism~\cite{SimmonsDuffin:2012uy,Pasterski:2017kqt}. See appendix~\ref{app:embeddingspace}.
As discussed in that appendix, we use a sign convention that matches the 2D shadow~\cite{upcoming2,PPP}, in contrast to~\cite{Pasterski:2017kqt}. This will be particularly convenient for shadow/descendancy relations that appear in section~\ref{sec:prim_desc}.
}
\be\begin{array}{ll}\label{SHCPWs}
\widetilde{\psi}_{\Delta,J=-\frac{1}{2}}={-}(-X^2)^{\Delta-\frac{3}{2}}\sqrt{2}\iota\varphi^\Delta\,,&~~~\widetilde{{\overline\psi}}_{\Delta,J=+\frac{1}{2}}={-}(-X^2)^{\Delta-\frac{3}{2}}\sqrt{2}{\bar \iota}\varphi^\Delta\,,\\
    \widetilde{A}_{\Delta,J=-1}={-}(-X^2)^{\Delta-1}\bar{m}\varphi^\Delta\,, &~~~\widetilde{A}_{\Delta,J=+1}={-}(-X^2)^{\Delta-1}m\varphi^\Delta\,,\\
    \widetilde{\chi}_{\Delta,J=-\frac{3}{2}}={+}(-X^2)^{\Delta-\frac{3}{2}}\sqrt{2}\iota{\bar m} \varphi^\Delta\,,&~~~\widetilde{{\overline\chi}}_{\Delta,J=+\frac{3}{2}}={+}(-X^2)^{\Delta-\frac{3}{2}}\sqrt{2}{\bar \iota}{m}\varphi^\Delta\,,\\
    \widetilde{h}_{\Delta,J=-2}\,={+}(-X^2)^{\Delta-1}\bar{m}\bar{m}\varphi^\Delta\,, &~~~ \widetilde{h}_{\Delta,J=+2}\,={+}(-X^2)^{\Delta-1}{m} {m}\varphi^\Delta\,.
\end{array}
\ee
These conformal shadow primaries have conformal dimension $\Delta$ and SL(2,$\mathbb{C}$) spin $J$ flipped relative to the 4D helicity $\pm s$. Under the shadow transform, radiative conformal primaries with $\Delta\in1+i\mathbb{R}$ get mapped to conformal shadow primaries with the same range of conformal dimension which therefore also form a conformal basis~\cite{Pasterski:2017kqt}. Again we can analytically continue to the complex $\Delta$ plane. Tables~\ref{table:shadowGoldstone} and~\ref{table:shadowgold} give the shadow transforms of tables~\ref{table:Goldstone} and~\ref{table:gold}.
\vspace{1em}

\begin{table}[ht!]
\renewcommand*{\arraystretch}{1.3}
\centering
\begin{tabular}{|c|c|c|cc|}
\hline
  & \multicolumn{1}{c|}{${\tA}_{\Delta,J=\pm1}$} & \multicolumn{1}{c|}{$\widetilde{\chi}_{\Delta,J=-\frac{3}{2}}$}  & \multicolumn{2}{c|}{${\th}_{\Delta,J=\pm2}$}\\
  \hline
 $\Delta$ &1 &$\frac{3}{2}$ &  1&  2 \\
 symmetry & large U(1) & large SUSY  & supertranslation & superrotation\\
\hline
\end{tabular}
\caption{Shadow Goldstone modes of spontaneously broken asymptotic symmetries for $1\le s\le2$.  
}
 \label{table:shadowGoldstone}
 \vspace{1em}
\begin{tabular}{|c|c|c|c|c|}
\hline
  & ${\tpsi}_{\Delta,J=-\frac{1}{2}}$ &   ${\tA}_{\Delta,J=\pm1}$ & $\tchi_{\Delta,J=-\frac{3}{2}}$  & ${\th}_{\Delta,J=\pm2}$\\
  \hline
 $\Delta$ &$\frac{3}{2}$ &2 &$\frac{5}{2}$ &  3 \\
\hline
\end{tabular}
\caption{Soft theorems without conformal shadow Goldstones for $\frac{1}{2}\le s\le2$.
}
 \label{table:shadowgold}
 \end{table}

 \pagebreak 
 
The conformally soft primary wavefunctions of tables~\ref{table:Goldstone} to \ref{table:shadowgold} will play an important role in the discussion of SL(2,$\mathbb{C}$) primary descendants where they will correspond to the left and right corners of `celestial diamonds'. Their primary descendants will fill the bottom corners. We adopt the notation $\Gold$ for the pure gauge primaries of tables~\ref{table:Goldstone} and~\ref{table:shadowGoldstone}, and $\gold$ for the further subleading conformally soft modes of tables~\ref{table:gold} and~\ref{table:shadowgold}. 

\subsubsection*{Canonical Pairings}

With respect to a suitable inner product~\cite{Pasterski:2017kqt} radiative wavefunctions with $\Delta\in1+i\mathbb{R}_+$ are canonically paired with radiative wavefunctions with $\Delta\in1+i\mathbb{R}_-$, i.e. they have a $\delta$-function normalizable inner product akin to that of plane waves which they are Mellin transforms of~\cite{Pasterski:2017ylz}.
This can be extended to the complex $\Delta$ plane~\cite{Donnay:2020guq}.
For example, in the free theory, one can expand Heisenberg picture operators in terms of the radiative data
\be\label{BulkOexpansion}
\begin{array}{rl}
O^{s}(X^\mu)=\sum\limits_{J=\pm s}\int d^2 w \int_{1-i\infty}^{1+i\infty}(-id\Delta)& \Big[{\cal N}^+_{2-\Delta,s}\Phi_{2-\Delta,-J}(X_+^\mu;w,\bw)a_{\Delta,J}(w,\bw)\\
&~+{\cal N}^-_{\Delta,s}\Phi_{\Delta,J}(X_-^\mu;w,\bw)a_{\Delta,J}(w,\bw)^\dagger\Big]\,.
\end{array}
\ee
Then the commutation relations of these modes are~\cite{Donnay:2020guq}
\be
\label{commutation_modes}
[a_{\Delta,J}(w,\bw),a_{\Delta',J'}(w',\bw')^\dagger]=\delta_{JJ'}\delta^{(2)}(w-w')\boldsymbol{\delta}(i(\Delta+\Delta'^*-2))\,,
\ee
for integer spin, and the anti-commutations relations with $[\cdot,\cdot]\rightarrow\{\cdot,\cdot\}$ for half-integer spin,
where the distribution $\boldsymbol{\delta}$ reduces to the ordinary Dirac delta function for $\Delta,\Delta'\in 1+i\mathbb{R}$ on the principal series and such that for generic $\Delta\in\mathbb{C}$ and $|J|=s$, the operator~\eqref{qdelta} generates the shift
 \begin{equation}\label{shift1}
[\O^{s}_{\Delta,J}(w,\bw),O^s(X^\mu)]=i\Phi_{\Delta,J}(X_-^\mu;w,\bw)\,.
\end{equation}
Effectively,~\cite{Donnay:2020guq} addressed the question of the completeness of the principal series for radiative modes of the free theory.  This question will need to be revisited once we introduce generalized non-radiative primaries in the next section. 

By tweaking the choice of {\it in}/{\it out} combinations or representation for the distribution $\boldsymbol{\delta}$, one can also define pairings between conformally soft operators. These were explored in~\cite{Donnay:2018neh,Ball:2019atb} for bosonic $\Gold$ modes and will be revisited for subleading $\Gold$ and $\gold$ modes in~\cite{upcoming3}. Goldstone primaries of dimension $\Delta\in \frac{1}{2}\mathbb{Z}$ are paired with memory primaries of dimension $2-\Delta$.  We will denote the canonical partners of the $\Gold$ modes by $\Memo$ and the canonical partners of the $\gold$ modes by $\memo$.

\subsubsection*{Logarithmic and Distributional Wavefunctions}
An important subtlety arises for wavefunctions with $\Delta=1$ corresponding to the Goldstone modes for the leading soft theorems in gauge theory and gravity: they degenerate with their $\Delta=1$ shadows and are thus naively missing a canonical partner. This was remedied in~\cite{Donnay:2018neh} by noticing a hidden logarithmic branch of the solution space that is obtained from a combination of primaries and their shadows in a careful limit, series expanding around $\Delta\to1$. This reveals the existence of an additional logarithmic $\Delta=1$ scalar primary~\cite{Pasterski:2020pdk}
\begin{equation}\label{logphi}
    \lim_{\Delta\to1} \partial_\Delta (\varphi^\Delta-\tvarphi^\Delta)=-\log(-X^2) \varphi^1\equiv \varphi^{\log}\,,
\end{equation}
as well as logarithmic $\Delta=1$ vector and metric primaries~\cite{Donnay:2018neh}
\begin{equation}\label{Ahlog}
    A^{\log}_{1,+1}=m \varphi^{\log}\,, \quad h^{\log}_{1,\pm2}=m m \varphi^{\log}\,,
\end{equation}
with similar expressions for $J\mapsto -J$ and $m\mapsto \bar{m}$.
Note that this limit commutes with all gauge conditions obeyed by the $\Delta=1$ vector and metric Goldstone modes and their shadows but, unlike them, the logarithmic modes~\eqref{Ahlog} are no longer pure gauge.

Note that the $\Delta=1$ degeneracy of the wavefunctions manifests itself in a degeneracy of the soft theorems~\cite{He:2014cra} -- opposite helicity modes have soft theorems related by a conformal shadow~\cite{ss}.  Besides this, soft theorems exhibit another degeneracy: between incoming and outgoing soft modes.  This degeneracy in the wavefunctions is not obvious after reinstating the $i\varepsilon$ regulator. 
Combinations of incoming and outgoing logarithmic $\Delta\to 1$ modes in the limit $\varepsilon \to 0$ have been used in~\cite{Donnay:2018neh} to define conformally soft spin-1 and spin-2 wavefunctions\footnote{While the appearance of the logarithmic branch in the solutions space is tied to the degeneracy at $\Delta=1$, for the other special values of $\Delta$ corresponding to the conformally soft gravitino and the subleading conformally soft graviton we can also take combinations of incoming and outgoing $i \varepsilon$-regulated solutions. Their relation to memory effects will be discussed in~\cite{upcoming3}.}
\begin{equation}\label{AhCS}
    A^{\CS}_{1,+1}=m \varphi^{\CS}\,,\quad  h^{\CS}_{1,+2}=m m \varphi^{\CS}\,,
\end{equation}
and similarly for $J\mapsto -J$ and $m\mapsto \bar{m}$,
with the conformally soft scalar introduced in~\cite{Pasterski:2020pdk}
\begin{equation}\label{phiCS}
\varphi^{\CS}=\left[\Theta(X^2)+{\rm log}(X^2)(q\cdot X)\delta(q\cdot X)\right]\varphi^1\,.
\end{equation}

Once the $\varepsilon \to 0$ limit is taken, there are
other wavefunctions which satisfy the Klein Gordon equation almost everywhere (one can expect sources where the regulator was avoiding a singularity).  In the notation of~\cite{Pasterski:2020pdk} we have the scalar wavefunctions
\be\label{pcsp}
\varphi^{\CS'}\equiv \Theta(X^2)\varphi^1\,, \quad \varphi^{\CS''}\equiv\log(X^2)\delta(q\cdot X)\,,
\ee
from which the corresponding vector and metric wavefunctions are obtained as above via the classical double copy with the Kerr-Schild vectors $m$ and $\bar{m}$. It is worth emphasizing that the $\CS'$ and $\CS''$ wavefunctions individually no longer satisfy the source-free equations of motion. 
These wavefunctions naturally fit in the framework of {\it generalized} conformal primary wavefunctions introduced in~\cite{Pasterski:2020pdk} which we now review.

\subsection{Generalized Conformal Primaries}
\label{sec:gencpw}

In~\cite{Pasterski:2020pdk} we introduced a generalized notion of conformal primaries which have definite SL(2,$\mathbb{C}$) conformal dimensions $\Delta$ and spins $J$ which need not a priori satisfy the equations of motion or the same gauge fixing as conformal primary wavefunctions and have $|J|\leq s$. The definitions given there can be succinctly written for arbitrary spin.\footnote{See~\cite{Law:2020tsg} for a spin-$s$ construction for massive spinning bosons on the celestial sphere.}\vspace{1em}

\noindent{\bf Definition:} A {\it generalized conformal primary}  is a wavefunction on $\mathbb{R}^{1,3}$, which transforms under SL(2,$\mathbb{C}$) as a 2D conformal primary of conformal dimension $\Delta$ and spin $J$, and a 4D (spinor-) tensor field of spin-$s$, namely:
\begin{equation}\label{Defgenprim}
    \badat{2}
\Phi^{gen,s}_{\Delta,J}\Big(\Lambda^{\mu}_{~\nu} X^\nu;\frac{a w+b}{cw+d},\frac{{\bar a} \bw+{\bar b}}{{\bar c}\bw+{\bar d}}\Big)=(cw+d)^{\Delta+J}({\bar c}\bw+{\bar d})^{\Delta-J}D_s(\Lambda)\Phi^{gen,s}_{\Delta,J}(X^\mu;w,\bw)\,,
\eadat
\end{equation}
where $D_s(\Lambda)$ is the 3+1D spin-$s$ representation of the Lorentz algebra. \vspace{1em}

\noindent We have already encountered generalized primaries above: the members of the tetrad and spin frame correspond to generalized vectors and spinors, respectively, with $(\Delta,J)$ given in table~\ref{table:tetradspinframe}. They are thus natural building blocks for generalized conformal primary wavefunctions. Starting from the definition~\eqref{Defgenprim} there are many natural constraints one can impose at each spin. The spin $s=0,1,2$ examples were originally considered in~\cite{Pasterski:2020pdk}.  Here we review those constructions and extend them to spin $s=\frac{1}{2},\frac{3}{2}$.

\subsubsection*{Generalized Scalar}

For the generalized conformal primary scalar $D_0(\Lambda)=\mathbb{1}$. In the following, we will consider analytic solutions, of the form
\be\label{vargen}
\varphi^{gen}_\Delta=f(X^2)\varphi^\Delta\,,
\ee
which obey
\be
\Box \varphi^{gen}=4[(2-\Delta) f'(X^2)+X^2 f''(X^2)]\varphi^\Delta\,.
\ee
Demanding that this is zero for generic $\Delta$ lands us on the radiative spin-0 conformal primary wavefunction~\eqref{varphi2d} and its shadow~\eqref{SHvarphi2d} for any $\Delta$.

\subsubsection*{Generalized Weyl Spinor}

For a left-handed generalized conformal primary Weyl spinor spinor $D_\frac{1}{2}(\Lambda)=M$ where $M$ is the  $(\frac{1}{2},0)$ representation of the Lorentz algebra. Generalized primary spinors with $|J|\leq s=\frac{1}{2}$ take the form:
\be
\psi^{gen}_{\Delta,\frac{1}{2}}=o\varphi_{\Delta}^{gen}\,,
\ee
and
\be
\psi^{gen}_{\Delta,-\frac{1}{2}}=\iota\varphi_{\Delta}^{gen}\,.
\ee
The Dirac equation reduces to the Weyl equation, and for analytic choices of $\varphi^{gen}_\Delta$, of the form~\eqref{vargen}, we get 
\be
\bar{\sigma}^\mu\p_\mu \psi^{gen}_{\Delta,\frac{1}{2}}=-2\sqrt{2}f'(X^2)\bar{\iota}\varphi^\Delta\,,
\ee
and
\be
\bar{\sigma}^\mu\p_\mu \psi^{gen}_{\Delta,-\frac{1}{2}}=-\sqrt{2}\left[\left(\frac{3}{2}-\Delta\right)f(X^2)+X^2f'(X^2)\right]\bar{o}\varphi^\Delta\,.
\ee
Setting these to zero lands us on the left-handed radiative spin-$\frac{1}{2}$ conformal primary wavefunction~\eqref{CPWs} and its shadow~\eqref{SHCPWs} for any $\Delta$.

\subsubsection*{Generalized Vector}

For a generalized primary vector $D_1(\Lambda)=\Lambda$.  We constructed such wavefunctions for integer spins $|J|\leq s=1$ in~\cite{Pasterski:2020pdk} as
\be\label{agen1}
A^{gen}_{\Delta,+1;\mu}=m_\mu \varphi_{\Delta}^{gen}\,,
\ee
and
\be\label{agen0}
A^{gen}_{\Delta,0;\mu}= l_\mu \varphi_{\Delta}^{gen,1}+ n_\mu \varphi_{\Delta}^{gen,2}\,,
\ee
while the opposite spin vectors are obtained by $m_\mu \mapsto \bar{m}_\mu$.
In the following we will focus on the analytic form~\eqref{vargen} of the generalized scalar primaries, $\varphi^{gen}_\Delta=f(X^2)\varphi^\Delta$ and $\varphi_{\Delta}^{gen,i} \equiv f_i(X^2) \varphi^{\Delta}$ for a priori arbitrary functions $f$ and $f_i$, and list in table~\ref{table:Aconst} various constraints one can impose on the form of these functions.
\begin{table}[hb!]
\renewcommand*{\arraystretch}{1.3}
\centering
    \begin{tabular}{l|l|l|l}
         & $X^\mu A_\mu$ &  $\nabla^\mu A_\mu$ & $\Box A_\mu$ \\   
         \hline
        $A^{gen}_{\Delta,+1}$& 0 & 0 &$4[(2-\Delta)f'+X^2f'']m_\mu\varphi^\Delta$\\
      $A^{gen}_{\Delta,0}$& $[-f_1+\frac{X^2}{2}f_2]\varphi^\Delta$ & $[-2f'_1+(3-\Delta)f_2+X^2f'_2]\varphi^\Delta$&$\Big\{4[(1-\Delta)f_1'+X^2f_1''+\frac{1}{2}f_2]l_\mu$\\
        &&&~~~$+4[(3-\Delta)f_2'+X^2f_2'']n_\mu\Big\}\varphi^\Delta$\\
    \end{tabular}
    \caption{Various constraints one can impose on a generalized conformal primary vector.}
    \label{table:Aconst}
\end{table}

Enforcing all of these conditions simultaneously and demanding that $\varphi^{gen,i}_{\Delta}$ are analytic, lands us on the following spectra:
\be\label{rad1}
\Delta\in \mathbb{C},~~~J=\pm1,
\ee
corresponding to the radiative solutions for which $\Delta\in 1+i\mathbb{R}$ is a basis, as well as a discrete set of solutions with
\be\label{B1}
\Delta=2,~~~J=0.
\ee
There are two such solutions, a $\Gold$ and $\Memo$ mode, at each of these $(\Delta,J)$. For~\eqref{rad1} the two solutions for fixed~$\Delta$ are radiative spin-1 conformal primary wavefunctions~\eqref{CPWs} and their shadows~\eqref{SHCPWs}. The $\Gold$ and $\Memo$ wavefunctions for~\eqref{B1} will correspond to primary descendants at the bottom corner of the celestial photon diamonds discussed in section~\ref{sec:prim_desc}.

\subsubsection*{Generalized Gravitino}

For a left-handed generalized conformal primary gravitino $D_\frac{3}{2}(\Lambda)=M\otimes\Lambda$ is the spinor-vector representation where $M$ is the  $(\frac{1}{2},0)$ representation of the Lorentz algebra. We can construct such wavefunctions for half-integer spins $|J|\leq s=\frac{3}{2}$. These are given by:
\be
\chi^{gen}_{\Delta,+\frac{3}{2};\mu}=m_\mu o\varphi_{\Delta}^{gen}\,,
\ee
\be\label{genp12}
\chi^{gen}_{\Delta,+\frac{1}{2};\mu}=l_\mu o\varphi_{\Delta}^{gen,1}+n_\mu o\varphi_{\Delta}^{gen,2}+m_\mu \iota\varphi_{\Delta}^{gen,3}\,,
\ee
\be\label{genn12}
\chi^{gen}_{\Delta,-\frac{1}{2};\mu}=l_\mu \iota\varphi_{\Delta}^{gen,1}+n_\mu \iota\varphi_{\Delta}^{gen,2}+\bar{m}_\mu o\varphi_{\Delta}^{gen,3}\,,
\ee
\be
\chi^{gen}_{\Delta,-\frac{3}{2};\mu}=\bar{m}_\mu\iota\varphi_{\Delta}^{gen}\,.
\ee
For analytic generalized scalar primaries, $\varphi_{\Delta}^{gen} \equiv f(X^2) \varphi^{\Delta} $ and $\varphi_{\Delta}^{gen,i} \equiv f_i(X^2) \varphi^{\Delta} $,
table~\ref{gravitino} gives the radial and harmonic gauge constraints one can impose on the generalized primary gravitinos, while 
the Rarita-Schwinger equations reduce to 
\begin{table}[ht]
\renewcommand*{\arraystretch}{1.3}
\centering\scalebox{0.9}{
    \begin{tabular}{l|l|l|l}
         & $X^\mu \chi_\mu$ &  $\nabla^\mu \chi_\mu$ &$\bar{\sigma}^\mu\chi_\mu$\\
         \hline
        $\chi^{gen}_{\Delta,+\frac{3}{2}}$& 0 & 0 &0\\ 
      $\chi^{gen}_{\Delta,+\frac{1}{2}}$& $[-f_1+\frac{X^2}{2}f_2]o\varphi^\Delta$ & $[-2f_1'+(\frac{5}{2}-\Delta)f_2+X^2 f'_2+f_3]o\varphi^\Delta$ & $[f_3-f_2]\sqrt{2}\bar{\iota}\varphi^\Delta$ \\
       $\chi^{gen}_{\Delta,-\frac{1}{2}}$& $[-f_1+\frac{X^2}{2}f_2]\iota\varphi^\Delta$& 
       $[-2f_1'+(\frac{7}{2}-\Delta)f_2+X^2 f'_2]\iota\varphi^\Delta$ &
       $[f_1-f_3]\sqrt{2}\bar{o}\varphi^\Delta$\\
        $\chi^{gen}_{\Delta,-\frac{3}{2}}$& 0 & 0 &0\\ 
    \end{tabular}}
    \caption{Various constraints one can impose on a generalized conformal primary gravitino.}
    \label{gravitino}
\end{table}
\be\badat{3}
\varepsilon^{\mu\nu\rho\kappa}\bar{\sigma}_\nu\nabla_\rho \chi^{gen}_{\Delta,+\frac{3}{2};\kappa}&=-2\sqrt{2}if'm^\mu\bar{\iota}\varphi^\Delta\,,\\
\varepsilon^{\mu\nu\rho\kappa}\bar{\sigma}_\nu\nabla_\rho \chi^{gen}_{\Delta,+\frac{1}{2};\kappa}&=-\sqrt{2}i [2f'_1+({\textstyle\frac{1}{2}}-\Delta)f_2+X^2f'_2-({\textstyle \frac{3}{2}}-\Delta)f_3-X^2 f'_3]m^\mu \bar{o}\varphi^\Delta\\
&~~~-\sqrt{2}i[({\textstyle \frac{5}{2}}-\Delta)f_3+X^2f'_3]l^\mu\bar{\iota}\varphi^\Delta-2\sqrt{2}if'_3 n^\mu\bar{\iota}\varphi^\Delta\,,\\
\varepsilon^{\mu\nu\rho\kappa}\bar{\sigma}_\nu\nabla_\rho \chi^{gen}_{\Delta,-\frac{1}{2};\kappa}&=-\sqrt{2}i[f_1+({\textstyle \frac{1}{2}}-\Delta)f_3+X^2f'_3]l^\mu\bar{o}\varphi^\Delta+\sqrt{2}i[f_2-2f'_3]n^\mu\bar{o}\varphi^\Delta\\
&~~~-\sqrt{2}i[2f'_1+({\textstyle \frac{5}{2}} -\Delta)f_2+X^2f'_2-2f'_3]\bar{m}^\mu\bar{\iota}\varphi^\Delta\,,\\
\varepsilon^{\mu\nu\rho\kappa}\bar{\sigma}_\nu\nabla_\rho \chi^{gen}_{\Delta,-\frac{3}{2};\kappa}&=-\sqrt{2}i[({\textstyle \frac{3}{2}}-\Delta)f+X^2f']\bar{m}^\mu\bar{o}\varphi^\Delta.\\
\eadat\ee

If we enforce all of these conditions simultaneously and demand that $\varphi^{gen,i}_{\Delta}$ are analytic, we land on the following spectra:
\be\label{rad32}
\Delta\in \mathbb{C},~~~J=\pm\frac{3}{2},
\ee
corresponding to the radiative solutions for which $\Delta\in 1+i\mathbb{R}$ is a basis, as well as a discrete set of solutions with
\be\label{B32}
\Delta=\frac{5}{2},~~~J=\pm\frac{1}{2}.
\ee
There is one such left-handed solution at each of these $(\Delta,J)$. For~\eqref{rad32} these correspond to the left-handed radiative spin-$\frac{3}{2}$ conformal primary wavefunction~\eqref{CPWs} and its shadow~\eqref{SHCPWs}, while the $\Gold$ and $\Memo$ wavefunctions for~\eqref{B32} will correspond to primary descendants at the bottom corner of the celestial gravitino diamonds discussed in section~\ref{sec:prim_desc}.

\subsubsection*{Generalized Metric}
For a generalized primary metric $D_2(\Lambda)=\Lambda\otimes\Lambda$. We will also demand it is is symmetric under exchange of the $3+1$D indices. We can construct such rank-two tensors for integer spins $|J|\leq s=2$.  These are given by\cite{Pasterski:2020pdk}:
\be\label{hgen2}
h^{gen}_{\Delta,+2;\mu\nu}=m_\mu m_\nu \varphi_{\Delta}^{gen}\,,
\ee
\be\label{hgen1}
h^{gen}_{\Delta,+1;\mu\nu}=(l_\mu m_\nu+m_\mu l_\nu) \varphi_{\Delta}^{gen,1}+(n_\mu m_\nu+m_\mu n_\nu) \varphi_{\Delta}^{gen,2}\,,
\ee
\be\label{hgen0}
h^{gen}_{\Delta,0;\mu\nu}= l_\mu l_\nu \varphi_{\Delta}^{gen,1}+ n_\mu n_\nu \varphi_{\Delta}^{gen,2}+(l_\mu n_\nu+n_\mu l_\nu) \varphi_{\Delta}^{gen,3}+\eta_{\mu\nu}\varphi_{\Delta}^{gen,4}\,,
\ee
while the opposite spin metrics are obtained by $m_\mu \mapsto \bar{m}_\mu$.
Focusing again on the analytic form~\eqref{vargen} of the generalized scalar primaries, $\varphi_{\Delta}^{gen} \equiv f(X^2) \varphi^{\Delta} $ and $\varphi_{\Delta}^{gen,i} \equiv f_i(X^2) \varphi^{\Delta} $, we list in table~\ref{hconst} various gauge constraints one can impose  on the generalized primary metrics.
\begin{table}[hb!]
\renewcommand*{\arraystretch}{1.3}
    \centering
    \scalebox{0.9}{
    \begin{tabular}{l|l|l|l}
         & $\eta^{\mu\nu}h_{\mu\nu}$& $X^\mu h_{\mu\nu}$ &  $\nabla^\mu h_{\mu\nu}$ \\
         \hline
          $h^{gen}_{\Delta,+2}$&0& 0 & 0 \\
        $h^{gen}_{\Delta,+1}$&0&$[-f_1+\frac{X^2}{2}f_2]m_\nu\varphi^\Delta$  &  $[-2f'_1+(4-\Delta)f_2+X^2 f'_2] m_{\nu}\varphi^\Delta$\\
        $h^{gen}_{\Delta,0}$&$[-2f_3+4 f_4]\varphi^\Delta$&$\Big\{[-f_1+\frac{X^2}{2}(f_3-f_4)] l_\nu$&$\Big\{[-2f'_1+(2-\Delta)f_3+X^2 (f'_3-f'_4)+\Delta f_4] l_\nu $\\
        &&~~$+[\frac{X^2}{2}f_2-(f_3-f_4)] n_\nu\Big\}\varphi^\Delta$&$~~+[(4-\Delta)f_2+X^2 f'_2-2(f'_3-f'_4)]n_\nu\Big\}\varphi^\Delta $\\
    \end{tabular}
    }
    \caption{Various constraints one can impose on a generalized conformal primary metric.}
    \label{hconst}
\end{table}
The equations of motion can be evaluated using table~\ref{hconst} and
\begin{equation}
\label{hgenbox}
\scalemath{0.88}{
  \badat{4}
\Box h^{gen}_{\Delta,+2;\mu\nu}&=4[(2-\Delta)f'+X^2 f'']m_\mu m_\nu \varphi^{\Delta}\,,\\
\Box h^{gen}_{\Delta,+1;\mu\nu}&=4[(1-\Delta)f'_1+X^2 f''_1+f_2](l_\mu m_\nu+m_\mu l_\nu) \varphi^{\Delta}+4[(3-\Delta)f'_2+X^2 f''_2](n_\mu m_\nu+m_\mu n_\nu) \varphi^{\Delta}\,,\\
\Box h^{gen}_{\Delta,+0;\mu\nu}&=4[-\Delta f'_1+X^2 f''_1+f_3]l_\mu l_\nu \varphi^{\Delta}+4[(4-\Delta) f'_2+X^2 f''_2]n_\mu n_\nu \varphi^{\Delta}\\
&~~~+4[f_2+(2-\Delta) f'_3+X^2 f''_3](l_\mu n_\nu+n_\mu l_\nu) \varphi^{\Delta}+4[{\textstyle \frac{1}{2}}f_2+(2-\Delta) f'_4+X^2 f''_4]\eta_{\mu\nu} \varphi^{\Delta}\,.
\eadat
}
\end{equation}
Enforcing all of these conditions simultaneously and demanding that $\varphi^{gen,i}_{\Delta}$ are analytic, lands us on the following spectra:
\be\label{rad2}
\Delta\in \mathbb{C},~~~J=\pm2,
\ee
corresponding to the radiative solutions for which $\Delta\in 1+i\mathbb{R}$ is a basis, as well as a discrete set of solutions with
\be\label{B2}
\Delta=3,~~~J=\{-1,0,1\}.
\ee
There are two such solutions at each of these $(\Delta,J)$.  For ~\eqref{rad2} the two solutions for fixed~$\Delta$ are the radiative spin-2 conformal primary wavefunctions~\eqref{CPWs} and their shadows~\eqref{SHCPWs}, while the $\Gold$ and $\Memo$ wavefunctions for~\eqref{B2} will correspond to primary descendants at the bottom corners of the celestial graviton diamonds discussed in section~\ref{sec:prim_desc}.

\section{Global Primary Descendants in 2D CFT}
\label{sec:global_primary_desc}

In this section we classify global primary descendant operators for generic 2D CFTs. We will then apply and adapt this classification to 4D wavefunctions transforming as 2D conformal primaries in section~\ref{sec:prim_desc}.

\subsubsection*{Conditions for Primary Descendants}

Consider an SL(2,$\mathbb{C}$) primary state $|h,\bar h \rangle$ which, by definition, is annihilated by both $L_1$ and $\bar{L}_1$. Often in the $d=2$ CFT literature such a state would be called quasi-primary. Its SL(2,$\mathbb{C}$)-descendants are built by acting with $L_{-1}$ and $\bar{L}_{-1}$ an arbitrary number of times.  A primary descendant is a state in this module that is also annihilated by  $L_{1}$ and $\bar{L}_{1}$. Since the algebra is divided in two sets of mutually commuting generators, we can examine the holomorphic and antiholomorphic descendants separately. Let us focus on states of the form $(L_{-1})^\n |h,\bar h \rangle $. 
Using 
\begin{eqnarray}
    [ L_{1}, (L_{-1})^\n]
     &=\n (L_{-1})^{\n-1} (2L_0+\n-1) \ ,
\end{eqnarray}
we obtain
\begin{equation}
\label{condition:primary_descendants}
    L_{1} (L_{-1})^\n |h,\bar h \rangle = \n (2h+\n-1) (L_{-1})^{\n-1} |h,\bar
    h \rangle \, .
\end{equation}
Here we see that $(L_{-1})^\n |h,\bar h \rangle$ is a primary descendant when $h=\frac{1-\n}{2}$ for $\n \in \mathbb{Z}_>$.  The primary descendant has dimension $h'=\frac{1+\n}{2}$, which corresponds to a reflection of the weight 
\be
h\to 1-h.
\ee
Repeating the same computation for the antiholomorphic algebra, one finds that
$({\bar L}_{-1})^{\bar \n} |h,\bar h \rangle$ is a primary descendant when ${\bar h}=\frac{1- \bar \n}{2}$ for $\bar \n \in \mathbb{Z}_>$. This descendant has dimension ${\bar h}'=\frac{1+{\bar \n}}{2}$. 

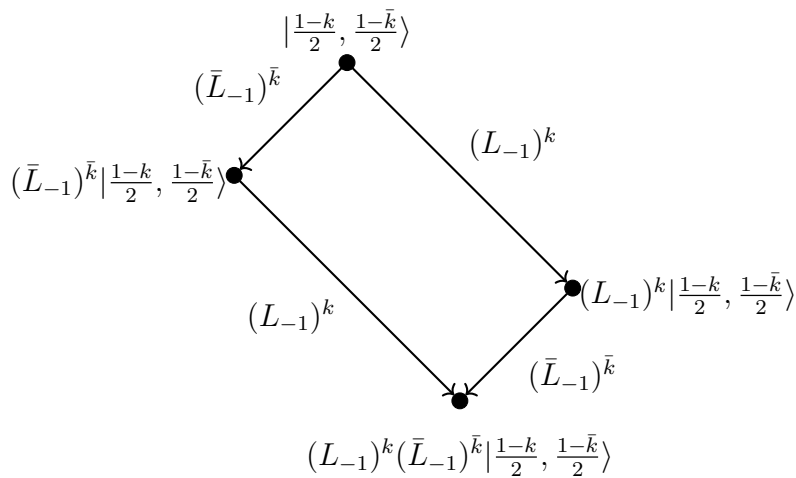
\begin{figure}[t]
\centering
\vspace{-0.5em}
\begin{tikzpicture}[scale=1.5]
\definecolor{darkgreen}{rgb}{.0, 0.5, .1};
\draw[thick](0,2)node[above]{$| \frac{1-\n}{2}, \frac{1-\bar \n}{2}\rangle $} ;
\draw[thick,->](0,2)--node[above left]{$(\bar L_{-1})^{ \bar \n}$} (-1+.05,1+.05);
\draw[thick,->](0,2)--node[above right]{$( L_{-1})^{  \n}$} (2-.05,0+.05);
\draw[thick,->] (-1+.05,1-.05)node[left]{$ (\bar L_{-1})^{\bar \n} | \frac{1-\n}{2}, \frac{1-\bar \n}{2}\rangle $} --node[below left]{$( L_{-1})^{ \n}$} (1-.05,-1+.05) ;
\draw[thick,->] (2-.05,0-.05)node[right]{$ ( L_{-1})^{ \n} | \frac{1-\n}{2}, \frac{1-\bar \n}{2}\rangle $} --node[below right]{$(\bar L_{-1})^{ \bar \n}$} (1.05,-1+.05) ;
\draw[thick](1,-1.2)node[below]{$(L_{-1})^{\n} (\bar L_{-1})^{\bar \n} | \frac{1-\n}{2}, \frac{1-\bar \n}{2}\rangle $};
\filldraw[black] (0,2) circle (2pt) ;
\filldraw[black] (2,0) circle (2pt) ;
\filldraw[black] (-1,1) circle (2pt) ;
\filldraw[black] (1,-1) circle (2pt) ;
\end{tikzpicture}
\caption{Diamond illustrating the nested submodule structure.  The SL(2,$\mathbb{C}$) module of the primary $|\frac{1-\n}{2},\frac{1-\bar \n}{2}\rangle$ contains three submodules associated with primary descendant operators. }
\label{Nested_submodules}
\end{figure}

Let us now consider the case when both of these conditions are satisfied, namely
\be
\label{condition_diamond_hhb}
h=\frac{1-\n}{2}
\, , \qquad\qquad
\bar{h}=\frac{1-\bar{\n}}{2}\, ,
\ee
where $\n,\bar{\n}\in\mathbb{Z}_{>}$.  We then have a third primary descendant:
\be\label{pdec}
(L_{-1})^\n({\bar L}_{-1})^{\bar \n} |h,\bar h \rangle
\ee
with conformal weights $(h',\bar{h}')=\frac{1}{2}(1+\n,1+\bar{\n})$. 
As summarized in figure~\ref{Nested_submodules},  the two submodules intersect at the position of this primary descendant~\eqref{pdec} forming a nested structure that we shall refer to as `diamond'.

\subsubsection*{Classification of Primary Descendants}

Following the notations\footnote{Let us draw attention to some differences between our definitions and those of~\cite{Penedones:2015aga}. In section 3.1 of~\cite{Penedones:2015aga} states of types I, II, III were also defined (and used e.g. in \cite{Banerjee:2019aoy}). Those definitions differ slightly because they were considering primaries in the symmetric and traceless representation of $SO(d)$ for CFT in $d$ dimensions. That classification was strictly speaking inherited from odd $d$ and analytically continued to generic (even non integer) $d$. In even $d$  subtleties arise. Section 6 of~\cite{Penedones:2015aga} introduced a type V primary descendant which exists only in even $d$ and replaces the type III defined in odd $d$. What we call type III in the following is actually the type V of \cite{Penedones:2015aga}. Moreover the split of type I into the type Ia and type Ib defined here is a feature of even dimensions.
}
of~\cite{Penedones:2015aga}, we classify primary descendants into three types I, II, III which we define below.
It will be convenient to rephrase the above conditions for primary descendants in terms of $\Delta=h+\bar h$ and $J=h-\bar h$.
In what follows $\O_{\Delta,\pm|J|}(w,\bw)$ is a primary operator with dimension $\Delta$ and spin $\pm|J|$. Its primary descendants have  dimensions $\Delta'$ and spin $J'$. The types of primary descendant operators are summarized in table~\ref{table:types}.
\vspace{1em}
 
\begin{table}[ht!]
\renewcommand*{\arraystretch}{1.3}
\centering
\begin{tabular}{c|c|c|c|c}
Type& range & level & $\Delta$ &  $(\Delta',|J'|)$  \\
\hline
Ia&$n\in \mathbb{Z}_>$ & $n$ &$1-|J|-n$ & $(1-|J|,|J|+n)$  \\
Ib&$n\in \mathbb{Z}_>$ & $2|J|+n$ &$1-|J|-n$ & $(1+|J|,|J|+n)$ \\
II &$n \in [1, 2|J|-1]$& $n$ &$1+|J|-n$ & $(1+|J|,|J|-n)$\\
III & $-$ & $2|J|$ & $1-|J|$ & $(1+|J|,|J|)$  \\
\end{tabular}
\caption{
Types of primary descendant operators. 
A spin $J$ primary operator with conformal dimension $\Delta$ has a descendant which is also a primary in four circumstances. The level of the descendant, its dimension $\Delta'$ and the absolute value of its spin $J'$ are presented in the table. }
 \label{table:types}
\end{table}

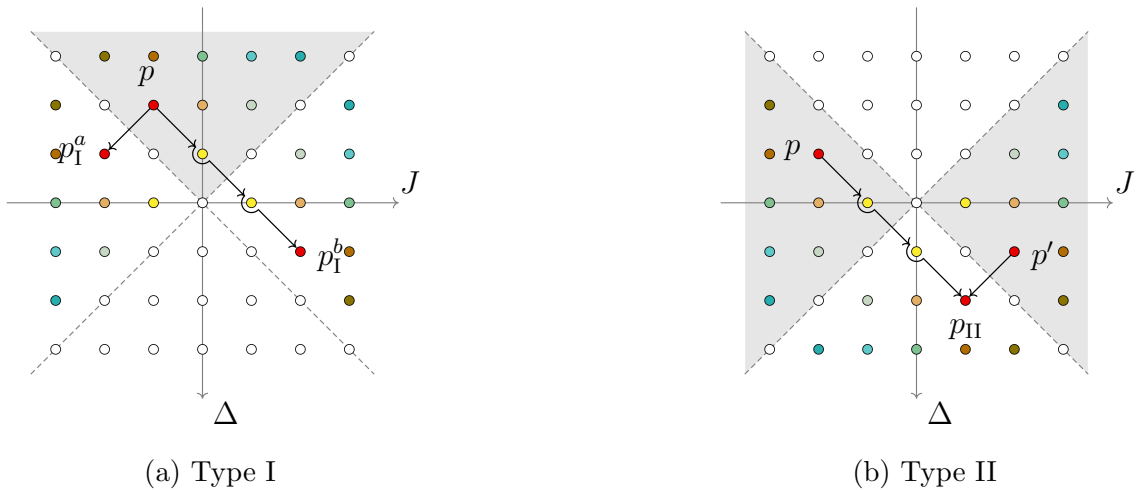
\begin{figure}[t!]
\begin{subfigure}[b]{0.45 \textwidth}
\centering
 \begin{tikzpicture}[scale=0.7]
\pgftext{	\includegraphics[width=1\linewidth]{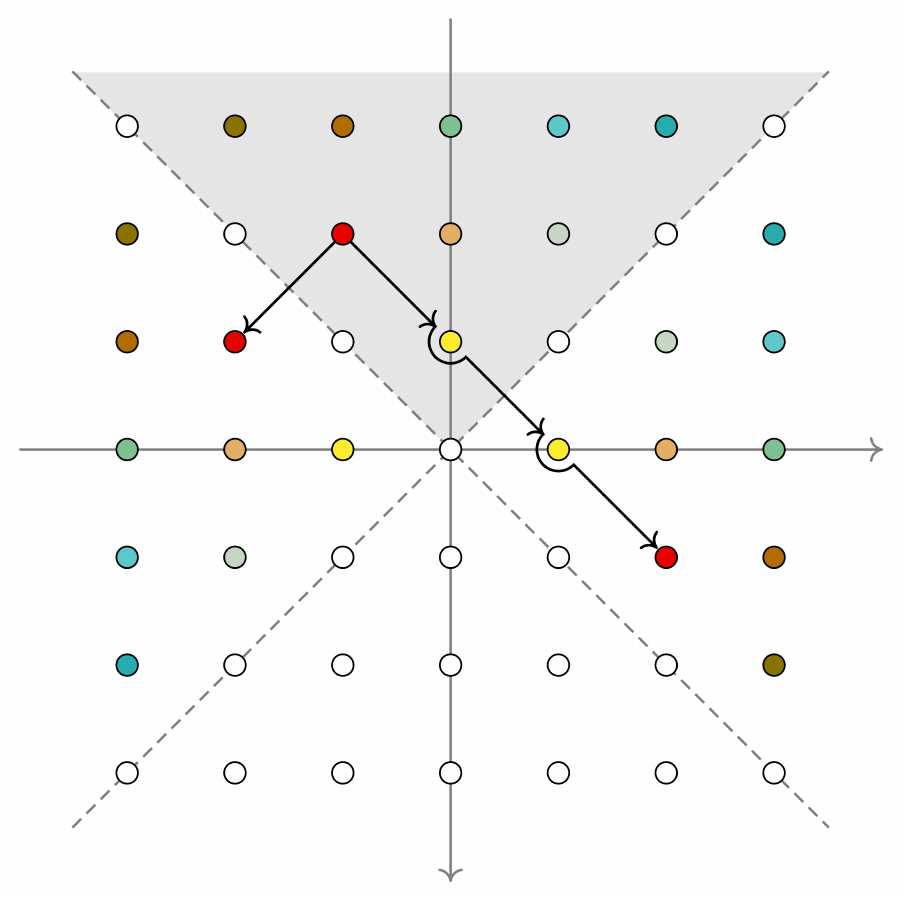}} at (0pt,0pt)
\draw[thick](0.5,-4.)node[]{$\Delta$} ;
\draw[thick](4,0.4)node[]{$J$} ;
\draw[thick](-1,2.4)node[]{$p$} ;
\draw[thick](-2.4,1)node[]{$p_{\textrm{I}}^{a}$} ;
\draw[thick](2.5,-1)node[]{$p_{\textrm{I}}^{b}$} ;
 \end{tikzpicture}
\caption{Type I
}
\label{TipeI_primdesc}
\end{subfigure}
\hfill
\begin{subfigure}[b]{0.45 \textwidth}
\centering
 \begin{tikzpicture}[scale=0.7]
\pgftext{	\includegraphics[width=1\linewidth]{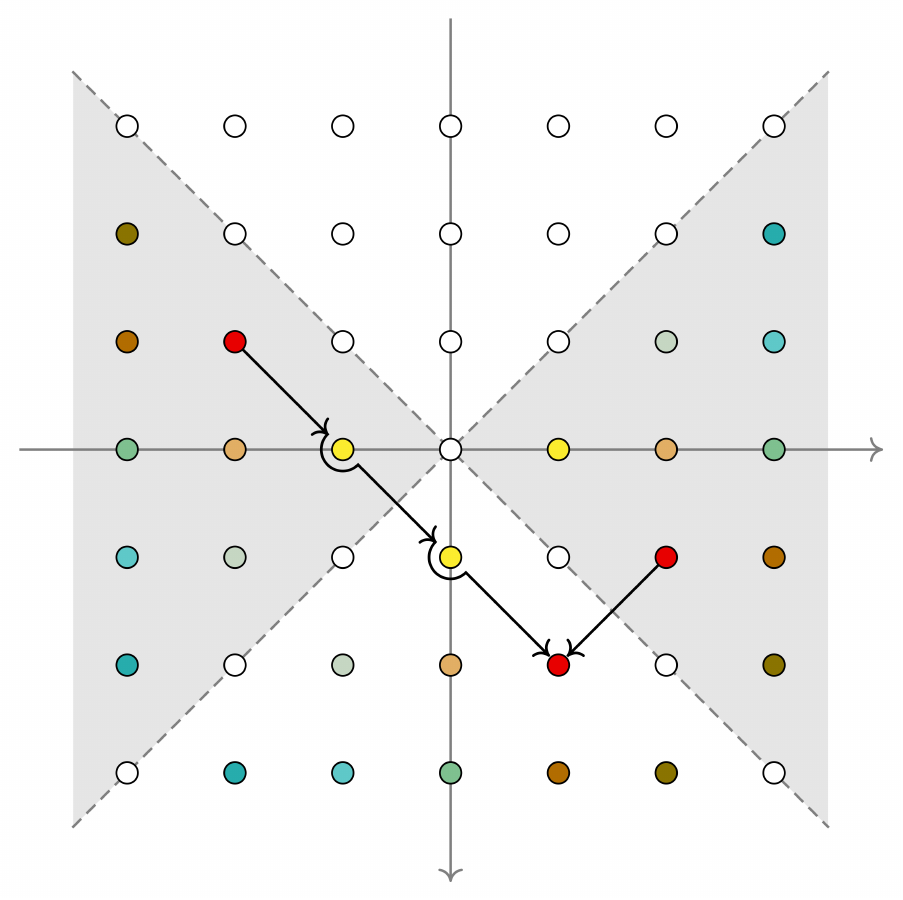}} at (0pt,0pt)
\draw[thick](0.5,-4.)node[]{$\Delta$} ;
\draw[thick](4,0.4)node[]{$J$} ;
\draw[thick](-2.3,1)node[]{$p$} ;
\draw[thick](2.45,-1)node[]{$p'$} ;
\draw[thick](1,-2.4)node[]{$p_{\textrm{II}}$} ;
 \end{tikzpicture}
\caption{Type II
}
\label{TipeII_primdesc}
\end{subfigure}
\caption{Primary descendants of type I and II. The dots define the integer lattice $\mathbb{Z}^{2}$ in the $J$-$\Delta$ plane. The origin is at $(\Delta,J)= (1,0)$.
Dots in the grey regions correspond to primaries that have primary descendants. The latter are obtained by reflecting across the dashed lines.
\label{Tipes_primdescI&II}
}
\end{figure}

\noindent{\bf Definition:} A {\it type~I primary descendant} is a primary descendant with  spin greater --~in absolute value~-- than that of the primary it descended from, namely $|J'|>|J|$. We further divided the type I into two sub-types Ia and Ib, where the former is a descendant at level~$n$, while the latter is a descendant at level~$2|J|+n$:
\be\label{DefTypeI}
\renewcommand{\arraystretch}{1.3}
\begin{array}{lll}
{\bf Ia} &\quad\partial_w^{n}{\cal O}_{\Delta,+|J|}, & \partial_\bw^{n}{\cal O}_{\Delta,-|J|}\\
{\bf Ib} &\quad \partial_w^{2|J|+n}{\cal O}_{\Delta,-|J|}, & \partial_\bw^{2|J|+n}{\cal O}_{\Delta,+|J|}
\end{array}\quad \Delta=1-|J|-n\,.
\ee
When $J=0$ the definition above is ambiguous, so we shall define type Ia being obtained by holomorphic derivatives, and Ib by  antiholomorphic ones. The relation between the quantum numbers $\Delta,J$ of the primary and the ones $\Delta',J'$ of the primary descendant is summarized in table~\ref{table:types} and illustrated in figure~\ref{TipeI_primdesc}. 
\vspace{1em}

\noindent{\bf Definition:} A {\it type~II primary descendant} is a primary descendant with  spin smaller --~in absolute value~-- than that of the primary it descended from,  $|J'|<|J|$,
\be\label{DefTypeII}
\begin{array}{lll}
{\bf II}~~~ &\partial_\bw^{n}{\cal O}_{\Delta,+|J|},~~ \partial_w^{n}{\cal O}_{\Delta,-|J|} &\quad \Delta=1+|J|-n\,,
\end{array}
\ee
where $n\in\{1,...,2|J|-1\}$ which implies $1-|J|<\Delta<1+|J|$.  These primary descendants only exists for $|J|\ge1$. The relation between the quantum numbers $\Delta,J$ of the primary and the ones $\Delta',J'$ of the primary descendant is summarized in table~\ref{table:types} and illustrated in figure~\ref{TipeII_primdesc}.
\vspace{1em}

\noindent{\bf Definition:} A {\it type~III primary descendant} is a primary descendant with the same spin --~in absolute value~-- as that of the primary it descended from,  $|J'|=|J|>0$,
\be\label{DefTypeIII}
\begin{array}{lll}
{\bf III}~~~ &\partial_\bw^{2|J|}{\cal O}_{\Delta,+|J|},~~ \partial_w^{2|J|}{\cal O}_{\Delta,-|J|} &\quad \Delta=1-|J|\,.
\end{array}
\ee
The descendant has its conformal spin flipped $J'=-J$ with respect to the parent primary.  The relation between the quantum numbers $\Delta,J$ of the primary and the ones $\Delta',J'$ of the primary descendant is summarized in table~\ref{table:types} and illustrated in figure~\ref{Tipes_primdescIII}.
\vspace{1em}

In appendix~\ref{App:rep_theory} we review how to find the primary descendants using techniques from representation theory. There we will also show that primaries and primary descendants are related through special reflections in weight space (Weyl reflections), as depicted in figure~\ref{Tipes_primdescI&II} and~\ref{Tipes_primdescIII} for the case of integer spin.  For half integer spins, the $\mathbb{Z}^2$ lattice gets shifted.
Since descendants have higher conformal dimension than their associated parent primary, the relevant reflections map points at the top of the pictures to points at the bottom.

In figure~\ref{TipeI_primdesc} there are two possible reflections corresponding to type Ia and  type Ib. 
In figure~\ref{TipeII_primdesc} there are two disconnected grey regions depending on the sign of $J$. Points of the same color in the two grey regions reflect to the same point at the bottom.
We also notice that 
the primary descendants of type Ia and Ib of figure~\ref{TipeI_primdesc}  are exactly at the position of the primaries in figure~\ref{TipeII_primdesc}. So when the condition for type I is satisfied, the full diamond of figure~\ref{Nested_submodules} appears -- indeed the condition for type I is equivalent to~\eqref{condition_diamond_hhb}.

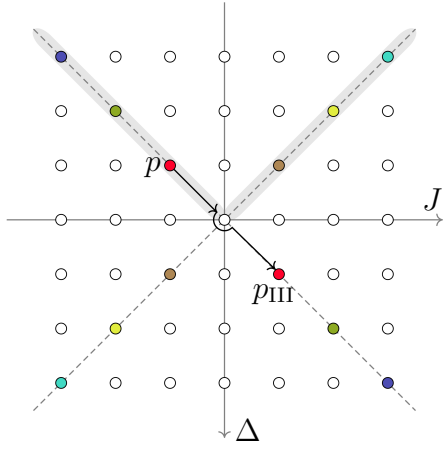
\begin{figure}[ht!]
\vspace{-.5em}
\centering
 \begin{tikzpicture}[scale=0.7]
\pgftext{	\includegraphics[width=.5\linewidth]{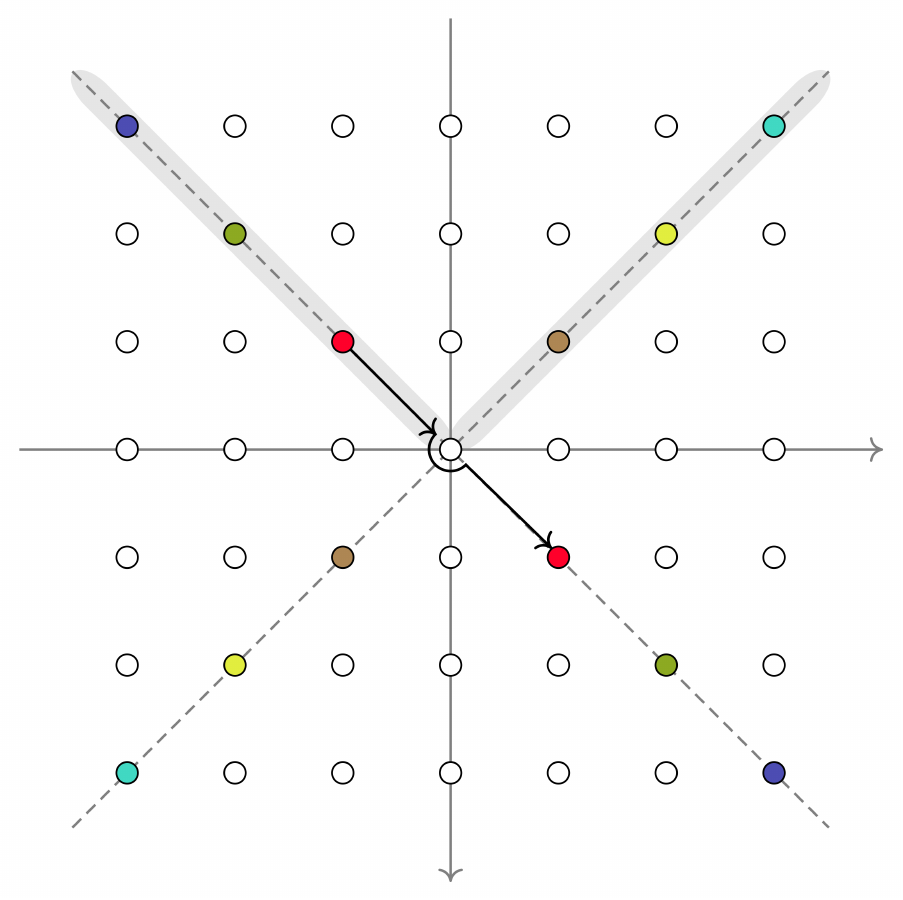}} at (0pt,0pt)
\draw[thick](0.5,-4.)node[]{$\Delta$} ;
\draw[thick](4,0.4)node[]{$J$} ;
\draw[thick](-1.3,1)node[]{$p$} ;
\draw[thick](1,-1.4)node[]{$p_{\textrm{III}}$} ;
 \end{tikzpicture}
\caption{Primary descendants of type III. The dots define the integer lattice $\mathbb{Z}^{2}$ in the $J$-$\Delta$ plane. The origin of the lattice is placed at $(\Delta,J)= (1,0)$.
\label{Tipes_primdescIII}
}
\end{figure}

In figure~\ref{Tipes_primdescIII} 
the $(\Delta,J)$ of the primaries are located on the dashed diagonals and their primary descendants $(\Delta',J')$ are obtained as a reflection with respect to the opposite diagonal. Notice that the condition for type III can be also written as $h=\frac{1-\n}{2}$, $\bar h=\frac{1}{2}$ or $\bar h=\frac{1-\bar \n}{2}$, $ h=\frac{1}{2}$. This is in turn equal to ~\eqref{condition_diamond_hhb} with $\n$ or $\bar \n$ equal to zero (even though $\n, \bar \n$ were defined to be in $\mathbb{Z}_>$).
Indeed, comparing the figures~\ref{Nested_submodules} and~\ref{Tipes_primdescIII}, one notices that type III may be formally obtained by shrinking to zero the area of the diamond, namely by taking $\n \bar{\n}=0$.

\subsubsection*{Shadows in the Diamond}

Looking at the weights appearing in the diamond in figure~\ref{Nested_submodules}, we see that antipodal points have shadow-related conformal dimensions and spins (indeed the shadow transform implements a Weyl reflection of the weights).
One natural question is if these primary descendants are related by a shadow transform,
\begin{equation}
\label{def:2dShadowTransform}
\badat{3}
 \widetilde{ \O_{h,\bar h}}(w,\bar w)= \frac{K_{h,\bar h}}{2\pi}  \int d^2w' \frac{ \O_{h,\bar h}(w',\bw')}{(w-w')^{2-2h}(\bw-\bw')^{2-2\bar{h}}}\, , \\
\eadat
\end{equation} 
where 
$K_{h,\bar h}=2\max\{h,\bar{h}\}-1$
is a normalization constant~\cite{Pasterski:2017kqt}\footnote{Note that this normalization differs from $K_{h,\bar h}=\frac{\Gamma(2-2 \bar h)}{ \Gamma(2 h-1)}$ used in~\cite{Osborn:2012vt}. 
} for which the shadow wavefunctions~\eqref{SHCPWs} take a nice form and
$\widetilde{ \widetilde{ \O_{h,\bar h}}}=(-1)^{2(h-\bar{h})} \O_{h,\bar h}$.
Indeed we can prove that, given an operator $\O_{\frac{1-\n}{2},\frac{1-\bar \n}{2}}$ at the top vertex of the diamond of figure~\ref{Nested_submodules}, then its primary descendants of type Ia and Ib are related by a shadow transform as follows, 
\begin{equation}
    \frac{1}{\bar \n!} \widetilde{\left(\p_{\bw}^{\bar \n} \O_{\frac{1-\n}{2},\frac{1-\bar \n}{2}}\right)}=  \frac{(-1)^{ \n}}{\n!} \p_w^\n \O_{\frac{1-\n}{2},\frac{1-\bar \n}{2}} \, , \qquad 
     \frac{1}{\n!} \widetilde{\left(\p_{w}^{ \n} \O_{\frac{1-\n}{2},\frac{1-\bar \n}{2}}\right)}=\frac{(-1)^{\bar \n}}{\bar \n!} 
    \p_{\wb}^{\bar \n} \O_{\frac{1-\n}{2},\frac{1-\bar \n}{2}} \, .
\end{equation}
To obtain these relations we integrate by parts within the shadow integrand~\eqref{def:2dShadowTransform} and use 
\begin{equation}
\label{shadow_Ia_Ib}
 \p_{\bw'}^{\bar \n} \frac{(\bw'-\bw)^{\bar{\n}-1}}{ (w'-w)^{\n+1}} 
   =2\pi (\bar{\n}-1)! \frac{(-1)^{\n}}{\n!} \p_{w'}^\n\delta^{(2)}(w'-w) \,,
\end{equation}
which follows from $\p_z \frac{1}{\bar{z}}=2\pi \delta^{(2)}(z)$. With a similar demonstration one can prove that if two primaries (not necessarily primary  descendants) of the left and right corner of figure \ref{TipeII_primdesc} are related by shadow transform, then they must descend to the same primary descendant,\footnote{While the weights of the top and bottom corners seem to satisfy $(h,\bh)\mapsto(1-h,1-\bh)$, the operators are not related by a conformal shadow transformation. This is not surprising 
as descendants typically lose information. 
} namely 
\begin{equation}
\label{shadow_II}
   \frac{1}{\bar \n!} \p_{\bw}^{\bar \n} \left( \widetilde{ \O_{\frac{1-\n}{2},\frac{1+\bar \n}{2}}} \right)=  \frac{(-1)^{ \n}}{\n!} \p_{w}^{\n}    \O_{\frac{1-\n}{2},\frac{1+\bar \n}{2}} \, , \qquad 
      \frac{1}{ \n!} \p_{w}^{\n} \left( \widetilde{
      \O_{\frac{1+\n}{2},\frac{1-\bar \n}{2}}
      } \right)=  \frac{(-1)^{\bar \n}}{\bar \n!} \p_{\wb}^{\bar \n}   \O_{\frac{1+\n}{2},\frac{1-\bar \n}{2}} \, .
\end{equation}
For type III the same shadow relations \eqref{shadow_Ia_Ib} and \eqref{shadow_II} hold (and degenerate to the same relation) by setting either $\n=0$ or $\bar{\n}=0$.

\subsubsection*{Primary Descendants vs Null States}
\label{sec:quant}

The discussion so far only relied on the SL(2,$\mathbb{C}$) algebra, and not on the prescription for the dual Hilbert space. Let us now look at how different quantization choices  affect whether or not primary descendants are null states.  We will contrast the standard 2D CFT prescription of radial quantization to the one inherited by celestial CFTs from the 4D bulk. 

In radial quantization of 2D Euclidean CFTs, the Hilbert space is spanned by states 
\be |h, \bar{h}\rangle\equiv \O_{h,\bar{h}}(0) |0\rangle
\ee
in one to one correspondence with primary operators.  These states are annihilated by $L_1$ and $\bar{L}_1$. Conformal multiplets are obtained by acting with $L_{-1}$ and  $\bar{L}_{-1}$. Conjugation in radial quantization is implemented by inversion \be[\O(w,\bw)]^\dagger=w^{-2\bar{h}}\bw^{-2h}\O(1/\bw,1/w) \ee
and the \emph{out} states are defined as 
\be\label{eq:radout}\langle h, \bar{h}| \equiv \lim_{w,\bw\to 0}w^{-2\bar{h}}\bw^{-2h} \langle 0|\O(1/\bw,1/w). \ee
Using the definition of \emph{in} and \emph{out} states we can then compute  norms as $\langle h, \bar{h}|h, \bar{h}\rangle $.
In reasonably well behaved CFTs (e.g. unitary and with discrete spectrum) one can choose a basis of primary states which are unit-normalized.
The prescription for radial quantization implies the following Hermitian conjugation relations
\be
\label{radial_Conjugation}
 L_{n}^\dagger={L}_{-n} \, ,
 \qquad
 \bar  L_{n}^\dagger=\bar {L}_{-n} \, .
\ee
The rules above can then be used to demonstrate that a primary descendant is orthogonal to all primaries including itself. E.g. if we consider a primary $|h_{2},\bh_2\rangle$ with $h_2=\frac{1-k}{2}$ and we compute the inner product of its level-$k$ primary descendant with a generic primary $\langle h_1,\bh_1|$, we obtain,
\be 
\label{nullness_primdesc}
\langle h_1,\bh_1|\Big(L_{-1}^\n|h_{2},\bh_2\rangle\Big)=\Big(\langle h_1,\bh_1|{ L}_{1}^\dagger\Big)L_{-1}^{\n-1}|h_{2},\bh_2\rangle=0\,.
\ee
The same equation for generic $h_2$ implies arbitrary descendants are orthogonal to any primary.

Meanwhile \emph{in} and \emph{out}-states in celestial CFTs are defined in terms of data at early and late Cauchy slices in the 4D bulk. Single particle states can be defined as 
\begin{equation}
  |h,\bh,w,\bw \rangle=  \O^{s,+}_{\Delta,J}(w,\bw)|0\rangle \, , 
  \qquad 
  \langle h,\bh, w,\bw|=  \langle 0|  \O^{s,-}_{\Delta^{*},-J}(w,\bw) \, ,
\end{equation}
where $\O^{s,+}_{\Delta,J}(w,\bw)$ was defined in \eqref{qdelta}.  These states depend on a point on the sphere $w,\bw$ in addition to the conformal dimension and spin. Using \eqref{commutation_modes}, their inner product takes the form
\begin{equation}
\label{celestial_norm}
  \langle h,\bh, w,\bw 
  |h',\bh',w',\bw' \rangle 
  =\delta_{JJ'}\delta^{(2)}(w-w')\boldsymbol{\delta}(i(\Delta+\Delta'^*-2)) \, .
\end{equation}
In appendix~\ref{sec:CelestialSL2C} we review the relation between performing Lorentz transformations of the spacetime coordinates and SL(2,$\mathbb{C}$) transformations of the reference direction $(w,\bw)$. 
Consider the single particle state with momentum pointing towards the north pole  $w,\bw=0$. The fact that massless states do not have continuous spin representations~\cite{Weinberg:1995mt,Banerjee:2018gce} means that that they are annihilated by $L_1, \bar{L}_1$
\begin{equation}
 L_1   |h,\bh,0,0 \rangle = 0 \, ,
 \qquad 
  \bar{L}_1   |h,\bh,0,0 \rangle = 0 \, .
\end{equation}
The generic $w,\bw$ is obtained by exponentiation of $L_{-1}$ and $\bar{L}_{-1}$, or equivalently defining primaries with respect to rotated generators.  The state $|h,\bh,0,0 \rangle $ is similar to the usual primary states in a 2D CFT: it depends only on the weights $h,\bh$, it is annihilated by  $L_1,\bar{L}_1$, the generators  $ L_0,\bar{L}_0$ read off its weights, and $L_{-1},\bar{L}_{-1}$ create descendants. However the inner product is not the same as the one in radial quantization.

In celestial CFTs we would like to keep the property that 4D scattering \emph{in}-states are conjugate to the \emph{out}-states.  Demanding that the representation of Poincar\'e on the bulk Hilbert space is unitary forces on us the following conjugation relations (see appendix \ref{sec:CelestialSL2C})
\be
L_{n}^\dagger=-\bar{L}_n
\ee
which differ from the ones obtained in radial quantization \eqref{radial_Conjugation}.
 In particular  the conjugate of  ${L}_{-1}$ is not $L_1$ but rather ${\bar L}_{-1}$. 
 Thus we cannot use an equation of the form \eqref{nullness_primdesc} to prove that primary descendant states are orthogonal to all primaries.
 Indeed
\be 
\langle h_1,\bh_1,w_1,\bar{w}_1|\Big(L_{-1}^\n|h_{2},\bh_2,w_2,\bar{w}_2\rangle\Big)=-\Big(\langle h_1,\bh_1,w_1,\bar{w}_1|{\bar L}_{-1}^\dagger\Big)L_{-1}^{\n-1}|h_{2},\bh_2,w_2,\bar{w}_2\rangle\,,
\ee
is non-vanishing even when we take $h_2=\frac{1-k}{2}$ and fix $w_2,\bar{w}_2=0$.
We thus see that, by using the most conventional form of the inner product, celestial primary descendants need not be null. Let us however stress that the classification of reducible modules and their shadow relations is independent of the inner product, thus it is valid no matter how we define it.

A final observation is that it is possible to mimic radial quantization also in celestial CFTs (see e.g. the discussion of~\cite{Fan:2021isc}).  Indeed, since the inner product \eqref{celestial_norm} is delta normalized, we can obtain a two-point function power-like behavior by taking a 2D shadow transform of the out states.\footnote{This idea has been floating around for a while, e.g.~\cite{ss,BP}. For example, the norm~\eqref{celestial_norm} of any given state is divergent but can be tamed using the shadow transform.} The standard form of the two point function automatically gives rise to inner products, as in radial quantization, by sending the points to zero and infinity.
In terms of this inner product one would thus conclude that the primary descendant states should actually be null. 
It is however still missing in the literature an argument which would suggest that the shadowed inner product is the natural one from the 4D perspective. We leave this question for future investigation.

\section{Celestial Diamonds}
\label{sec:prim_desc}

Making use of the general discussion of the previous section and the fact that all 4D wavefunctions relevant to massless scattering of spin-$s$ particles have $\{|J|,|J'|\}\le s$, we can now classify and construct all global conformal multiplets relevant in 2D celestial CFT -- namely, those for which radiative primaries appear in the SL(2,$\mathbb{C}$) module. These are organized into `celestial diamonds'. Radiative primaries can appear in two forms: either as the parent primary ($|J|=s$), or as a primary descendant ($|J'|=s$). We illustrate the two cases in tables~\ref{table:types1} and~\ref{table:types2}. 

\begin{table}[ht!]
\renewcommand*{\arraystretch}{1.3}
\centering
\begin{tabular}{c|c|c|c|c}
Type& range & level & $\Delta$ &  $(\Delta',|J'|)$  \\
\hline
Ia&$n\in \mathbb{Z}_>$ & $n$ &$1-s-n$ & $(1-s,s+n)$  \\
Ib&$n\in \mathbb{Z}_>$ & $2s+n$ &$1-s-n$ & $(1+s,s+n)$ \\
II &$n \in [1, 2s-1]$& $n$ &$1+s-n$ & $(1+s,s-n)$\\
III & $-$ & $2s$ & $1-s$ & $(1+s,s)$  \\
\end{tabular}
\caption{
Primary descendants of radiative fields with $|J|=s$.  We can only have non-zero wavefunctions if $|J'|\le s$, hence the type I descendants are trivially null.  We thus see that all of the nontrivial descendants of radiative primaries are at $\Delta'=1+s$.
}
 \label{table:types1}
\end{table}

Starting from a radiative conformal primary wavefunction with $(\Delta,J=\pm s)$, the abstract representation theoretic argument of the previous section determines the values of $(\Delta',J')$ at which a primary descendant occurs. Upon taking the appropriate $w$ and $\bw$ derivatives we are guaranteed to land on a wavefunction that satisfies the same equations of motion and gauge fixing as the parent radiative primary. (See appendix~\ref{sec:cpwdec} for the explicit form these descendants take.) 
Table~\ref{table:types1} tells us where a primary descendant will occur.

\begin{table}[ht!]
\renewcommand*{\arraystretch}{1.3}
\centering
\begin{tabular}{c|c|c|c|c}
Type& range & level & $(\Delta,J)$ &  $\Delta'$  \\
\hline
Ia&$n\in \mathbb{Z}_>$ & $n$ &$(1-s,s-n)$ & $1-s+n$  \\
Ib&$n\in \mathbb{Z}_>$ & $2s-n$ &$(1-s,s-n)$ & $1+s-n$ \\
III & $-$ & $2s$ & $(1-s,s)$ & $1+s$  \\
\end{tabular}
\caption{
Spectrum of primaries that can descend to radiative primary descendants with $|J'|=s$. We can only have non-zero wavefunctions if $|J|\le s$, which excludes the type II case.  We see that the only parents which descend to radiative primaries have $\Delta=1-s$.
}
 \label{table:types2}
\end{table}

For the case where a radiative primary is a primary descendent with $(\Delta',J'=\pm s)$, representation theory again determines the $(\Delta,J)$ of the parent primary as illustrated in table~\ref{table:types2}. However, now we are no longer guaranteed that the parent primary satisfies the same gauge fixing and equations of motion as the radiative primary (and in general they won't) and, moreover, we need to verify that such a parent wavefunction can indeed be constructed and transforms covariantly under SL(2,$\mathbb{C}$). Parents of radiative conformal primaries thus seem to correspond to states in celestial CFT that one has to add.

Let us summarize the group-theoretic predictions from tables~\ref{table:types1} and~\ref{table:types2} for the global conformal multiplets in 2D celestial CFT. For fixed spin~$s$ there are two distinct loci in the $(\Delta,J)$ plane where we expect to find SL(2,$\mathbb{C}$) primary descendants: 1) at $(\Delta,J)=(1-s-n,\pm s)$ for $n\in \mathbb{Z}_>$ we expect type I primary descendants which must be identically zero since $|J'|>s$; and 2) the square formed by the union of
$(\Delta,J)\in([1-s,...,1+s],\pm s)$ and $(\Delta,J)\in(1\pm s,[-s,...,+s])$ captures all non-vanishing primary descendants of type I,~II and~III, as well as their parents, and forms the corners of a set of celestial diamonds.  We will now examine these two cases in turn.

\subsection{Trivial Null States}
\label{sec:NullStates}

In celestial CFT, primary descendant wavefunctions of type I arising from radiative conformal primaries $\Phi_{\Delta,J}$ and $\tPhi_{\Delta,J}$ with $|J|=s$ take the form
\be
\renewcommand{\arraystretch}{1.3}
\label{def:typeIa_wfns}
{\bf Ia} \quad \begin{array}{ll}
\partial_w^{n}\Phi_{\Delta,+s}, & \partial_\bw^{n}\Phi_{\Delta,-s}\\
\partial_w^{n}\widetilde{\Phi}_{\Delta,+s}, & \partial_\bw^{n}\widetilde{\Phi}_{\Delta,-s}
\end{array}\quad \Delta=1-s-n\,,
\ee
and
\be
\renewcommand{\arraystretch}{1.3}
\label{def:typeIb_wfns}
{\bf Ib} \quad \begin{array}{ll}
 \partial_w^{2s+n}\Phi_{\Delta,-s}, & \partial_\bw^{2s+n}\Phi_{\Delta,+s}\\
\partial_w^{2s+n}\widetilde{\Phi}_{\Delta,-s}, & \partial_\bw^{2s+n}\widetilde{\Phi}_{\Delta,+s}
\end{array}\quad \Delta=1-s-n\,,
\ee
for $n\in \mathbb{Z}_{>}$. 
As in the general classification~\eqref{DefTypeI}, when $J=0$ we define type Ia primary descendant wavefunctions via holomorphic derivatives and type Ib primary descendant wavefunctions via  antiholomorphic derivatives.
Notice that the type I primary descendants~\eqref{def:typeIa_wfns} and~\eqref{def:typeIb_wfns} vanish identically, as can be checked directly from the 4D radiative wavefunctions of section~\ref{sec:CPW}. 
For example, when $J=0$ the conditions~\eqref{def:typeIa_wfns}-\eqref{def:typeIb_wfns} give $\partial_w^{n} (-q\cdot X)^{n-1}=0$ for type~Ia and $\partial_\bw^{n} (-q\cdot X)^{n-1}=0$  for Ib, which both trivially vanish since we are deriving $n$ times a polynomial of order $n-1$ in $w$ and $\bw$. This is expected from the fact that there are no generalized conformal primaries with spin $|J'|>s$.

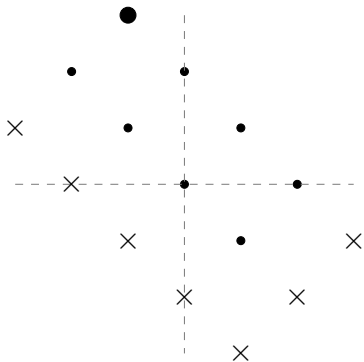
\begin{figure}[ht!]
\centering
\begin{tikzpicture}[scale=1.5]
\usetikzlibrary{patterns}
\definecolor{darkgreen}{rgb}{.8, 0.8, .8};
\definecolor{blue}{rgb}{.5, 0.5, .5};
\filldraw[black] (-2/2,6/2) circle (2pt) ;
\filldraw[] (0,1) circle (1pt) ;
\filldraw[] (0,2) circle (1pt) ;
\filldraw[] (-1,2) circle (1pt) ;
\filldraw[] (1/2,1.5) circle (1pt) ;
\filldraw[] (-1/2,1.5) circle (1pt) ;
\filldraw[] (-1/2,2.5) circle (1pt) ;
\filldraw[] (-3/2,2.5) circle (1pt) ;
\draw[dashed,blue] (-.5,3) --(-0.5,0);
\draw[dashed,blue] (-2,1.5) --(1,1.5);
\node[] at (0,0) {$\times$};
\node[] at (1/2,1/2) {$\times$};
\node[] at (2/2,2/2) {$\times$};
\node[] at (-1/2,1/2) {$\times$};
\node[] at (-2/2,2/2) {$\times$};
\node[] at (-3/2,3/2) {$\times$};
\node[] at (-4/2,4/2) {$\times$};
\end{tikzpicture}
\caption{Example type I module in celestial CFT. For a radiative state (\tikzcircle{2pt}) with $\Delta=1-|J|-n$ where $n\in\mathbb{Z_{>}}$, the predicted primary descendants and their descendants are identically zero~($\times$). The radiative primary (\tikzcircle{2pt}) and its descendants (\tikzcircle{1pt}) thus form an $(n-1)(2|J|+n-1)$ dimensional vector space. The axes intersect at $(\Delta,J)=(1,0)$.}
\label{typeIquotient}
\end{figure}

These vanishing wavefunctions correspond to states that are thus null in the trivial sense. 
Indeed given a vanishing wavefunction, its associated 2D operator -- obtained through equation~\eqref{qdelta} -- automatically vanishes.
There are thus infinitely many modules with type I null states 
in celestial CFT descending from radiative states
with spin $|J|=s$ and conformal dimension $\Delta\leq 1-|J|-n$ for $n\in \mathbb{Z}_{>}$. 
Since all their descendants vanish as well, we are left with a finite dimensional representation. This is illustrated in an example in figure~\ref{typeIquotient}. Starting from a radiative parent primary of spin $|J|=s$ and conformal dimension $\Delta=1-|J|-n$ for $n\in\mathbb{Z_{>}}$ at the top, we have a total of $(n-1)(2|J|+n-1)$ states arising from SL(2,$\mathbb{C}$) descendants. All but one corner of the resulting celestial diamond are null.  

\vspace{2em}

\subsection{Primary Descendants in Celestial CFT}
\label{sec:diamond}

Besides the infinitely many celestial type~I primary descendants of section~\ref{sec:NullStates} which give rise to
vanishing wavefunctions, there is a finite set of primary descendants 
of types~I,~II and~III which are not trivially null. 
These primary descendants and their parent primaries form the corners of non-trivial celestial diamonds.

\subsubsection*{Finite-Area Celestial Diamonds}

In the general discussion of global primary descendants of SL(2,$\mathbb{C}$) in section~\ref{sec:global_primary_desc} we alluded to a diamond structure in figure~\ref{Nested_submodules}, where starting from a conformal primary operator at the top of the diamond we may descend to primary operators at the left, right and bottom corners for special values of the primary operator's SL(2,$\mathbb{C}$) quantum numbers. In celestial CFT, it is most natural to center this discussion around radiative solutions. We will now outline the general structure of celestial diamonds before discussing the spin-$s$ examples of interest in detail below.
\vspace{1em}

\noindent \underline{\it Left and Right Corners}~~
Radiative conformal primary wavefunctions $\Phi_{\Delta,J}$ or $\tPhi_{\Delta,J}$ with conformally soft SL(2,$\mathbb{C}$) dimensions $\Delta$ form the left (negative $J$) and right (positive $J$) corners of the diamond structure in figure~\ref{Nested_submodules}.
There are two types of diamonds: a Goldstone~($\Gold$) and a Memory~($\Memo$) diamond. The left and right corners of the first correspond to the wavefunctions in tables~\ref{table:Goldstone} and~\ref{table:shadowGoldstone} of section~\ref{sec:CPW} which are Goldstone modes of spontaneously broken asymptotic symmetries whose Ward identities are equivalent to (conformally) soft theorems. Their canonical partners related to memory effects form the left and right corners of the second diamond. Moreover, from section~\ref{sec:global_primary_desc} we know that the left and right corners of each diamond are related by a conformal shadow transform. For the leading soft theorem in gauge theory and gravity this relates opposite helicity Goldstone modes and explains their degeneracy in soft theorems.
\vspace{1em}

\noindent \underline{\it Bottom Corners}~~
Since radiative conformal primary wavefunctions at the left and right corner of a given celestial diamond obey the shadow relation~\eqref{shadow_II}, it follows that they must descend to the same primary descendant which is filling the bottom corner of that diamond.
From the definition~\eqref{DefTypeII}, we see that these are type~II primary descendants of the form
\be
\label{bottom_corner_wfs}
\renewcommand{\arraystretch}{1.3}
{\bf II}\quad \begin{array}{lll}
\partial_\bw^{n}\Phi_{\Delta,+s},~~ \partial_w^{n}\Phi_{\Delta,-s}\\
\partial_\bw^{n}\widetilde{\Phi}_{\Delta,+s},~~ \partial_w^{n}\widetilde{\Phi}_{\Delta,-s}
\end{array}\quad \Delta=1+s-n
\ee
where $|J|=s$ and $n\in\{1,...,2s-1\}$ which implies $1-s<\Delta<1+s$ and only exist for $s\ge1$. 
As explained in section~\ref{sec:quant}, type II primary descendants of radiative modes {\it do not vanish} identically from the point of view of 4D bulk wavefunctions (though they fall off faster than radiative order at null infinity away from isolated points). Instead they turn out to be generalized conformal primaries that give contact term contributions to celestial amplitudes. This can be seen from the fact that the soft theorems are meromorphic, and when hit by a suitable amount of $\p_w$ and $\p_\bw$ derivatives give rise to contact terms. In an upcoming paper~\cite{PPT2}, we will show that the bottom corners of celestial diamonds define soft operators in terms of which soft charges are most naturally expressed.\footnote{The authors of~\cite{Banerjee:2019aoy,Banerjee:2019tam} previously studied null states in celestial CFT whose decoupling was interpreted as effectively reducing the number of polarization states of soft particle and found to be crucial in deriving soft-theorems from the Ward identities of asymptotic symmetries. In our language these null states would correspond to taking differences of primaries descending from the left and right corners of the diamond. The connection between null states and soft charges was also discussed in~\cite{Banerjee:2019tam}.}

Curiously, the special spectrum of $(\Delta',J')$ for type II primary descendants
is precisely predicted by the discussion in section~\ref{sec:gencpw}.
Recall that wavefunctions that descend from radiative conformal primaries obey the same gauge conditions (since $\p_w$ and $\p_\bw$ commute with them) as the latter. The primary descendants filling the bottom corners thus precisely correspond to the finite set of generalized primaries identified explicitly in section~\ref{sec:gencpw}.  
\vspace{1em}

\noindent \underline{\it Top Corners}~~
Conformally soft primary wavefunctions turn out to be primary descendants themselves and so we can complete the diamond structure at the top corners for each spin-$s$ case. These are given by another set of generalized conformal primary wavefunctions which obey more relaxed gauge conditions than the radiative primary wavefunctions they descend to. These parent primaries give rise to  type I primary descendants given by
\be
\label{TypesI_Top}
\renewcommand{\arraystretch}{1.3}
\begin{array}{lll}
{\bf Ia} &\quad\partial_w^{n}\Phi^{gen,s}_{\Delta,+|J|}, & \partial_\bw^{n}\Phi^{gen,s}_{\Delta,-|J|}\\
{\bf Ib} &\quad \partial_w^{2|J|+n}\Phi^{gen,s}_{\Delta,-|J|}, & \partial_\bw^{2|J|+n}\Phi^{gen,s}_{\Delta,+|J|}, 
\end{array}\quad \Delta=1-s\,,~~|J|=s-n
\ee
for $n\in \mathbb{Z}_{>0}$. Hence, the conformally soft radiative wavefunctions at the left and right corners of the celestial diamonds are, respectively, primary descendants of type Ia and Ib. 

Unlike for the bottom corners of celestial diamonds, the discussion of section~\ref{sec:gencpw} does not identify the particular spectrum   $(\Delta',J')$ or functional form for the wavefunctions at the top corners.
However the spectrum can be easily determined using table~\ref{table:types1} and imposing that their descendants are the known radiative primary descendants. With the relaxed gauge fixing, there is an ambiguity in the parents which we will return to at the end of this section. In the tables that follow we will pick convenient representatives of the parent primaries.  {In upcoming work~\cite{PPT2} we will show that the top corners of celestial diamonds are intimately tied to conformal Faddeev-Kulish dressings which render celestial amplitudes infrared finite. }

\vspace{1em}

We now discuss the celestial diamonds for spin $s=1,\frac{3}{2}$ and $2$ in turn, and comment on the much simpler primary descendant structure for spin $s=0$ and~$\frac{1}{2}$ below.

\vspace{1em}

\pagebreak

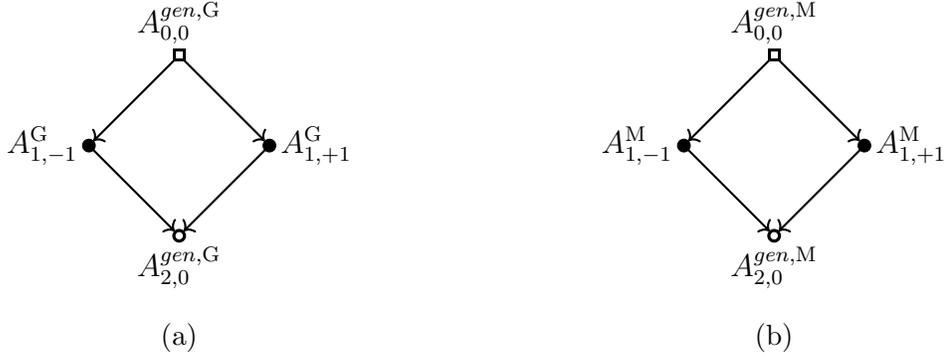
\begin{figure}[ht!]
\begin{subfigure}{.45\linewidth}
\centering
\begin{tikzpicture}[scale=1.2]
\definecolor{darkgreen}{rgb}{.0, 0.5, .1};
\draw[thick,->] (-1,1) node [below=4mm]{
} --   (0-.05,0+.05) ;
\draw[thick,->] (1,1) node [below=4mm]{
} -- (.05,.05);
\filldraw[black] (0,0) circle (2pt);
\filldraw[black] (-1,1) circle (2pt) ;
\filldraw[black] (1,1) circle (2pt) ;
\draw[thick,->] (0,2)-- (1-.05,1+.05) node [above=4mm]{
} ;
\draw[thick,->] (0,2)-- (-1+.05,1.05) node [above=4mm]{
} ;
\node[fill=black,regular polygon, regular polygon sides=4,inner sep=1.6pt] at (0,2) {};
\node[fill=white,regular polygon, regular polygon sides=4,inner sep=.8pt] at (0,2) {};
\draw[thick](0,2) node [above]{$A^{gen,\Gold}_{0,0}$};
\draw[thick](0,0) node [below]{$A^{gen,\Gold}_{2,0}$};
\filldraw[white] (0,0) circle (1pt);
\draw[thick](1,1) node [right]{$A^{\Gold}_{1,+1}$};
\draw[thick](-1,1) node [left]{$A^{\Gold}_{1,-1}$};
\end{tikzpicture}
\caption{}
\end{subfigure}
\begin{subfigure}{.45\linewidth}
\centering
\begin{tikzpicture}[scale=1.2]
\definecolor{darkgreen}{rgb}{.0, 0.5, .1};
\draw[thick,->] (-1,1) node [below=4mm]{
} --   (0-.05,0+.05) ;
\draw[thick,->] (1,1) node [below=4mm]{
} -- (.05,.05);
\filldraw[black] (0,0) circle (2pt);
\filldraw[black] (-1,1) circle (2pt) ;
\filldraw[black] (1,1) circle (2pt) ;
\draw[thick](0,2) node [above]{$A^{gen,\Memo}_{0,0}$};
\draw[thick,->] (0,2)-- (1-.05,1+.05) node [above=4mm]{
} ;
\draw[thick,->] (0,2)-- (-1+.05,1.05) node [above=4mm]{
} ;
\node[fill=black,regular polygon, regular polygon sides=4,inner sep=1.6pt] at (0,2) {};
\node[fill=white,regular polygon, regular polygon sides=4,inner sep=.8pt] at (0,2) {};
\draw[thick](0,0) node [below]{$A^{gen,\Memo}_{2,0}$};
\filldraw[white] (0,0) circle (1pt);
\draw[thick](1,1) node [right]{$A^{\Memo}_{1,+1}$};
\draw[thick](-1,1) node [left]{$A^{\Memo}_{1,-1}$};
\end{tikzpicture}
\caption{}
\end{subfigure}
\caption{Goldstone (a) and memory (b) diamonds for the leading soft photon theorem.}
\label{fig:photondiamond}
\end{figure}

\begin{center}
{\bf---}~$\diamond$~{\bf Photon Diamonds}~$\diamond$~{\bf---}
\end{center}
\noindent The celestial diamonds relevant for the leading conformally soft theorem in gauge theory are summarized in figure~\ref{fig:photondiamond}. 
Table~\ref{table:Gphotons} gives the elements of the Goldstone diamond, for which 
\be
A^{\Gold}_{\Delta,J;\mu}=\nabla_\mu\Lambda_{\Delta,J}\,,
\ee
while table~\ref{table:Mphotons} gives elements of the memory diamond. While we have picked the log mode wavefunctions for table~\ref{table:Mphotons}, one can replace them with the other conformally soft modes described in section~\ref{sec:CPWs}.

\vspace{1em}
 \noindent 
 \underline{\it Left and Right Corners}~~
The leading conformally soft photon theorem~\cite{Fan:2019emx,Nandan:2019jas,Pate:2019mfs} arises from the spin-1 conformal primary wavefunctions with $\Delta=1$ and $J=\pm 1$. These are canonically paired with memory modes of the same dimension and spin. The pairings with various conformally soft modes have been examined in~\cite{Donnay:2018neh} and~\cite{Arkani-Hamed:2020gyp}. The operators selected by the Goldstone modes generate Kac-Moody symmetries in the celestial CFT~\cite{He:2015zea,Nande:2017dba}.

\vspace{1em}
\noindent
\underline{\it Bottom Corners}~~
Both the left and right corners descend to the same $\Delta=2,J=0$ generalized primary which is a type~II primary descendant
\be
    \p_w A^{\Gold/\Memo}_{1,-1}=A^{gen,\Gold/\Memo}_{2,0}=\p_\bw A^{\Gold/\Memo}_{1,+1}\,.
\ee
Note that the above descendancy relations for the $\Delta=1$ wavefunctions display the degeneracy of the two helicities in the soft theorems~\cite{ss}. 
 In celestial correlators, descendants of the radiative currents reduce to contact terms supported at the locations of other charged operators.  

\vspace{1em}
\noindent
\underline{\it Top Corners}~~To complete the celestial diamonds relevant for the leading conformally soft photon theorem, we augment the Hilbert space with a pair of $(\Delta,J)=(0,0)$ generalized photons whose type~I primary descendants at level~1 land us on the spin-1 radiative wavefunctions
\begin{equation}\label{DescAgen00G}
   \partial_\bw A^{gen,\Gold/\Memo}_{0,0} = A^{\Gold/\Memo}_{1,-1}\,, \quad \partial_w A^{gen,\Gold/\Memo}_{0,0} = A^{\Gold/\Memo}_{1,+1}\,.   
\end{equation}
These parent primaries need not obey the same gauge fixing as the other corners. While for the Goldstone modes this is not a problem, for
the memory mode it would be natural to ask whether one should allow this parent in the phase space.  If not, then the Kac-Moody currents are no longer primary descendants, just primaries. However, we can always define this parent in terms of an appropriate Green's function, as will be explored in~\cite{PPT2}. We also point out that one can create distributional solutions that formally satisfy all of the gauge conditions and have isolated sources
\be
 A^{gen,\CS''}_{0,0;\mu}={\textstyle \frac{1}{\sqrt{2}}} q_\mu {\rm log}(X^2) \delta(q\cdot X)\,,
 \ee
 which we recognize as the electromagnetic analogue of the Aichelburg-Sexl ultraboost~\cite{Pasterski:2020pdk}.

\pagebreak 

\vspace{1em}
\begin{table}[h!]
\renewcommand*{\arraystretch}{1.3}
\centering
    \begin{tabular}{l|l|l|l|l}

     Corner & $\Delta$  & $J$ 
       &  $A^{\Gold}_{\Delta,J}$ & $\Lambda_{\Delta,J}$\\
         \hline
         Top&0&0
         & {$\frac{1}{\sqrt{2}}l_\mu$}
         & $-\frac{1}{\sqrt{2}}\log\varphi^{-1}$\\
          Left&1&$-1$
          &$\bar{m}_\mu\varphi^{1}$&$\p_\bw\Lambda_{0,0}$\\
           Right&1&$+1$
           &${m}_\mu\varphi^{1}$&$\p_w\Lambda_{0,0}$\\
           Bottom&2&0
           &${\sqrt{2}}\left(\frac{X^2}{2}l_\mu  +n_\mu\right)\varphi^2$ 
           &$\p_w\p_\bw\Lambda_{0,0}$\\
    \end{tabular}
    \caption{Elements of the celestial diamond corresponding to large U(1) gauge symmetry. 
    }
    \label{table:Gphotons}
\end{table}
\vspace{1em}
\begin{table}[h!]
\renewcommand*{\arraystretch}{1.3}
\centering
    \begin{tabular}{l|l|l|l}
     Corner & $\Delta$  & $J$   &  $A^{\log}_{\Delta,J}$ \\
         \hline
         Top&0&0
         &${\frac{1}{\sqrt{2}}l_\mu\log(X^2)}$\\ 
          Left&1&$-1$
          &$\bar{m}_\mu\log(X^2)\varphi^{1}$\\
           Right&1&$+1$
           &${m}_\mu\log(X^2)\varphi^{1}$\\
           Bottom&2&0
           &${\sqrt{2}}\left(\frac{X^2}{2}l_\mu  +n_\mu\right)\log(X^2)\varphi^2$\\
    \end{tabular}
    \caption{Elements of the celestial diamond corresponding to electromagnetic memory.}
    \label{table:Mphotons}
\end{table}

\clearpage
\pagebreak

\begin{center}
{\bf---}~$\diamond$~{\bf Gravitino Diamonds}~$\diamond$~{\bf---}
\end{center}

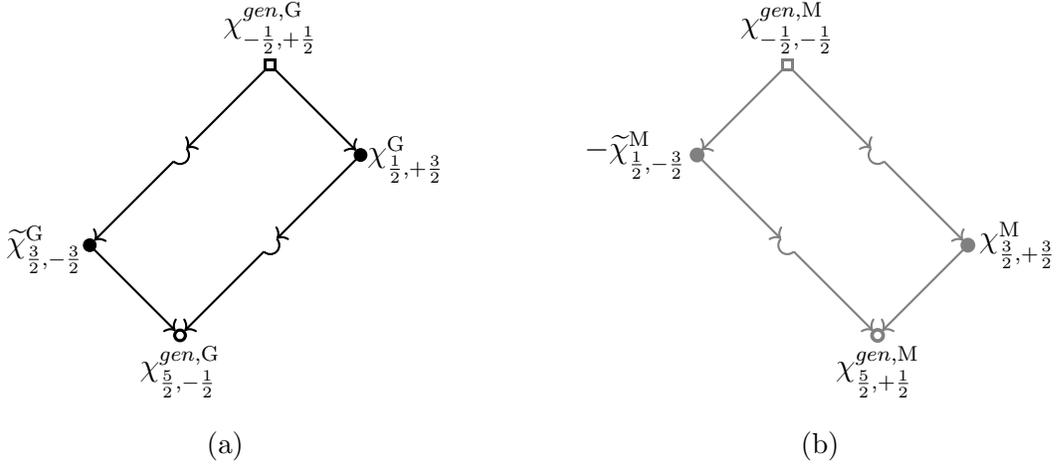
\begin{figure}[ht!]
\begin{subfigure}{.45\linewidth}
\centering
\begin{tikzpicture}[scale=1.2]
\definecolor{red}{rgb}{.5, 0.5, .5};
\draw[thick,->] (-1+.05,1-.05)node[left]{${\tchi}^{\Gold}_{\frac{3}{2},-\frac{3}{2}}$} --node[below left]{} (-.05,.05) ;
\draw[thick,->] (2-.05,2-.05) node[right]{${\chi}^{\Gold}_{\frac{1}{2},+\frac{3}{2}}$} --node[below right]{} (1+.1414/2,1+.1414/2);
\draw[thick,->] (1-.1414/2,1-.1414/2)-- (.05,.05);
\filldraw[black] (0,0) circle (2pt) node[below]{$\chi^{gen,\Gold}_{\frac{5}{2},-\frac{1}{2}}$};
\filldraw[white] (0,0) circle (1pt) ;
\filldraw[black] (-1,1) circle (2pt) ;
\filldraw[black] (2,2) circle (2pt) ;
\draw[thick] (1+.1414/2,1+.1414/2) arc (45:-135:.1);
\draw[thick] (0+.1414/2,2+.1414/2) arc (45:-135:.1);
\draw[thick,->] (1,3)  node[above,black]{$\chi^{gen,\Gold}_{-\frac{1}{2},+\frac{1}{2}}$} -- (2-.05,2+.05);
\draw[thick,->] (0-.1414/2,2-.1414/2)-- (-1+.05,1.05);
\draw[thick,->] (1,3) -- (0+.1414/2,2+.1414/2);
\node[fill=black,regular polygon, regular polygon sides=4,inner sep=1.6pt] at (1,3) {};
\node[fill=white,regular polygon, regular polygon sides=4,inner sep=.8pt] at (1,3) {};
\end{tikzpicture}
\caption{}
\end{subfigure}
\begin{subfigure}{.45\linewidth}
\centering
\begin{tikzpicture}[scale=1.2]
\definecolor{red}{rgb}{.5, 0.5, .5};
\filldraw[red,thick] (1,0) circle (2pt);
\filldraw[white] (1,0) circle (1pt) node[below,black]{$\chi^{gen,\Memo}_{\frac{5}{2},+\frac{1}{2}}$};
\filldraw[red,thick] (-1,2) circle (2pt) node[left,black]{$-{\tchi}^{\Memo}_{\frac{1}{2},-\frac{3}{2}}$};
\draw[red, thick] (0-.1414/2,1+.1414/2) arc (135:315:.1);
\draw[red, thick] (1-.1414/2,2+.1414/2) arc (135:315:.1);
\draw[->,red,thick] (-1,2) --  (0-.1414/2,1+.1414/2);
\draw[->,red,thick] (0+.1414/2,1-.1414/2) -- (1-.05,0.05);
\filldraw[red,thick] (2,1) circle (2pt) ;
\draw[->,red,thick] (2,1)  node[right,black]{${\chi}^{\Memo}_{\frac{3}{2},+\frac{3}{2}}$} -- (1+.05,0.05);
\draw[->,red,thick] (0,3) node[above,black]{${\chi}^{gen,\Memo}_{-\frac{1}{2},-\frac{1}{2}}$}  --  (1-.1414/2,2+.1414/2);
\draw[->,red,thick] (1+.1414/2,2-.1414/2) --  (2-.05,1+.05);
\draw[->,red,thick] (0,3) --  (-1+.05,2+.05);
\node[fill=red,regular polygon, regular polygon sides=4,inner sep=1.6pt] at (0,3) {};
\node[fill=white,regular polygon, regular polygon sides=4,inner sep=.8pt] at (0,3) {};
\end{tikzpicture}
\caption{}
\end{subfigure}
\caption{Goldstone (a) and memory (b) diamonds for the leading soft gravitino theorem.}
\label{fig:gravitinodiamond}
\end{figure}

\noindent The celestial diamonds for the leading conformally soft gravitino theorem are chiral and we focus here on the left-handed ones which are summarized in figure~\ref{fig:gravitinodiamond}. Table~\ref{table:Ggravitinos} gives the elements of the Goldstone diamond, for which 
\be
\chi^{\Gold}_{\Delta,J;\mu}=\nabla_\mu\lambda_{\Delta,J}\,.
\ee
The physical interpretation of memory effects for fermionic modes is still an open question~\cite{Avery:2015iix,Lysov:2015jrs}, though we expect them to be related by supersymmetry to memory effects that are well understood. The modes in table~\ref{table:Ggravitinos} are expected to be relevant based on their conformal dimensions. Similar expressions can be obtained for the right-handed gravitinos and their celestial diamonds.

\vspace{1em}

\noindent \underline{\it Left and Right Corners}~~
The conformally soft gravitino theorem~\cite{Fotopoulos:2020bqj} appears at $\Delta=\frac{1}{2}$.  The conformal shadow of this mode appears at $\Delta=\frac{3}{2}$. These are summarized in table~\ref{table:Ggravitinos}. Non-gauge modes with canonically paired conformal dimensions are given in table~\ref{table:Mgravitinos}. The operators selected by the Goldstone modes generate large supersymmetry transformations~\cite{PPP,Avery:2015iix,Lysov:2015jrs}.

\vspace{1em}

\noindent \underline{\it Bottom Corners}~~ For each diamond, the left and right corners descend to the same generalized primary, at the levels expected for a type II primary descendant
\begin{equation}
\p_w \tchi^\Gold_{\frac{3}{2},-\frac{3}{2}}= \chi^{gen,\Gold}_{\frac{5}{2},-\frac{1}{2}}=\frac{1}{2!}\partial_\bw^2 \chi^\Gold_{\frac{1}{2},+\frac{3}{2}}\,,~~~ -\frac{1}{2!}\partial_w^2 \tchi^\Memo_{\frac{1}{2},-\frac{3}{2}}=\chi^{gen,\Memo}_{\frac{5}{2},+\frac{1}{2}}=\p_\bw \chi^\Memo_{\frac{3}{2},+\frac{3}{2}}.
\end{equation}
The relative minus sign arises from the shadow of fermions.  Descendants of the radiative currents reduce to contact terms in celestial correlators which we expect to correspond to a soft charge.

\vspace{1em}

\noindent \underline{\it Top Corners}~~
 To complete the celestial diamond relevant for the conformally soft gravitino theorem we augment the Hilbert space with a $(\Delta,J)=(-\frac{1}{2},+\frac{1}{2})$ generalized gravitino, which is pure gauge, whose level~1 and level~2 primary descendants land us on the spin-$\frac{3}{2}$ Goldstino wavefunctions with $\Delta=\frac{3}{2}$ and $\Delta=\frac{1}{2}$
\begin{equation}
\frac{1}{2!}\partial_\bw^2  \chi^{gen,\Gold}_{-\frac{1}{2},+\frac{1}{2}} =\tchi^{\Gold}_{\frac{3}{2},-\frac{3}{2}}\,, \quad \partial_w  \chi^{gen,\Gold}_{-\frac{1}{2},+\frac{1}{2}} =\chi^{\Gold}_{\frac{1}{2},+\frac{3}{2}}\,.
\end{equation}
We also add a 
$(\Delta,J)=(-\frac{1}{2},-\frac{1}{2})$ generalized gravitino, which is not pure gauge, whose level~1 and level~2 primary descendants land us on the spin-$\frac{3}{2}$ conformally soft wavefunctions with $\Delta=\frac{1}{2}$ and $\Delta=\frac{3}{2}$
\begin{equation}
\partial_\bw  \chi^{gen,\Memo}_{-\frac{1}{2},-\frac{1}{2}} =-\widetilde{\chi}^\Memo_{\frac{1}{2},-\frac{3}{2}}\,, \quad \frac{1}{2!}\partial_w^2  \chi^{gen,\Memo}_{-\frac{1}{2},-\frac{1}{2}} =\chi^\Memo_{\frac{3}{2},+\frac{3}{2}}\,.
\end{equation}

\pagebreak

\vspace{1em}
\begin{table}[h!]
\renewcommand*{\arraystretch}{1.3}
\centering
    \begin{tabular}{l|l|l|l|l}
     Corner & $\Delta$  & $J$
       &  $\chi^{\Gold}_{\Delta,J}$ & $\lambda_{\Delta,J}$\\
         \hline
         Top&$-\frac{1}{2}$&$+\frac{1}{2}$
         & $\frac{1}{\sqrt{2}}  ol_\mu\varphi^{-\frac{1}{2}}$&$- \frac{1}{\sqrt{2}}o\varphi^{-\frac{1}{2}}\log\varphi^{-1}$\\
          Left&$~~\frac{3}{2}$&$-\frac{3}{2}$
          &$\sqrt{2}\iota{\bar m}_\mu \varphi^{\frac{3}{2}}$&$\frac{1}{2!}\p_\bw^2\lambda_{-\frac{1}{2},\frac{1}{2}}$\\
           Right&$~~\frac{1}{2}$&$+\frac{3}{2}$
           &$o m_\mu \varphi^\frac{1}{2}$&$\p_w\lambda_{-\frac{1}{2},\frac{1}{2}}$\\
           Bottom&$~~\frac{5}{2}$&$-\frac{1}{2}$
           &${2}\left[\left(\frac{X^2}{2}l_\mu+n_\mu\right) \iota +\frac{X^2}{2} o\bar{m}_\mu\right] \varphi^{\frac{5}{2}}$\,
           &$\frac{1}{2!}\p_w\p_\bw^2\lambda_{-\frac{1}{2},\frac{1}{2}}$\\
    \end{tabular}
    \caption{Elements of the celestial diamond corresponding to large supersymmetry.}
    \label{table:Ggravitinos}
\end{table}
\vspace{1em}
\begin{table}[h!]
\renewcommand*{\arraystretch}{1.3}
\centering
    \begin{tabular}{l|l|l|l}
     Corner & $\Delta$  & $J$ 
       &  $\chi^{\Memo}_{\Delta,J}$ \\
         \hline
         Top&$-\frac{1}{2}$&$-\frac{1}{2}$
         & $\frac{2}{X^4}  \iota n_\mu \varphi^{-\frac{1}{2}}$\\
          Left&$~~\frac{1}{2}$&$-\frac{3}{2}$
          &$\frac{\sqrt{2}}{X^2}\iota{\bar m}_\mu \varphi^{\frac{1}{2}}$\\
           Right&$~~\frac{3}{2}$&$+\frac{3}{2}$
           &$o{m}_\mu\varphi^{\frac{3}{2}}$\\
           Bottom&$~~\frac{5}{2}$&$+\frac{1}{2}$
           &${\sqrt{2}}\left[\left(\frac{X^2}{2}l_\mu +n_\mu\right) o  + \iota m_\mu \right] \varphi^{\frac{5}{2}}$\\
    \end{tabular}
    \caption{Elements of the celestial diamond corresponding to a supergravity memory effect.}
    \label{table:Mgravitinos}
\end{table}

\pagebreak

\begin{center}
 {\bf---}~$\diamond$~{\bf Leading Graviton Diamonds}~$\diamond$~{\bf---}
\end{center}

\begin{figure}[h]
\begin{subfigure}{.45\linewidth}
\centering
\begin{tikzpicture}[scale=1.2]
\filldraw[black] (-2,1) circle (2pt)  node [left]{$h^{\Gold}_{1,-2}$};
\filldraw[black] (2,1) circle (2pt) node [right]{$h^{\Gold}_{1,+2}$};
\draw[thick] (1+.1414/2,0+.1414/2) arc (45:-135:.1);
\draw[thick] (-1+.1414/2,2+.1414/2) arc (45:-135:.1);
\draw[thick] (0,3) node [above]{$h^{gen,\Gold}_{-1,0}$} ;
\draw[->,thick] (0,3) --  (1-.1414/2,2+.1414/2) node [above=4mm]{
};
\draw[->,thick] (1+.1414/2,2-.1414/2) -- (2-.05,1+.05) node [above=4mm]{
};
\draw[->,thick] (0,3) --  (-1+.07,2+.07) node [above=4mm]{
};
\draw[->,thick] (-1-.07,2-.07)-- (-2+.05,1+.05) node [above=4mm]{
};
\draw[->,thick] (1-.07,0-.07)-- (0+.05,-1+.05) ;
\node[fill=black,regular polygon, regular polygon sides=4,inner sep=1.6pt] at (0,3) {};
\node[fill=white,regular polygon, regular polygon sides=4,inner sep=.8pt] at (0,3) {};
\filldraw[black] (0,-1) circle (2pt)  node [below]{$h^{gen,\Gold}_{3,0}$};
\filldraw[white] (0,-1) circle (1pt);
\draw[->,thick] (2,1) node [below=4mm]{
} --  (1+.07,0+.07) node [below=4mm]{
} ;
\draw[->,thick] (-2,1) node [below=4mm]{
} --  (-1-.1414/2,0+.1414/2) node [below=4mm]{
};
\draw[thick] (-1-.1414/2,0+.1414/2)  arc (135:315:.1);
\draw[->,thick] (-1+.1414/2,0-.1414/2)  -- (0-.05,-1+.05)  ;
\draw[thick] (1-.1414/2,2+.1414/2) arc (135:315:.1);
\end{tikzpicture}
\caption{}
\end{subfigure}
\begin{subfigure}{.45\linewidth}
\centering
\begin{tikzpicture}[scale=1.2]
\filldraw[black] (-2,1) circle (2pt)  node [left]{$h^{\Memo}_{1,-2}$};
\filldraw[black] (2,1) circle (2pt) node [right]{$h^{\Memo}_{1,+2}$};
\draw[thick] (1+.1414/2,0+.1414/2) arc (45:-135:.1);
\draw[thick] (-1+.1414/2,2+.1414/2) arc (45:-135:.1);
\draw[thick] (0,3) node [above]{$h^{gen,\Memo}_{-1,0}$} ;
\draw[->,thick] (0,3) --  (1-.1414/2,2+.1414/2) node [above=4mm]{
};
\draw[->,thick] (1+.1414/2,2-.1414/2) -- (2-.05,1+.05) node [above=4mm]{
};
\draw[->,thick] (0,3) --  (-1+.07,2+.07) node [above=4mm]{
};
\draw[->,thick] (-1-.07,2-.07)-- (-2+.05,1+.05) node [above=4mm]{
};
\draw[->,thick] (1-.07,0-.07)-- (0+.05,-1+.05) ;
\node[fill=black,regular polygon, regular polygon sides=4,inner sep=1.6pt] at (0,3) {};
\node[fill=white,regular polygon, regular polygon sides=4,inner sep=.8pt] at (0,3) {};
\filldraw[black] (0,-1) circle (2pt)  node [below]{$h^{gen,\Memo}_{3,0}$};
\filldraw[white] (0,-1) circle (1pt);
\draw[->,thick] (2,1) node [below=4mm]{
} --  (1+.07,0+.07) node [below=4mm]{
} ;
\draw[->,thick] (-2,1) node [below=4mm]{
} --  (-1-.1414/2,0+.1414/2) node [below=4mm]{
};
\draw[thick] (-1-.1414/2,0+.1414/2)  arc (135:315:.1);
\draw[->,thick] (-1+.1414/2,0-.1414/2)  -- (0-.05,-1+.05)  ;
\draw[thick] (1-.1414/2,2+.1414/2) arc (135:315:.1);
\end{tikzpicture}
\caption{}
\end{subfigure}
\caption{Goldstone (a) and memory (b) diamonds for the leading soft graviton theorem.}
\label{fig:gravitondiamond}
\end{figure}
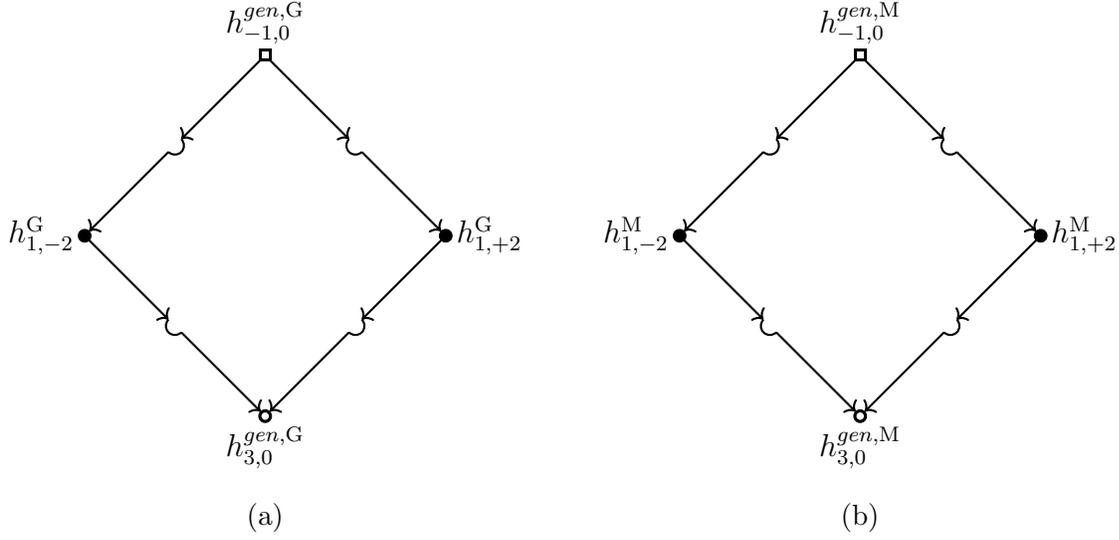

\noindent The celestial diamonds corresponding to the leading conformally soft graviton theorem are summarized in figure~\ref{fig:gravitondiamond}.
Table~\ref{table:Ggravitons} gives the elements of each Goldstone diamond, for which 
\be
\label{hGold}
h^{\Gold}_{\Delta,J;\mu\nu}=\nabla_\mu\xi_{\Delta,J;\nu}+\nabla_\nu\xi_{\Delta,J;\mu}=\nabla_\mu\nabla_\nu \Lambda_{\Delta,J}\,,
\ee
while table~\ref{table:Mgravitons} gives elements of the memory diamond. While we have picked the log mode wavefunctions for table~\ref{table:Mgravitons}, one can replace these with the other conformally soft modes described in section~\ref{sec:CPWs}. 
\vspace{1em}

\noindent \underline{\it Left and Right Corners}~~
The leading conformally soft graviton
theorem~\cite{Adamo:2019ipt,Puhm:2019zbl,Guevara:2019ypd} arises from the spin-2 conformal primary wavefunction with $\Delta=1$ and $J=\pm 2$.  These are canonically paired with memory modes of the same dimension and spin. The pairings with various conformally soft modes have been examined in~\cite{Donnay:2018neh} and~\cite{Ball:2019atb}.  The operators selected by the Goldstone modes generate BMS supertranslations~\cite{Donnay:2018neh}.  

\vspace{1em}
\noindent \underline{\it Bottom Corners}~~
Both the left and right hand corners descend to the same $\Delta=3,J=0$ generalized primary which is a type~II primary descendant
\begin{equation}
   \frac{1}{2!} \p_w^2 h^{\Gold/\Memo}_{1,-2}=h^{gen,{\Gold/\Memo}}_{3,0}= \frac{1}{2!}\p_\bw^2 h^{\Gold/\Memo}_{1,+2}\,.
\end{equation}
Note that the above descendancy relations for the $\Delta=1$ wavefunctions display the degeneracy of the two helicities in the soft theorems~\cite{ss}. 
In celestial correlators, descendants of the radiative currents reduce to contact terms supported at the locations of other operators. 

\vspace{1em}
\noindent \underline{\it Top Corners}~~To complete the celestial diamonds relevant for the leading conformally soft graviton theorem, we augment the Hilbert space by a pair of $(\Delta,J)=(-1,0)$ generalized gravitons whose descendants at level~2 land us on the spin-2 radiative wavefunctions
\begin{equation}\label{Deschgenm10G}
    \partial_\bw^2 h^{gen,\Gold/\Memo}_{-1,0} = h^{\Gold/\Memo}_{1,-2}\,, \quad \partial_w^2 h^{gen,\Gold/\Memo}_{-1,0} = h^{\Gold/\Memo}_{1,+2}\,.
\end{equation}
These parents need not obey the same gauge fixing as the other corners.  For the Goldstone modes this is not a problem. For the memory mode it would be natural to ask whether one should allow this parent in the phase space. Regardless, one can always define this parent in terms of an appropriate Green's function~\cite{PPT2}.  Note that one can create distributional solutions that formally satisfy all of the gauge conditions and have isolated sources
\be
h^{gen,\CS''}_{-1,0;\mu\nu}={\textstyle \frac{1}{4}} q_\mu q_\nu {\rm log}(X^2)\delta(q\cdot X)
 \ee
 which we recognize as the Aichelburg-Sexl ultraboost~\cite{Pasterski:2020pdk}.

\clearpage
\pagebreak

\vspace{1em}
\begin{table}[h!]
\renewcommand*{\arraystretch}{1.3}
\centering
    \begin{tabular}{l|l|l|l|l|l}
        Corner & $\Delta$  & $J$  
       &  $h^{\Gold}_{\Delta,J}$ & $\xi_{\Delta,J} $& $\Lambda_{\Delta,J}$\\
         \hline
         Top&$-1$&$~~0$
         & $\frac{1}{2}l_\mu l_\nu \varphi^{-1}$&$-\frac{1}{2}l^\mu\varphi^{-1} \log\varphi^{-1}$&$\frac{1}{2}\varphi^{-1}\log\varphi^{-1}$\\
          Left&$~~1$&$-2$
          &$\bar{m}_\mu\bar{m}_\nu\varphi^{1}$&$\frac{1}{2!}\p_w^2\xi^\mu_{-1,0}$&$\frac{1}{2!}\p_\bw^2\Lambda_{-1,0}$\\
           Right&$~~1$&$+2$
           &${m}_\mu{m}_\nu\varphi^{1}$&$\frac{1}{2!}\p_\bw^2\xi^\mu_{-1,0}$&$\frac{1}{2!}\p_w^2\Lambda_{-1,0}$\\
           Bottom&$~~3$&$~~0$
           &${2}\Big[ \Big(\frac{X^2}{2}\Big)^2l_\mu l_\nu + n_\mu n_\nu +\frac{X^2}{2}\eta_{\mu\nu}$\,&$\frac{1}{(2!)^2}\p_w^2\p_\bw^2\xi^\mu_{-1,0}$&$\frac{1}{(2!)^2}\p_w^2\p_\bw^2\Lambda_{-1,0}$\\
           &&&$~~~~~~+X^2(l_\mu n_\nu+n_\mu l_\nu) \Big]\varphi^3$&&\\
    \end{tabular}
    \caption{Elements of the celestial diamond corresponding to supertranslation symmetry.}
    \label{table:Ggravitons}
\end{table}
\begin{table}[h!]
\renewcommand*{\arraystretch}{1.3}
\centering
    \begin{tabular}{l|l|l|l}
     Corner & $\Delta$  & $J$ 
       &  $h^{\log}_{\Delta,J}$ \\
         \hline
         Top&$-1$&$~~0$
         & $\frac{1}{2}l_\mu l_\nu\log(X^2) \varphi^{-1}$\\
          Left&$~~1$&$-2$
          &$\bar{m}_\mu\bar{m}_\nu\log(X^2)\varphi^{1}$\\
           Right&$~~1$&$+2$
           &${m}_\mu{m}_\nu\log(X^2)\varphi^{1}$\\
           Bottom&$~~3$&$~~0$
           &${2}\Big[ \Big(\frac{X^2}{2}\Big)^2l_\mu l_\nu + n_\mu n_\nu +\frac{X^2}{2}\eta_{\mu\nu}$\\
           &&&$~~+X^2(l_\mu n_\nu+n_\mu l_\nu) \Big]\log(X^2)\varphi^3$\\
    \end{tabular}
    \caption{Elements of the celestial diamond corresponding to gravitational memory.}
    \label{table:Mgravitons}
\end{table}

\pagebreak


 \begin{center}
  {\bf---}~$\diamond$~{\bf Subleading Graviton Diamonds}~$\diamond$~{\bf---}
 \end{center}

\begin{figure}[ht!]
\begin{subfigure}{.45\linewidth}
\centering
\begin{tikzpicture}[scale=1.2]
\definecolor{darkgreen}{rgb}{.0, 0.5, .1};
\definecolor{darkgreen}{rgb}{.8, 0.8, .8};
\definecolor{blue}{rgb}{.5, 0.5, .5};
\filldraw[blue] (-2,2) circle (2pt) ;
\filldraw[blue] (2,0) circle (2pt)  node [right,black]{${-}\th^{\Gold}_{2,+2}$};
\filldraw[blue] (1,-1) circle (2pt) node [below,black]{$h^{gen,\Gold}_{3,+1}$};
\filldraw[white] (1,-1) circle (1pt) ;
\filldraw[darkgreen] (2,2) circle (2pt)  node [right,black]{$h^{\Gold}_{0,+2}$};
\filldraw[darkgreen] (-2,0) circle (2pt) node [left,black]{${-}\th^{\Gold}_{2,-2}$};
\filldraw[darkgreen] (-1,-1) circle (2pt) node [below,black]{$h^{gen,\Gold}_{3,-1}$};
\filldraw[white] (-1,-1) circle (1pt) ;
\draw[thick,->,blue] (2,0)-- (1+.05,-1+.05);
\draw[thick,->,blue] (-1,3)-- (-2+.05,2+.05);
\draw[thick,darkgreen] (0+.1414/2,0+.1414/2) arc (45:-135:.1);
\draw[thick,darkgreen] (1+.1414/2,1+.1414/2) arc (45:-135:.1);
\draw[thick,darkgreen] (-1+.1414/2,1+.1414/2) arc (45:-135:.1);
\draw[thick,darkgreen] (0+.1414/2,2+.1414/2) arc (45:-135:.1);
\draw[thick,blue] (0-.1414,0+.1414) arc (135:315:.2);
\draw[thick,blue] (1-.1414,1+.1414) arc (135:315:.2);
\draw[thick,blue] (0-.1414,2+.1414) arc (135:315:.2);
\draw[thick,blue] (-1-.1414,1+.1414) arc (135:315:.2);
\draw[->,thick,blue] (-2,2)  node [left,black]{$h^{\Gold}_{0,-2}$} --  (-1-.1414,1+.1414);
\draw[->,thick,blue] (-1,3) node [above,black]{$h^{gen,\Gold}_{-1,-1}$} --  (0-.1414,2+.1414);
\draw[->,thick,blue] (-1+.1414,1-.1414) --  (0-.1414,0+.1414);
\draw[->,thick,blue] (0+.1414,0-.1414) -- (1-.05,-1+.05);
\draw[->,thick,blue] (0+.1414,2-.1414) --  (1-.1414,1+.1414);
\draw[->,thick,blue] (1+.1414,1-.1414) -- (2-.05,0+.05);
\draw[->,thick,darkgreen] (1,3) node [above,black]{$h^{gen,\Gold}_{-1,+1}$} -- (2-.05,2+.05);
\draw[->,thick,darkgreen] (1,3) -- (0+.07,2+.07);
\draw[->,thick,darkgreen] (-2,0) -- (-1-.05,-1+.05);
\draw[->,thick,darkgreen] (2,2) -- (1+.07,1+.07);
\draw[->,thick,darkgreen] (1-.07,1-.07) -- (0+.07,0+.07);
\draw[->,thick,darkgreen] (0-.07,2-.07) -- (-1+.07,1+.07);
\draw[->,thick,darkgreen] (-1-.07,1-.07) -- (-2+.05,0+.05);
\draw[->,thick,darkgreen] (0-.07,0-.07) -- (-1+.05,-1+.05);
\node[fill=blue,regular polygon, regular polygon sides=4,inner sep=1.6pt] at (-1,3) {};
\node[fill=white,regular polygon, regular polygon sides=4,inner sep=.8pt] at (-1,3) {};
\node[fill=blue,regular polygon, regular polygon sides=4,inner sep=1.6pt] at (1,3) {};
\node[fill=white,regular polygon, regular polygon sides=4,inner sep=.8pt] at (1,3) {};
\end{tikzpicture}
\caption{}
\end{subfigure}
\begin{subfigure}{.45\linewidth}
\centering
\begin{tikzpicture}[scale=1.2]
\definecolor{darkgreen}{rgb}{.0, 0.5, .1};
\definecolor{darkgreen}{rgb}{.8, 0.8, .8};
\definecolor{blue}{rgb}{.5, 0.5, .5};
\filldraw[blue] (-2,2) circle (2pt) ;
\filldraw[blue] (2,0) circle (2pt)  node [right,black]{$h^{\Memo}_{2,+2}$};
\filldraw[blue] (1,-1) circle (2pt) node [below,black]{$h^{gen,\Memo}_{3,+1}$};
\filldraw[white] (1,-1) circle (1pt) ;
\filldraw[darkgreen] (2,2) circle (2pt)  node [right,black]{$-\th^{\Memo}_{0,+2}$};
\filldraw[darkgreen] (-2,0) circle (2pt) node [left,black]{$h^{\Memo}_{2,-2}$};
\filldraw[darkgreen] (-1,-1) circle (2pt) node [below,black]{$h^{gen,\Memo}_{3,-1}$};
\filldraw[white] (-1,-1) circle (1pt) ;
\draw[thick,->,blue] (2,0)-- (1+.05,-1+.05);
\draw[thick,->,blue] (-1,3)-- (-2+.05,2+.05);
\draw[thick,darkgreen] (0+.1414/2,0+.1414/2) arc (45:-135:.1);
\draw[thick,darkgreen] (1+.1414/2,1+.1414/2) arc (45:-135:.1);
\draw[thick,darkgreen] (-1+.1414/2,1+.1414/2) arc (45:-135:.1);
\draw[thick,darkgreen] (0+.1414/2,2+.1414/2) arc (45:-135:.1);
\draw[thick,blue] (0-.1414,0+.1414) arc (135:315:.2);
\draw[thick,blue] (1-.1414,1+.1414) arc (135:315:.2);
\draw[thick,blue] (0-.1414,2+.1414) arc (135:315:.2);
\draw[thick,blue] (-1-.1414,1+.1414) arc (135:315:.2);
\draw[->,thick,blue] (-2,2)  node [left,black]{$-\th^{\Memo}_{0,-2}$} --  (-1-.1414,1+.1414);
\draw[->,thick,blue] (-1,3) node [above,black]{$h^{gen,\Memo}_{-1,-1}$} --  (0-.1414,2+.1414);
\draw[->,thick,blue] (-1+.1414,1-.1414) --  (0-.1414,0+.1414);
\draw[->,thick,blue] (0+.1414,0-.1414) -- (1-.05,-1+.05);
\draw[->,thick,blue] (0+.1414,2-.1414) --  (1-.1414,1+.1414);
\draw[->,thick,blue] (1+.1414,1-.1414) -- (2-.05,0+.05);
\draw[->,thick,darkgreen] (1,3) node [above,black]{$h^{gen,\Memo}_{-1,+1}$} -- (2-.05,2+.05);
\draw[->,thick,darkgreen] (1,3) -- (0+.07,2+.07);
\draw[->,thick,darkgreen] (-2,0) -- (-1-.05,-1+.05);
\draw[->,thick,darkgreen] (2,2) -- (1+.07,1+.07);
\draw[->,thick,darkgreen] (1-.07,1-.07) -- (0+.07,0+.07);
\draw[->,thick,darkgreen] (0-.07,2-.07) -- (-1+.07,1+.07);
\draw[->,thick,darkgreen] (-1-.07,1-.07) -- (-2+.05,0+.05);
\draw[->,thick,darkgreen] (0-.07,0-.07) -- (-1+.05,-1+.05);
\node[fill=blue,regular polygon, regular polygon sides=4,inner sep=1.6pt] at (-1,3) {};
\node[fill=white,regular polygon, regular polygon sides=4,inner sep=.8pt] at (-1,3) {};
\node[fill=blue,regular polygon, regular polygon sides=4,inner sep=1.6pt] at (1,3) {};
\node[fill=white,regular polygon, regular polygon sides=4,inner sep=.8pt] at (1,3) {};
\end{tikzpicture}
\caption{}
\end{subfigure}
\caption{Goldstone (a) and memory (b) diamonds for the subleading soft graviton theorem.}
\label{fig:subgravitondiamond}
\end{figure}
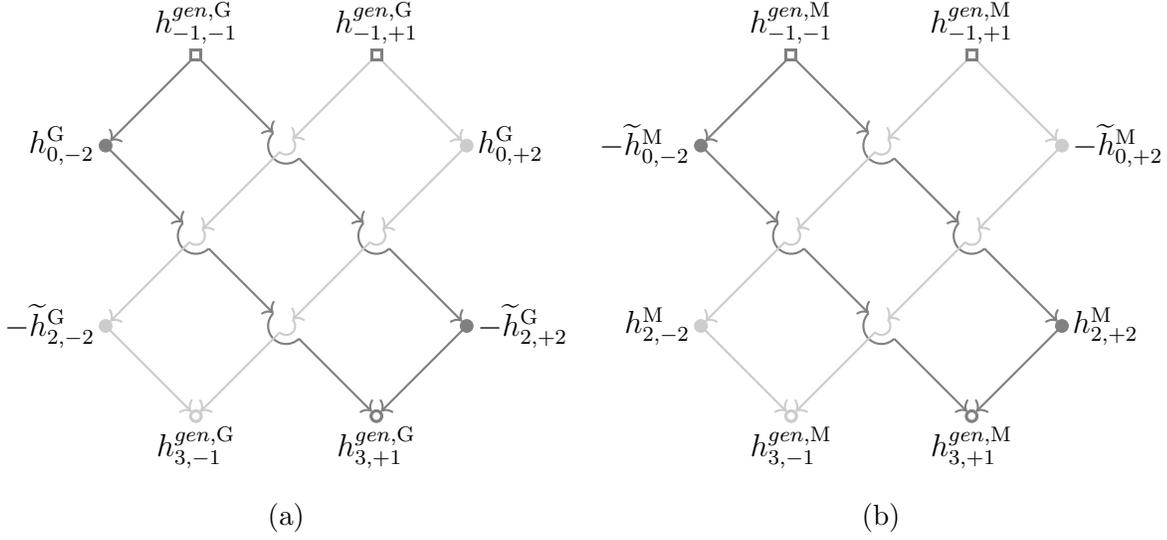

\noindent The celestial diamonds relevant for the subleading soft graviton theorem are chiral, i.e. for the Goldstone and memory modes there are two diamonds each with opposite self-duality property. The dark grey diamonds figure~\ref{fig:subgravitondiamond}a) and~b) correspond to self-dual Goldstone and anti-self dual memory modes while the ones in light grey correspond to the opposite self-duality.
Table~\ref{table:Gsubgravitons} gives the elements of the self-dual Goldstone diamond, for which 
\be
h^{\Gold}_{\Delta,J;\mu\nu}=\nabla_\mu\xi_{\Delta,J;\nu}+\nabla_\nu\xi_{\Delta,J;\mu}
\ee
while table~\ref{table:Msubgravitons} gives elements of the anti-self dual memory diamond. 
\vspace{1em}

\noindent \underline{\it Left and Right Corners}~~
The subleading conformally soft graviton theorem arises from the spin-2 conformal primary wavefunction with $\Delta=0$ and $J=\pm2$ which extends the BMS group to Diff($S^2$) transformations while the shadow of this mode at $\Delta=2$ corresponds to superrotations~\cite{Donnay:2020guq}. The non-diffeo modes with canonically paired conformal dimensions are given by the $\Delta=0$ shadow gravitons and the $\Delta=2$ gravitons.
Their $i\varepsilon$--regulated combinations of in-coming and out-going modes will be addressed in~\cite{upcoming3}. 
\vspace{1em}

\noindent \underline{\it Bottom Corners}~~~Within each diamond, the left and right corners descend to the same generalized primaries at $\Delta=3,J=\pm 1$, which are type~II primary descendants, as follows
\begin{equation}\label{hsubsoftspinrelationG}
   \frac{1}{3!}\partial_w^3 h^{{\Gold/\Memo}}_{0,-2}=h^{gen,{\Gold/\Memo}}_{3,+1}=- \partial_\bw \th^{\Gold/\Memo}_{2,+2}\,,~~~-\partial_w \th^{\Gold/\Memo}_{2,-2}=h^{gen,{\Gold/\Memo}}_{3,-1}=\frac{1}{3!}\partial_\bw^3 h^{\Gold/\Memo}_{0,+2}\,.
\end{equation}
The explicit expression for the primary descendant $h^{gen,\Gold/\Memo}_{3,+1}$ is given in tables~\ref{table:Gsubgravitons} and~\ref{table:Msubgravitons} from which the expressions for $h^{gen,\Gold/\Memo}_{3,-1}$ can be obtained by replacing $J\mapsto -J$ and $m \mapsto \bar{m}$. In figure~\ref{fig:subgravitondiamond} the generalized metric $h^{gen,\Gold/\Memo}_{3,+1}$ fills the bottom corner of the dark grey diamonds, while $h^{gen,\Gold/\Memo}_{3,-1}$ fills the bottom corner of the light grey diamonds.

\vspace{1em}

\noindent \underline{\it Top Corners}~~
The Goldstone modes of spontaneously broken Virasoro and Diff($S^2$) superrotation symmetry obey descendancy relations~\cite{Cheung:2016iub,Donnay:2020guq} which hint at the existence of a parent primary. Indeed, we find that we need to augment the Hilbert space with a pair of generalized primary metrics with $(\Delta,J)=(-1,\pm1)$. The level~1 and level~3 primary descendants of the $J=-1$ generalized metric land us on the spin-2 wavefunction with $\Delta=0$ and its shadow with $\Delta=2$
\begin{equation}
    \partial_\bw h^{gen,\Gold/\Memo}_{-1,-1}=h^{\Gold/\Memo}_{0,-2}\,, \quad \frac{1}{3!}\partial_w^3 h^{gen,\Gold/\Memo}_{-1,-1}=-\th^{\Gold/\Memo}_{2,+2}\,.
\end{equation}
Similarly, the primary descendants of the $J=+1$ generalized metric land us on the opposite helicity spin-2 wavefunctions
\begin{equation}
    \frac{1}{3!}\partial_\bw^3 h^{gen,\Gold/\Memo}_{-1,+1}=-\th^{\Gold/\Memo}_{2,-2}\,, \quad \partial_w h^{gen,\Gold/\Memo}_{-1,+1}=h^{\Gold/\Memo}_{0,+2}\,.
\end{equation}

\clearpage
\pagebreak

\vspace{1em}
\begin{table}[h!]
\renewcommand*{\arraystretch}{1.3}
\centering
    \begin{tabular}{l|l|l|l|l}
     Corner & $\Delta$  & $J$ 
       &  $h^{\Gold}_{\Delta,J}$ & $\xi_{\Delta,J}$ \\
         \hline
         Top&$-1$&$-1$
         & $\frac{1}{2\sqrt{2}}(l_\mu \bar{m}_\nu+\bar{m}_\mu l_\nu)\varphi^{-1}$&$- \frac{1}{2\sqrt{2}}\bar{m}^\mu \varphi^{-1}\log\varphi^{-1} $\\
          Left&$~~0$&$-2$
          &$\bar{m}_\mu\bar{m}_\nu$&$\p_\bw\xi^\mu_{-1,-1}$\\
           Right&$~~2$&$+2$
           &$X^2{m}_\mu{m}_\nu\varphi^{2}
           $            &$\frac{1}{3!}\p_w^3\xi^\mu_{-1,-1}$
           \\
           Bottom&$~~3$&$+1$
           &${\sqrt{2}} X^2\Big[\frac{X^2}{2} (l_\mu m_\nu+m_\mu l_\nu)$\,&$\frac{1}{3!}\p_w^3\p_\bw\xi^\mu_{-1,-1}$\\
           &&&$~~~~~~~+(n_\mu m_\nu+m_\mu n_\nu)\Big]\varphi^3$&\\
    \end{tabular}
     \caption{Elements of the celestial diamond corresponding to superrotation symmetry.} 
    \label{table:Gsubgravitons}
\end{table}

\begin{table}[h!]
\renewcommand*{\arraystretch}{1.3}
\centering
    \begin{tabular}{l|l|l|l}
     Corner & $\Delta$  & $J$ 
       &  $h^{\Memo}_{\Delta,J}$ \\
         \hline
         Top&$-1$&$-1$
         & $\frac{1}{2\sqrt{2}X^2}(l_\mu \bar{m}_\nu+\bar{m}_\mu l_\nu)\varphi^{-1}$\\
          Left&$~~0$&$-2$
          &$\frac{1}{X^2}\bar{m}_\mu\bar{m}_\nu 
          $\\
           Right&$~~2$&$+2$
           &${m}_\mu{m}_\nu \varphi^{2}$\\
           Bottom&$~~3$&$+1$
           &${\sqrt{2}}\Big[\frac{X^2}{2} (l_\mu m_\nu+m_\mu l_\nu)$\\
           &&&$~~~+(n_\mu m_\nu+m_\mu n_\nu)\Big]\varphi^3$\\
    \end{tabular}
     \caption{Elements of the celestial diamond corresponding to spin memory.}
    \label{table:Msubgravitons}
\end{table}

\subsubsection*{Zero-Area Celestial Diamonds}
\label{sec:degeneratediamond}

From the general classification~\eqref{DefTypeIII} it follows that there are only two type~III primary descendant wavefunctions for a given SL(2,$\mathbb{C}$) spin $J$ which is flipped with respect to that of the primaries they descended from:
\be
\partial_\bw^{2|J|}\Phi_{\Delta,+|J|},~~ \partial_w^{2|J|}\Phi_{\Delta,-|J|} \quad \Delta=1-|J|\,,
\ee
and
\be
\partial_\bw^{2|J|}\widetilde{\Phi}_{\Delta,+|J|},~~ \partial_w^{2|J|}\widetilde{\Phi}_{\Delta,-|J|} \quad \Delta=1-|J|\,.
\ee
Type III radiative conformal primaries correspond to the wavefunctions of tables~\ref{table:gold} and~\ref{table:shadowgold} in section~\ref{sec:CPW} denoted by $\gold$ which are not pure gauge but give rise to conformally soft theorems, and the canonically paired wavefunctions labeled by $\memo$. 
The conformally soft photino theorem~\cite{Fotopoulos:2020bqj,PPP} arises from the spin-$\frac{1}{2}$ primary with $\Delta=\frac{1}{2}$; the spin-1 primary with $\Delta=0$ gives rise to the subleading conformally soft photon theorem~\cite{Adamo:2019ipt,Guevara:2019ypd,Fotopoulos:2020bqj,upcoming2} (see also~\cite{Lysov:2014csa,Himwich:2019dug}); the subleading conformally soft gravitino theorem~\cite{Liu:2014vva,Fotopoulos:2020bqj} arises from the spin-$\frac{3}{2}$ primary with $\Delta=-\frac{1}{2}$; the spin-$2$ primary with $\Delta=-1$ gives rise to the subsubleading conformally soft graviton theorem~\cite{Guevara:2019ypd}. Their primary descendants, as well as those of their canonically paired wavefunctions, are summarized in table~\ref{table:typeIIIgm}.

\begin{table}[ht!]
    \centering
    \begin{tabular}{l|l|l|l}
   \multicolumn{2}{c|}{$\gold$ wavefunctions} & \multicolumn{2}{c}{$\memo$ wavefunctions} \\
     \hline
     \rule{0pt}{4ex}  
      $\p_\bw {\psi}_{\frac{1}{2},+\frac{1}{2}}={-}\widetilde{\psi}_{\frac{3}{2},-\frac{1}{2}}$ &$\p_w \overline{\psi}_{\frac{1}{2},-\frac{1}{2}}={-}\widetilde{\overline\psi}_{\frac{3}{2},+\frac{1}{2}}$&$\p_\bw \widetilde{\overline\psi}_{\frac{1}{2},+\frac{1}{2}}={\overline{\psi}}_{\frac{3}{2},-\frac{1}{2}}$ &$\p_w \widetilde{\psi}_{\frac{1}{2},-\frac{1}{2}}={\psi}_{\frac{3}{2},+\frac{1}{2}}$\\\rule{0pt}{4ex}  
       $\frac{1}{2!}\p_\bw^2A_{0,+1}=\tA_{2,-1}$ & $\frac{1}{2!}\p_w^2A_{0,-1}=\tA_{2,+1}$ & $\frac{1}{2!}\p_\bw^2\tA_{0,+1}=A_{2,-1}$ & $\frac{1}{2!}\p_w^2\tA_{0,-1}=A_{2,+1}$\\\rule{0pt}{4ex} 
$\frac{1}{3!}\p_\bw^3 {\chi}_{-\frac{1}{2},+\frac{3}{2}}=-\widetilde{ \chi}_{\frac{5}{2},-\frac{3}{2}}$  & $\frac{1}{3!}\p_w^3 \overline{\chi}_{-\frac{1}{2},-\frac{3}{2}}=-\widetilde{\overline \chi}_{\frac{5}{2},+\frac{3}{2}}$      & $\frac{1}{3!}\p_\bw^3 {\widetilde{\overline{\chi}}}_{-\frac{1}{2},+\frac{3}{2}}={\overline{\chi}}_{\frac{5}{2},-\frac{3}{2}}$
     & $\frac{1}{3!}\p_w^3 \widetilde{\chi}_{-\frac{1}{2},-\frac{3}{2}}={\chi}_{\frac{5}{2},+\frac{3}{2}}$\\\rule{0pt}{4ex} 
     $\frac{1}{4!}\p_\bw^4 h_{-1,+2}=\th_{3,-2}$ & $\frac{1}{4!}\p_w^4 h_{-1,-2}=\th_{3,+2}$ & $\frac{1}{4!}\p_\bw^4 \th_{-1,+2}=h_{3,-2} $ & $\frac{1}{4!}\p_w^4 \th_{-1,-2}=h_{3,+2} $
    \end{tabular}
    \caption{Type III primary descendant wavefunctions.  
    }
    \label{table:typeIIIgm}
\end{table}

Type~III primary descendant wavefunctions at level~$2|J|$ are related to the $\Delta=1-|J|$ parent primaries they descended from via a conformal shadow transform. Hence, type~III primary descendant wavefunctions are not null in the trivial sense. Instead, they are null in the same sense as type II: rather than vanishing identically, hitting the soft theorems with the same $2|J|$ derivatives needed to land on the primary descendant yields a contact term contribution.  
Notice that primary descendant wavefunctions of type~III correspond to a degenerate limit of both type~I and type~II: The celestial diamond shrinks to zero area and, because the primary descendant is the parent primary's shadow, no generalized primaries need to be introduced to complete the `diamond'. This assigns a special status to these most subleading soft theorems.


\pagebreak

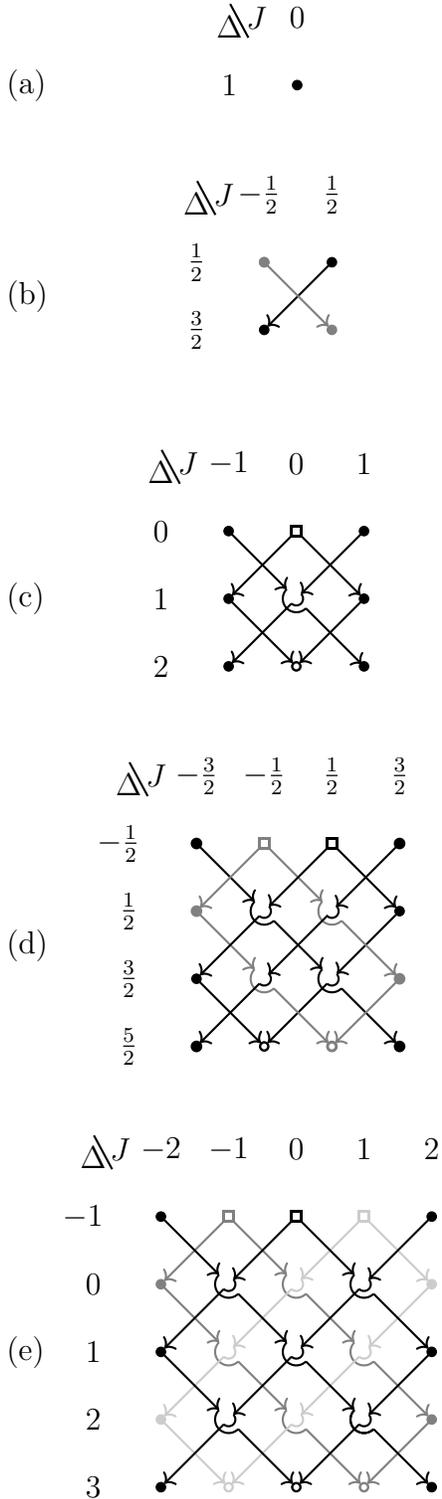
\begin{figure}[hp!]
\adjustbox{}{
\begin{minipage}[t]{.42\linewidth}
\vspace{0pt}
\begin{subfigure}{\linewidth}
\begin{tikzpicture}[scale=.9]
\definecolor{darkgreen}{rgb}{.0, 0.5, .1};
\filldraw[black] (0,2) circle (2pt) ;
\node at (-1,2) {$1$};
\node at (0,3) {$0$};
\node at (-0.6,3) {$J$};
\node at (-1,2.87) {$\tiny{\Delta}$};
\draw[thick] (-1+.01,3+.2) -- (-1+.28,3-.3);
\node at (-4,2) {(a)};
\node at (0,1) {};
\end{tikzpicture}
\label{scalardiamond}
\end{subfigure}
\vspace{1em}
\begin{subfigure}{\linewidth}
\begin{tikzpicture}[scale=.9]
\definecolor{darkgreen}{rgb}{.0, 0.5, .1};
\definecolor{red}{rgb}{.5, 0.5, .5};
\filldraw[red] (1,1) circle (2pt) ;
\filldraw[red] (0,2) circle (2pt) ;
\filldraw[black] (0,1) circle (2pt) ;
\filldraw[black] (1,2) circle (2pt) ;
\draw[thick,->] (1,2)-- (0+.05,1+.05);
\draw[thick,red,->] (0,2)-- (1-.05,1+.05);
\filldraw[red] (0,2) circle (2pt) ;
\node at (-1,1) {$\frac{3}{2}$};
\node at (-1,2) {$\frac{1}{2}$};
\node at (-0.6,3) {$J$};
\node at (-1,2.87) {$\tiny{\Delta}$};
\draw[thick] (-1+.01,3+.2) -- (-1+.28,3-.3);
\node at (0,3) {$-\frac{1}{2}~$};
\node at (1,3) {$\frac{1}{2}$};
\node at (-3.5,1.5) {(b)};
\node at (0,0) {};
\end{tikzpicture}
\end{subfigure}

\begin{subfigure}{\linewidth}
\begin{tikzpicture}[scale=.9]
\definecolor{darkgreen}{rgb}{.0, 0.5, .1};
\draw[thick,->] (-1+.05,1-.05)node[left]{
} --node[below left]{
} (-.05,.05) ;
\draw[thick,->] (1,1)-- (.05,.05);
\filldraw[black] (0,0) circle (2pt);
\filldraw[black] (-1,1) circle (2pt) ;
\draw[thick] (0+.1414/2,1+.1414/2) arc (45:-135:.1);
\filldraw[black] (1,1) circle (2pt) ;
\filldraw[black] (0,2) circle (2pt) ;
\draw[thick,->] (0,2)-- (1-.05,1+.05);
\node at (-2,0) {$2$};
\node at (-2,1) {$1$};
\node at (-2,2) {$0$};
\node at (-1.6,3) {$J$};
\node at (-2,2.87) {$\tiny{\Delta}$};
\draw[thick] (-2+.01,3+.2) -- (-2+.28,3-.3);
\node at (-1,3) {$-1$};
\node at (0,3) {$0$};
\node at (1,3) {$1$};
\filldraw[black] (1,0) circle (2pt) ;
\filldraw[black] (-1,0) circle (2pt) ;
\filldraw[black] (1,2) circle (2pt) ;
\filldraw[black] (-1,2) circle (2pt) ;
\draw[->,thick] (1,2) --  (0+.07,1+.07);
\draw[->,thick]  (0-.07,1-.07) -- (-1+.05,.05);
\draw[thick] (0-.1414,1+.1414) arc (135:315:.2);
\draw[->,thick] (-1,2) --  (0-.1414,1+.1414);
\draw[->,thick] (0+.1414,1-.1414) -- (1-.03,0.03);
\draw[thick,->] (0,2)-- (-1+.05,1.05);
\filldraw[white] (0,0) circle (1pt) ;
\node at (-4,1) {(c)};
\node[fill=black,regular polygon, regular polygon sides=4,inner sep=1.6pt] at (0,2) {};
\node[fill=white,regular polygon, regular polygon sides=4,inner sep=.8pt] at (0,2) {};
\node at (0,-1) {};
\end{tikzpicture}
\end{subfigure}

\begin{subfigure}{\linewidth}
\begin{tikzpicture}[scale=.9]
\definecolor{red}{rgb}{.5, 0.5, .5};
\draw[thick,->] (-1+.05,1-.05)node[left]{
} --node[below left]{
} (-.05,.05) ;
\draw[thick,->] (2-.05,2-.05)node[above]{
} --node[below right]{
} (1+.1414/2,1+.1414/2);
\draw[thick,->] (1-.1414/2,1-.1414/2)-- (.05,.05);
\filldraw[black] (0,0) circle (2pt);
\filldraw[black] (-1,1) circle (2pt) ;
\filldraw[black] (2,2) circle (2pt) ;
\draw[thick] (1+.1414/2,1+.1414/2) arc (45:-135:.1);
\draw[thick] (0+.1414/2,2+.1414/2) arc (45:-135:.1);
\node at (-2,0) {$\frac{5}{2}$};
\node at (-2,1) {$\frac{3}{2}$};
\node at (-2,2) {$\frac{1}{2}$};
\node at (-2,3) {$-\frac{1}{2}~~$};
\node at (-1.6,4) {$J$};
\node at (-2,3.87) {$\tiny{\Delta}$};
\draw[thick] (-2+.01,4+.2) -- (-2+.28,4-.3);
\node at (-1,4) {$-\frac{3}{2}$};
\node at (0,4) {$-\frac{1}{2}$};
\node at (1,4) {$\frac{1}{2}$};
\node at (2,4) {$\frac{3}{2}$};
\filldraw[black] (1,0) circle (2pt) ;
\filldraw[red,thick] (1,0) circle (2pt) ;
\filldraw[red,thick] (-1,2) circle (2pt) ;
\draw[red, thick] (0-.1414,1+.1414) arc (135:315:.2);
\draw[red, thick] (1-.1414,2+.1414) arc (135:315:.2);
\draw[black, thick] (0-.1414,2+.1414) arc (135:315:.2);
\draw[black, thick] (1-.1414,1+.1414) arc (135:315:.2);
\draw[->,red,thick] (-1,2) --  (0-.1414,1+.1414);
\draw[->,red,thick] (0+.1414,1-.1414) -- (1-.05,0.05);
\filldraw[red,thick] (2,1) circle (2pt) ;
\filldraw[black,thick] (-1,0) circle (2pt) ;
\filldraw[black,thick] (-1,3) circle (2pt) ;
\filldraw[black,thick] (2,3) circle (2pt) ;
\filldraw[black,thick] (2,0) circle (2pt) ;
\draw[->,thick] (-1,3) -- (0-.1414,2+.1414);
\draw[->,thick] (1+.1414,1-.1414) -- (2-.05,0+.05);
\draw[->,thick] (0+.1414,2-.1414) -- (1-.1414,1+.1414);
\draw[black,thick] (0+.1414/2,1+.1414/2) arc (45:-135:.1);
\draw[black,thick] (1+.1414/2,2+.1414/2) arc (45:-135:.1);
\draw[->,thick] (0-.1414/2,1-.1414/2) -- (-1+.05,0+.05);
\draw[->,thick] (1-.06,2-.06) -- (0+.06,1+.06);
\draw[->,thick] (2,3) -- (1+.06,2+.06);
\draw[->,red,thick] (2,1) -- (1+.05,0.05);
\draw[->,red,thick] (0,3) --  (1-.1414,2+.1414);
\draw[->,red,thick] (1+.1414,2-.1414) --  (2-.05,1+.05);
\draw[->,red,thick] (0,3) --  (-1+.05,2+.05);
\draw[thick,->] (1,3)-- (2-.05,2+.05);
\filldraw[black] (1,3) circle (2pt) ;
\draw[thick,->] (0-.1414/2,2-.1414/2)-- (-1+.05,1.05);
\draw[thick,->] (1,3) -- (0+.1414/2,2+.1414/2);
\filldraw[red,thick] (0,3) circle (2pt) ;
\filldraw[white] (1,0) circle (1pt) ;
\filldraw[white] (0,0) circle (1pt) ;
\node[fill=red,regular polygon, regular polygon sides=4,inner sep=1.6pt] at (0,3) {};
\node[fill=white,regular polygon, regular polygon sides=4,inner sep=.8pt] at (0,3) {};
\node[fill=black,regular polygon, regular polygon sides=4,inner sep=1.6pt] at (1,3) {};
\node[fill=white,regular polygon, regular polygon sides=4,inner sep=.8pt] at (1,3) {};
\node at (-3.5,1.5) {(d)};
\node at (0,-1) {};
\end{tikzpicture}
\end{subfigure}
\vspace{2em}
\begin{subfigure}{\linewidth}
\begin{tikzpicture}[scale=.9]
\definecolor{darkgreen}{rgb}{.0, 0.5, .1};
\definecolor{darkgreen}{rgb}{.8, 0.8, .8};
\definecolor{blue}{rgb}{.5, 0.5, .5};
\filldraw[black] (-2,1) circle (2pt) ;
\filldraw[black] (2,1) circle (2pt) ;
\filldraw[blue] (-2,2) circle (2pt) ;
\filldraw[blue] (2,0) circle (2pt) ;
\filldraw[blue] (1,-1) circle (2pt) ;
\filldraw[darkgreen] (2,2) circle (2pt) ;
\filldraw[darkgreen] (-2,0) circle (2pt) ;
\filldraw[darkgreen] (-1,-1) circle (2pt) ;
\draw[thick,->,blue] (2,0)-- (1+.05,-1+.05);
\draw[thick,->,blue] (-1,3)-- (-2+.05,2+.05);
\draw[thick] (0+.1414/2,1+.1414/2) arc (45:-135:.1);
\draw[thick] (1+.1414/2,2+.1414/2) arc (45:-135:.1);
\draw[thick] (-1+.1414/2,0+.1414/2) arc (45:-135:.1);
\draw[thick] (1+.1414/2,0+.1414/2) arc (45:-135:.1);
\draw[thick] (-1+.1414/2,2+.1414/2) arc (45:-135:.1);
\draw[thick,darkgreen] (0+.1414/2,0+.1414/2) arc (45:-135:.1);
\draw[thick,darkgreen] (1+.1414/2,1+.1414/2) arc (45:-135:.1);
\draw[thick,darkgreen] (-1+.1414/2,1+.1414/2) arc (45:-135:.1);
\draw[thick,darkgreen] (0+.1414/2,2+.1414/2) arc (45:-135:.1);
\draw[thick,blue] (0-.1414,0+.1414) arc (135:315:.2);
\draw[thick,blue] (1-.1414,1+.1414) arc (135:315:.2);
\draw[thick,blue] (0-.1414,2+.1414) arc (135:315:.2);
\draw[thick,blue] (-1-.1414,1+.1414) arc (135:315:.2);
\draw[->,thick,blue] (-2,2) --  (-1-.1414,1+.1414);
\draw[->,thick,blue] (-1,3) --  (0-.1414,2+.1414);
\draw[->,thick,blue] (-1+.1414,1-.1414) --  (0-.1414,0+.1414);
\draw[->,thick,blue] (0+.1414,0-.1414) -- (1-.05,-1+.05);
\draw[->,thick,blue] (0+.1414,2-.1414) --  (1-.1414,1+.1414);
\draw[->,thick,blue] (1+.1414,1-.1414) -- (2-.05,0+.05);
\draw[->,thick,darkgreen] (1,3) -- (2-.05,2+.05);
\draw[->,thick,darkgreen] (1,3) -- (0+.07,2+.07);
\draw[->,thick,darkgreen] (-2,0) -- (-1-.05,-1+.05);
\draw[->,thick,darkgreen] (2,2) -- (1+.07,1+.07);
\draw[->,thick,darkgreen] (1-.07,1-.07) -- (0+.07,0+.07);
\draw[->,thick,darkgreen] (0-.07,2-.07) -- (-1+.07,1+.07);
\draw[->,thick,darkgreen] (-1-.07,1-.07) -- (-2+.05,0+.05);
\draw[->,thick,darkgreen] (0-.07,0-.07) -- (-1+.05,-1+.05);
\draw[->,thick] (0,3) --  (1-.1414,2+.1414);
\draw[->,thick] (0,3) --  (-1+.07,2+.07);
\draw[->,thick] (-1-.07,2-.07)-- (-2+.05,1+.05);
\draw[->,thick] (1-.07,0-.07)-- (0+.05,-1+.05);
\filldraw[black] (-2,3) circle (2pt) ;
\filldraw[black] (2,3) circle (2pt) ;
\filldraw[black] (0,3) circle (2pt) ;
\filldraw[blue] (-1,3) circle (2pt) ;
\filldraw[white] (-1,-1) circle (1pt) ;
\filldraw[darkgreen] (1,3) circle (2pt) ;
\filldraw[white] (1,-1) circle (1pt) ;
\filldraw[black] (2,-1) circle (2pt) ;
\filldraw[black] (0,-1) circle (2pt) ;
\filldraw[black] (-2,-1) circle (2pt) ;
\filldraw[white] (0,-1) circle (1pt) ;
\node at (-3,-1) {$3$};
\node at (-3,0) {$2$};
\node at (-3,1) {$1$};
\node at (-3,2) {$0$};
\node at (-3,3) {$-1~~$};
\node at (-2.6,4) {$J$};
\node at (-3,3.87) {$\tiny{\Delta}$};
\node at (-2,4) {$-2$};
\draw[thick] (-3+.01,4+.2) -- (-3+.28,4-.3);
\node at (-1,4) {$-1$};
\node at (0,4) {$0$};
\node at (1,4) {$1$};
\node at (2,4) {$2$};
\draw[->,thick] (2,1) --  (1+.07,0+.07);
\draw[->,thick] (1-.07,2-.07) --  (0+.07,1+.07);
\draw[->,thick]  (0-.07,1-.07) -- (-1+.07,.07);
\draw[->,thick]  (2,3) -- (1+.07,2.07);
\draw[->,thick]  (-1-.07,0-.07) -- (-2+.05,-1+.05);
\draw[thick] (0-.1414,1+.1414) arc (135:315:.2);
\draw[thick] (1-.1414,0+.1414) arc (135:315:.2);
\draw[thick] (-1-.1414,2+.1414) arc (135:315:.2);
\draw[thick] (1-.1414,0+.1414) arc (135:315:.2);
\draw[->,thick] (-1+.1414,2-.1414) --  (0-.1414,1+.1414);
\draw[->,thick] (0+.1414,1-.1414) -- (1-.1414,.1414);
\draw[->,thick] (1+.1414,0-.1414) -- (2-.05,-1+.05);
\draw[->,thick] (-2,3) --  (-1-.1414,2+.1414);
\draw[->,thick] (-2,1) --  (-1-.1414,0+.1414);
\draw[thick] (-1-.1414,0+.1414) arc (135:315:.2);
\draw[->,thick] (-1+.1414,0-.1414) -- (0-.05,-1+.05);
\draw[->,thick] (1+.1414,2-.1414) -- (2-.05,1+.05);
\draw[thick] (1-.1414,2+.1414) arc (135:315:.2);
\node[fill=blue,regular polygon, regular polygon sides=4,inner sep=1.6pt] at (-1,3) {};
\node[fill=white,regular polygon, regular polygon sides=4,inner sep=.8pt] at (-1,3) {};
\node[fill=black,regular polygon, regular polygon sides=4,inner sep=1.6pt] at (0,3) {};
\node[fill=white,regular polygon, regular polygon sides=4,inner sep=.8pt] at (0,3) {};
\node[fill=darkgreen,regular polygon, regular polygon sides=4,inner sep=1.6pt] at (1,3) {};
\node[fill=white,regular polygon, regular polygon sides=4,inner sep=.8pt] at (1,3) {};
\node at (-4,1) {(e)};
\end{tikzpicture}
\end{subfigure}
\label{celestialdiamonds}

\end{minipage}\hfill
\begin{minipage}[t]{.53\linewidth}
\vspace{0pt}\raggedright
\caption{Celestial diamonds demonstrating non-trivial primary descendants. Solid dots correspond to radiative primaries while open dots and squares correspond to generalized primaries.  Skipped nodes correspond to non-primary descendants. Subfigures (a)-(e) cover the $s=0$ through $s=2$ examples in half-integer steps.  In each case the SL(2,$\mathbb{C}$) spin $J$ is bounded by $|J|\le s$.  
Radiative states lie at $J=\pm s$.  Operators corresponding to conformally soft dressings lie at $\Delta=1-s$~\cite{PPT2}.  The structure for generic $s$ is summarized in subfigure (f). The most subleading soft theorems correspond to zero-area diamonds which appear as the diagonals in each subfigure. In integer spin cases the leading soft theorems are non-chiral and the celestial diamond then implies a relation between soft theorems of opposite helicity. 
\label{celestialdiamonds2}}
\begin{subfigure}{\linewidth}
\centering
\begin{tikzpicture}[scale=.85]
\definecolor{darkgreen}{rgb}{.6, 0.6, .6};
\definecolor{red}{rgb}{.7, 0.7, .7};
\node[darkgreen] at (-3,0.5) {$-s~$};
\node[darkgreen] at (0,0.5) {$0$};
\draw [red,decorate,decoration={brace,amplitude=5pt,raise=4ex}]
  (-3,0) -- (3,0);
  \draw [red,decorate,decoration={brace,amplitude=5pt,raise=4ex}]
  (3.8,0) -- (3.8,-6);
\node[darkgreen] at (0,1.5) {$J$};
\node[darkgreen] at (5.3,-3) {$\Delta$};
\node[darkgreen] at (3,0.5) {$s$};
\draw[red] (-3,0)-- (3,-6);
\draw[red] (-2,0)-- (3,-5) -- (2,-6) -- (-3,-1) -- (-2,0);
\node at (0,1) {};
\draw[black] (-.2,-3)-- (.2,-3);
\draw[black] (0,-3+.2)-- (0,-3-.2);
\draw[dashed,red] (0,0) --(0,-6);
\draw[dashed,red] (-3,-3) --(3,-3);
\node[darkgreen,right] at (3,-3) {$~1$};
\node[darkgreen,right] at (3,0) {$~1-s$};
\node[darkgreen,right] at (3,-6) {$~1+s$};
\node[fill=black,regular polygon, regular polygon sides=4,inner sep=1.6pt] at (-2,0) {};
\node[fill=white,regular polygon, regular polygon sides=4,inner sep=.8pt] at (-2,0) {};
\node[fill=black,regular polygon, regular polygon sides=4,inner sep=1.6pt] at (2,0) {};
\node[fill=white,regular polygon, regular polygon sides=4,inner sep=.8pt] at (2,0) {};
\filldraw[black] (-3,0) circle (2pt) ;
\filldraw[black] (-2,-6) circle (2pt) ;
\filldraw[white] (-2,-6) circle (1pt) ;
\filldraw[black] (2,-6) circle (2pt) ;
\filldraw[white] (2,-6) circle (1pt) ;
\node at (2,-6.5) {II};
\node at (3,-6.5) {III};
\node at (3.5,-5) {Ib};
\node at (-3.5,-1) {Ia};
\node at (0,-7.5) {(f)};
\node at (-4.5,0) {};
\node at (0,3) {};
\filldraw[black] (-3,-1) circle (2pt) ;
\filldraw[black] (-3,-5) circle (2pt) ;
\filldraw[black] (3,-1) circle (2pt) ;
\filldraw[black] (3,-5) circle (2pt) ;
\filldraw[black] (-3,-3) circle (1pt) ;
\filldraw[black] (-3,-3.3) circle (1pt) ;
\filldraw[black] (-3,-2.7) circle (1pt) ;
\filldraw[black] (3,-3) circle (1pt) ;
\filldraw[black] (3,-3.3) circle (1pt) ;
\filldraw[black] (3,-2.7) circle (1pt) ;
\filldraw[black] (3,0) circle (2pt) ;
\filldraw[black] (3,-6) circle (2pt) ;
\filldraw[black] (-3,-6) circle (2pt) ;
\end{tikzpicture}
\end{subfigure}
\end{minipage}
}
\end{figure}

\subsection{Symmetries in Celestial Diamonds}
We end this section with a discussion of our results, a few comments contrasting the nature of the type I,~II and~III primary descendants of section~\ref{sec:diamond} with the trivial null states of section~\ref{sec:NullStates}, and their relation to global primary descendants in standard CFTs.

\subsubsection*{Celestial Primary Descendants}
Thus far we have focused on the corners of the diamond which are primaries.
For finite-area celestial diamonds, the radiative wavefunctions at the left and right corners give rise to celestial currents generating spontaneously broken asymptotic symmetries: large U(1) Kac-Moody symmetries ($s=1)$
, large supersymmetries ($s=\frac{3}{2}$)
, supertranslations and superrotations ($s=2$), related to the leading conformally soft photon and gravitino theorems and the leading and subleading conformally soft graviton theorems, respectively.
The radiative conformal primaries at the top corners of the zero-area celestial diamonds are associated to the most subleading universal soft theorems for $s=\{0,\frac{1}{2},1,\frac{3}{2},2\}$. While they may lack an obvious asymptotic symmetry interpretation they are related via the classical double copy or supersymmetry to the aforementioned primaries with symmetry interpretations. There are finitely many wavefunctions of this type for a given spin as discussed in section~\ref{sec:diamond}. 

This is in contrast to the infinitely many radiative conformal primary wavefunctions of section~\ref{sec:NullStates}. These generate an infinite tower of symmetries which is, however, captured by commutators of the symmetry generators associated to the finite- and zero-area celestial diamonds so that they give no new constraints on the $\S$-matrix~\cite{Guevara:2021abz}. Their type~I primary descendants are trivially null. 

The type~II primary descendants at the bottom corners of finite-area celestial diamonds are given by non-vanishing generalized conformal primaries which satisfy the same equations of motions and gauge fixings as their radiative parent primaries. The type~III primary descendants of zero-area celestial diamonds are special in the sense that they are radiative primaries which are related by a shadow transform to their radiative parent primaries. Zero-area celestial diamonds are thus completely described by radiative data. Completing the diamond structure associated to the more leading soft theorems, on the other hand, requires adding generalized conformal primaries at the top. Then, the radiative primaries at the left and right corners are themselves primary descendants of type~I.

Notice that this completion is ambiguous, due to the fact that some generalized wavefunctions belong to the kernel of the derivative operators that define the left and right corners. Let us illustrate this point for the example of the leading soft graviton diamond, where the left and right corners are given by the radiative wavefunctions $m_{\mu} m_{\nu} \varphi^{1}$ and $\bar m_{\mu} \bar m_{\nu} \varphi^{1}$ of table  \ref{table:Ggravitons}.
The top wavefunction must be a linear combination of the generalized conformal primary wavefunctions of $\Delta=-1$ and $J=0$,
\be
\label{top_lincomb}
[a_1 l_\mu l_\nu +a_2 \Big(\frac{2}{X^2}\Big)^2  n_\mu n_\nu+a_3\frac{4}{X^2}(n_\mu l_\nu+l_\mu n_\nu)+a_4\frac{2}{X^2}\eta_{\mu\nu}]\varphi^{-1} \, ,
\ee
for arbitrary functions $a_i(X^2)$.
The other requirement is that \eqref{top_lincomb} must descend to the radiative wavefunctions by applying $\p_w^2$ and $\p_\bw^2$. This does not uniquely fix the linear combination since all wavefunctions with  $a_1+a_2+4a_3=0$ belong to the kernel of $\{\p_w^2,\p_\bw^2\}$.

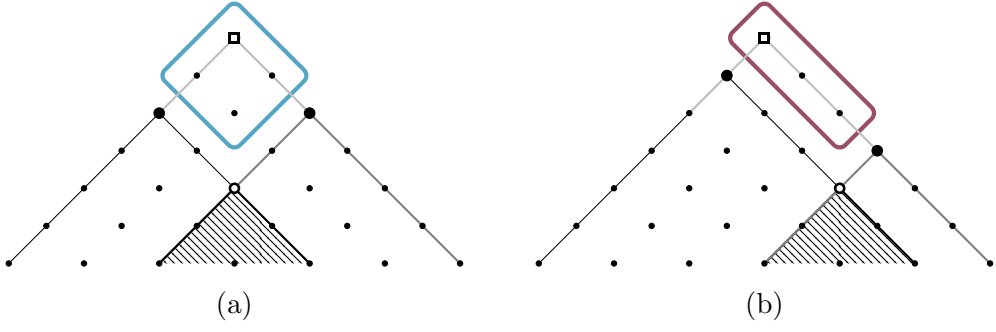
\begin{figure}[t!]
\centering
\begin{subfigure}{.4\linewidth}
\centering
\begin{tikzpicture}[scale=1]
\definecolor{darkgreen}{rgb}{.8, 0.8, .8};
\definecolor{blue}{rgb}{.5, 0.5, .5};
\filldraw[draw=cyan!50!gray, 
fill=white, rounded corners=3, ultra thick] (-1/2,3/2+.5) -- (0+.5,2/2) -- (-1/2,1/2-.5)
-- (-2/2-.5,2/2) -- cycle;
\filldraw[draw=none,pattern=north west lines] (-1/2,-1/2) -- (1/2,-3/2)
-- (-3/2,-3/2) -- cycle;
\draw[black] (-3/2,1/2) -- (1/2,-3/2);
\draw[black] (-3/2,1/2) -- (-7/2,-3/2);
\draw[gray, thick] (1/2,1/2) -- (-3/2,-3/2);
\draw[gray,thick] (1/2,1/2) -- (5/2,-3/2);
\draw[lightgray, thick] (-1/2,3/2) -- (1/2,1/2);
\draw[lightgray, thick] (-1/2,3/2) -- (-3/2,1/2);
\draw[black, thick] (-1/2,-1/2) -- (-3/2,-3/2);
\draw[black, thick] (-1/2,-1/2) -- (1/2,-3/2);
\filldraw[black] (-1/2,3/2) circle (2pt) ;
\filldraw[] (-1/2,1/2) circle (1pt) ;
\filldraw[] (-1/2,-3/2) circle (1pt) ;
\filldraw[] (0,-2/2) circle (1pt) ;
\filldraw[] (1/2,-3/2) circle (1pt) ;
\filldraw[] (-2/2,2/2) circle (1pt) ;
\filldraw[] (0,2/2) circle (1pt) ;
\filldraw[] (-2/2,0/2) circle (1pt) ;
\filldraw[] (-1/2,-1/2) circle (1pt) ;
\filldraw[] (1/2,1/2) circle (1pt) ;
\filldraw[] (2/2,0/2) circle (1pt) ;
\filldraw[] (1/2,-1/2) circle (1pt) ;
\filldraw[] (2/2,-2/2) circle (1pt) ;
\filldraw[] (3/2,-3/2) circle (1pt) ;
\filldraw[] (4/2,-2/2) circle (1pt) ;
\filldraw[] (5/2,-3/2) circle (1pt) ;
\filldraw[] (-3/2,-3/2) circle (1pt) ;
\filldraw[] (-5/2,-3/2) circle (1pt) ;
\filldraw[] (-7/2,-3/2) circle (1pt) ;
\filldraw[] (-6/2,-2/2) circle (1pt) ;
\filldraw[] (-4/2,-2/2) circle (1pt) ;
\filldraw[] (-2/2,-2/2) circle (1pt) ;
\filldraw[] (-5/2,-1/2) circle (1pt) ;
\filldraw[] (-4/2,0) circle (1pt) ;
\filldraw[] (-3/2,-1/2) circle (1pt) ;
\filldraw[] (0/2,0/2) circle (1pt) ;
\filldraw[black] (-3/2,1/2) circle (2pt) ;
\filldraw[black] (1/2,1/2) circle (2pt) ;
\filldraw[black] (-1/2,-1/2) circle (2pt) ;
\filldraw[white] (-1/2,-1/2) circle (1pt) ;
\filldraw[] (3/2,-1/2) circle (1pt) ;
\node[fill=black,regular polygon, regular polygon sides=4,inner sep=1.6pt] at (-1/2,3/2) {};
\node[fill=white,regular polygon, regular polygon sides=4,inner sep=.8pt] at (-1/2,3/2) {};
\end{tikzpicture}
\caption{}
\label{quotient}
\end{subfigure}
\begin{subfigure}{.4\linewidth}
\centering
\begin{tikzpicture}[scale=1]
\usetikzlibrary{patterns}
\definecolor{darkgreen}{rgb}{.8, 0.8, .8};
\definecolor{blue}{rgb}{.5, 0.5, .5};
\filldraw[draw=purple!40!gray, 
 fill=white, rounded corners=3, ultra thick] (-1/2,3/2+.5) -- (1/2+.5,1/2) -- (1/2,1/2-.5)
-- (-1/2-.5,3/2) -- cycle;
\filldraw[draw=none,pattern=north west lines] (-1/2+1,-1/2) -- (1/2+1,-3/2)-- (-3/2+1,-3/2) -- cycle;
\draw[black] (-3/2+.5,1/2+.5) -- (1/2+1,-3/2);
\draw[black] (-3/2+.5,1/2+.5) -- (-7/2,-3/2);
\draw[lightgray, thick] (-1/2,3/2) -- (1/2+.5,1/2-.5);
\draw[lightgray, thick] (-1/2,3/2) -- (-3/2,1/2);
\draw[black, thick] (-1/2+1,-1/2) -- (-3/2+1,-3/2);
\draw[black, thick] (-1/2+1,-1/2) -- (1/2+1,-3/2);
\draw[gray, thick] (1/2+.5,1/2-.5) -- (-3/2+1,-3/2);
\draw[gray,thick] (1/2+.5,1/2-.5) -- (5/2,-3/2);
\filldraw[black] (-1/2,3/2) circle (2pt) ;
\filldraw[] (-1/2,1/2) circle (1pt) ;
\filldraw[] (-3/2,1/2) circle (1pt) ;
\filldraw[] (-1/2,-3/2) circle (1pt) ;
\filldraw[] (0,-2/2) circle (1pt) ;
\filldraw[] (1/2,-3/2) circle (1pt) ;
\filldraw[] (-2/2,2/2) circle (1pt) ;
\filldraw[] (0,2/2) circle (1pt) ;
\filldraw[] (-2/2,0/2) circle (1pt) ;
\filldraw[] (-1/2,-1/2) circle (1pt) ;
\filldraw[] (1/2,1/2) circle (1pt) ;
\filldraw[] (2/2,0/2) circle (1pt) ;
\filldraw[] (1/2,-1/2) circle (1pt) ;
\filldraw[] (2/2,-2/2) circle (1pt) ;
\filldraw[] (3/2,-3/2) circle (1pt) ;
\filldraw[] (4/2,-2/2) circle (1pt) ;
\filldraw[] (5/2,-3/2) circle (1pt) ;
\filldraw[] (-3/2,-3/2) circle (1pt) ;
\filldraw[] (-5/2,-3/2) circle (1pt) ;
\filldraw[] (-7/2,-3/2) circle (1pt) ;
\filldraw[] (-6/2,-2/2) circle (1pt) ;
\filldraw[] (-4/2,-2/2) circle (1pt) ;
\filldraw[] (-2/2,-2/2) circle (1pt) ;
\filldraw[] (-5/2,-1/2) circle (1pt) ;
\filldraw[] (-4/2,0) circle (1pt) ;
\filldraw[] (-3/2,-1/2) circle (1pt) ;
\filldraw[] (0/2,0/2) circle (1pt) ;
\filldraw[black] (-3/2+.5,1/2+.5) circle (2pt) ;
\filldraw[black] (1/2+.5,1/2-.5) circle (2pt) ;
\filldraw[black] (-1/2+1,-1/2) circle (2pt) ;
\filldraw[white] (-1/2+1,-1/2) circle (1pt) ;
\filldraw[] (3/2,-1/2) circle (1pt) ;
\node[fill=black,regular polygon, regular polygon sides=4,inner sep=1.6pt] at (-1/2,3/2) {};
\node[fill=white,regular polygon, regular polygon sides=4,inner sep=.8pt] at (-1/2,3/2) {};
\end{tikzpicture}
\caption{}
\label{quotient2}
\end{subfigure}
\caption{Nested modules of the (a) leading and (b) subleading soft graviton. 
}
\label{grav_modules}
\end{figure}

Notice that the ambiguity in ascending from the left and right corners to the top corners is proportional to the following kernels which are associated to global symmetries:\footnote{
 For the zero-area diamonds, the analogous ambiguity is fixed by the shadow relations between corners. 
}
\begin{equation}
\begin{array}{rlll}
    &s=1:\quad \{\p_w,\p_\bw\} \quad  & \Rightarrow \quad \{1\}\quad &\text{U(1)}\\
    &{\textstyle s=\frac{3}{2}}:\quad \{\p_w,\p_\bw^2\} \quad  & \Rightarrow \quad \{1,\bw\} \quad &\text{supersymmetry}\\
    &s=2:\quad \{\p_w^2,\p_\bw^2\} \quad  & \Rightarrow \quad \{1,w,\bw,w\bw\}\quad &\text{translations}\\
    &\;\quad\quad\quad\quad\{\p_w,\p_\bw^3\} \quad  & \Rightarrow \quad \{1,\bw,\bw^2\}\quad &\text{rotations and boosts}.\\
\end{array}
\end{equation}
Similar expressions with $w\leftrightarrow \bw$ hold for the opposite chirality/helicity diamonds.
Different representatives of the top corners are thus related by the global symmetries of the associated diamond.
Once the top corners are added to the phase space, we get additional states. We exemplify this for the leading and subleading soft graviton diamonds in figure~\ref{grav_modules} where the extra states are encircled. The counting of these states matches that of the global Poincar\'e symmetries which are written as finite dimensional modules in appendix~\ref{sec:ConfModules}.

The type~I primary descendants of the top corners in figure~\ref{grav_modules}, i.e. the left and right corners, are full-fledged conformal primary wavefunctions and thus not null, and neither are their infinite tower of non-primary descendants (\tikzcircle{1pt}). Moreover, their type~II primary descendants, i.e. the bottom corners, give rise to contact terms in celestial amplitudes as do their tower of descendants captured by the shaded SL(2,$\mathbb{C}$) submodules in figure~\ref{grav_modules}.

\subsubsection*{Relation to standard CFTs}

Global primary descendants appear in standard CFTs as descendants of special `protected' operators. 
By `setting to zero' the primary descendants one defines shortening conditions which the protected operators must satisfy.

For example, any full-fledged CFT has a stress tensor $\T_{ww}$ and its antiholomorphic counterpart $\T_{\bw\bw}$  with spin equal to $|J|=2$  and dimension $\Delta=2$. From \eqref{DefTypeII} one finds that they have a level-1 primary descendant of type II \eqref{DefTypeII}. These are set to zero generating a shortening condition $\partial_{\bw} \T_{ww}=0=\partial_{w} \T_{\bw\bw}$. 
Similarly, conserved currents (and conserved tensors of any spin) have primary descendants of type II at level~1.
It is important to stress that these shortening conditions should be understood as operator equations only valid up to contact terms. Indeed, when inserted in correlation functions they define the Ward identities which, upon appropriate contour integration, give charges associated to the conserved operators.
The identity operator is also a simple example of a protected operator of $J=0$ and $\Delta=0$. It has two level~1 primary descendants of type Ia and Ib defined by~\eqref{DefTypeI} which are fully shortening its multiplet since derivatives of the identity trivially vanish.

We see a similar behavior for our conformal primary wavefunctions. The primary descendant wavefunctions of type~I in section~\ref{sec:NullStates} are zero in the trivial sense, while the type~II primary descendants at the bottom corners of the finite-area celestial diamonds of section~\ref{sec:diamond} are not identically zero. It is thus consistent for these to select operators which give contact terms in correlation functions, i.e. for their parents at the left and right corners to be operators which satisfy conservation equations with non-trivial sources. The only novelty (with respect to standard unitary CFTs) is that the conservation equations may involve higher derivatives, namely that the primary descendants may appear at level higher than one.

The role of the top corners of the finite-area diamonds of section~\ref{sec:diamond} is more subtle since the type I primary descendants at the left and right corner are full-fledged operators (they can be associated to states in the Hilbert space and do not just give contact terms when inserted in a correlation function). At first sight this situation looks puzzling, but it has a very simple CFT counterpart e.g. in the theory of a single free boson $\phi$. In free boson theory there are two currents $\p_w \phi$ and  $\p_\bw \phi$ which correspond to the left and right corner of the diamond. Their type~II primary descendant at the bottom corner vanishes up to contact terms, due to the equations of motion $\p_w \p_\bw \phi = 0$. One can further place the field $\phi$ itself at the top of the diamond. This is possible because $\phi$ is not a well defined primary operator, i.e. it cannot be associated to a state in the Hilbert space of the theory (so the argument of equation  \eqref{nullness_primdesc} does not apply). However, the action of $L_n$ on $\phi$ is the same as on any other primary operator, so the conclusions of section~\ref{sec:global_primary_desc} apply and formally one can complete the diamond. 
\vspace{2em}

It is important to  stress that in this example the top and bottom operators are not associated to states in the Hilbert space. Nevertheless, they determine interesting properties of the theory: the bottom corner defines the Ward identities (and thus the charges) associated to the conserved currents at the left and right corner, and the top corner contains information on the zero modes (and can be used to build vertex operators). In upcoming work~\cite{PPT2} we will further explore  the the role of the operators at top and bottom corners of the celestial diamonds.

\section*{Acknowledgements}

The work of S.P. is supported by the Princeton Center for Theoretical Science. The work of A.P. and E.T. is supported by the European Research Council (ERC) under the European Union’s Horizon 2020 research and innovation programme (grant agreement No 852386). 

\appendix{

\section{Representation Theory of 2D CFTs 
}
\label{App:rep_theory}
In section \ref{sec:global_primary_desc} we showed how to build the primary descendants for the $2$D global conformal algebra. 
Since the $2$D case is very simple, we can classify the modules through a straightforward computation. On the other hand, for a generic algebra the computation may be much more intricate and it may be useful to apply a more sophisticated mathematical technology (for a nice review of the subject see the book \cite{Humphreys}). 
An application of this technology to conformal representation theory in generic dimensions was e.g. given in~\cite{Penedones:2015aga, Bourget:2017kik}.
In the following we will use the notation of section 6 of \cite{Penedones:2015aga} and we exemplify how this can be applied to $d=2$.   
\subsubsection*{Conformal Algebra in the Cartan-Weyl Basis}
Let us consider the conformal group of a $d$ dimensional Euclidean CFT. This is isomorphic to  $SO(d+1,1)$. 
The generators of the algebra are $D,P_{a},K_{a},J_{a b}$ (where Latin indices take the values $1,\dots d$), which respectively define infinitesimal dilatations, translations, special conformal transformations and $SO(d)$ rotations. 
Their respective Killing vectors are
\begin{equation}
    k_{D}=-i x^a \partial_a \, , \quad
     k^{a}_{P}=-i \partial^a\,, \quad
    k^{a}_{K}=-i (2 x^a x^b \partial_b-x^2 \partial^a)\, ,\quad
k^{a b}_{J}=i (x^a  \partial^b-x^b  \partial^a)\, . \quad
\end{equation}
The generators satisfy the following  commutation relations  
\be \label{ConformalAlgebra}
\begin{array}{lllllll}
&[D,P_a] &=& i P_a \ , \qquad\qquad\qquad\qquad\qquad \ \, &
[P_a,J_{b r}] &=&  i(\eta_{ab}P_r-\eta_{ar}P_b) \ ,
\\
&
[D,K_a]&=&-iK_a 
 \ , 
 \qquad \qquad &[K_a,J_{br}]&=&i(\eta_{ab}K_r-\eta_{ar}K_b) \ ,   \\
& [K_a,P_b ]& =& 2i (\eta_{a b} D- J_{ab}) \ , 
 \qquad \qquad &
[J_{ab},J_{r s}]&=& i(\eta_{b r} J_{a s} \pm \textrm{perm}) \ .
\end{array}
\ee
A bosonic primary state can be written  as  $|\Delta , b_{1} \dots b_{s} \rangle$, where $\Delta $ is the conformal dimension and the $s$ indices correspond to an irreducible tensor of $SO(d)$ (possibly with mixed symmetry properties). 
In this tensor representation, the action of the generators is given by
\begin{align}
&D |\Delta , b_{1} \dots b_{s} \rangle = i \Delta |\Delta , b_{1} \dots b_{s} \rangle\ ,\qquad
K_a |\Delta , b_{1} \dots b_{s} \rangle =0\ ,
\label{eq:actionDK}\\
&J_{r s} | \Delta ,b_{1} \dots b_{s} \rangle =
\sum_{k=1}^s [\Sigma_{r s}]^{b_k}_{\ a}\ 
| \Delta ,b_{1} \dots b_{k-1} \, a \, b_{k+1} \dots b_{s}\rangle\ ,
\label{eq:actionJ}
\end{align}
where $[\Sigma_{r s}]^{b}_{\ a} = i \left(
\delta^{b}_s \eta_{r a} -\delta^{b}_r \eta_{s a}
\right)$.
The action of $P_a$ creates descendants.

When $d=2$ the algebra contains only $6$ generators, $K_1,K_2,P_1,P_2,D,J_{12}$.
To write them in the Cartan-Weyl basis we first define for a given vector $v_a$ the following combinations $v_w\equiv \frac{v_1-iv_2}{2}$ and $v_\bw \equiv \frac{v_1+iv_2}{2}$.
We can then formulate the conformal algebra in a Cartan-Weyl basis using the notation of \cite{Penedones:2015aga} 
\begin{align}
&
\begin{array}{c l}
E_{\alpha^{+ +}}&\equiv K_w=\frac{K_1-i K_2}{2} \ ,   \\
E_{\alpha^{+ -}}&\equiv K_\bw= \frac{K_1+i K_2}{2} 
\ ,   \\
E_{\alpha^{- +}}&\equiv P_{w}= \frac{P_1-i P_2}{2}   \ , \\
E_{\alpha^{- -}}&\equiv P_\bw= \frac{P_1+i P_2}{2}   \ , \\
\end{array}
 \qquad 
\begin{array}{c l}
H_0&\equiv i D  \ ,  \\
H_1&
\equiv  2 i  \;   J_{\bw w} = J_{12} 
\ ,  \\
\end{array}
\label{eq:CartanWeylbasis}
\end{align}
where we have associated generators to the four roots $\alpha^{\pm \pm}$ of $so(1,3)$, which are the following  $2$-dimensional vectors
\be
 \alpha^{++}=(+1,+1) \, , \quad 
 \alpha^{+-}=(+1,-1)  \, , \quad 
 \alpha^{-+}=(-1,+1)  \, , \quad 
 \alpha^{--}=(-1,-1) . \label{RootsSO2N+1}
\ee
We further divide the root system $\Phi$ into two parts $\Phi=\Phi_+ \cup \Phi_-$ where   $\Phi_+\equiv \{ \alpha^{++},\alpha^{+-} \}$ is the set of  positive roots, while $\Phi_-\equiv \{ \alpha^{-+},\alpha^{--} \}$ is the set of negative roots. 
The generators $H_0$ and $H_1$ commute among themselves and form the Cartan subalgebra.
The algebra in this basis satisfies the following commutation relations 
\be \label{CartanWeylAlgebra}
[H_k,E_\alpha]=(\alpha)^k E_\alpha\ , \qquad
[E_\alpha,E_{-\alpha}]=\frac{2}{\langle \alpha,\alpha \rangle}(\alpha)^k H_k\ .
\ee
where $\alpha$ are the root of $so(4)$, $(\alpha)^k$ denotes its $k+1$-th coordinate and the angle brakets define the standard scalar product $\langle \alpha, \beta \rangle= (\alpha)^0 (\beta)^0+(\alpha)^1 (\beta)^1$.
The algebra is now written in a more convenient form, however in $d=2$ a further simplification is possible. In particular, we see that the rules in \eqref{CartanWeylAlgebra} give
\begin{equation}
[E_{\alpha^{+-}},E_{\alpha^{-+}}]= H_0-H_1
\, , \qquad 
[E_{\alpha^{++}},E_{\alpha^{--}}]= H_0+H_1
\, . 
\end{equation}
It is thus natural to rearrange the Cartan subalgebra in order to make manifest the presence of two  mutually commuting subalgebras, which are indeed expected since  $so(1,3)^{\mathbb{C}} \cong sl(2,\mathbb{C}) \oplus sl(2,\mathbb{C})$. 
The following definition is the one which is most commonly used in $d=2$ CFTs,
\begin{equation}
\begin{array}{lll}
    L_1\equiv  -i E_{\alpha^{+ -}} \,, &
    L_{-1}\equiv -i E_{\alpha^{- +}} \,, &
    L_0 \equiv \frac{-H_0+H_1}{2} \,, \\
      \bar L_1\equiv -i E_{\alpha^{+ +}} \,, &
    \bar  L_{-1}\equiv  -i E_{\alpha^{- -}} \, ,&
   \bar L_0 \equiv \frac{-H_0-H_1}{2} \, ,
    \end{array}
\end{equation}
which correspond to the following Killing vectors
\begin{equation}
    k_{L_n}=- w^{n+1} \p_w \, ,
    \qquad
     k_{\bar L_n}=- \bw^{n+1} \p_\bw \, .
\end{equation}
This redefinition of generators is also useful for embedding the global conformal algebra into the Virasoro algebra
\begin{equation}
    [L_{m},L_{n}]=(m-n)L_{ n+m}+\frac{c}{12} (m^3-m) \delta_{n+m,0} \, . 
\end{equation}
 Notice that in our case the central extension does not play any role since for $m,n=-1,0,1$ $(m^3-m)$ is always zero.

Let us consider a primary tensor operator $\mathcal{O}_\Delta^{a_1 \dots a_J}
$ with $J$ traceless and symmetric indices and conformal dimension $\Delta $.
To see how the operators transform under the action of the  Cartan-Weyl algebra it is convenient to choose the directions $w$ and $\bw$. 
The resulting operators are 
\begin{equation}
\mathcal{O}_{\Delta, J}(x)\equiv \mathcal{O}_{\Delta, w \dots w}(x) 
 \, ,
\qquad 
\mathcal{O}_{\Delta, -J}(x)\equiv \mathcal{O}_{\Delta, \bw \dots \bw}(x)
\, .
\end{equation}
It is important to stress that if we use a mixed set of $w$ and $\bw$, the result vanishes because of the tracelessness condition, e.g. 
$4 \mathcal{O}_{\bw w}= \mathcal{O}_{1 1}+\mathcal{O}_{2 2}=0$.
Using   (\ref{eq:CartanWeylbasis}) and (\ref{eq:actionDK}-\ref{eq:actionJ}), it is easy to show that the states $
\mathcal{O}_{\Delta, J}(0)| 0 \rangle$ and $
\mathcal{O}_{\Delta, -J}(0)| 0 \rangle$ are annihilated by all positive roots
\begin{align}
 E_{\alpha^{+ \pm}} \mathcal{O}_{\Delta, J}(0)| 0 \rangle =0  \, ,
 \qquad
  E_{\alpha^{+ \pm}} \mathcal{O}_{\Delta, -J}(0)| 0 \rangle =0 
  \,
  .
\end{align}
Moreover, one can read off the weights by acting with the Cartan, 
\begin{equation}
\begin{array}{lr ll r} \label{CartanEigenvalues}
H_0& \mathcal{O}_{\Delta, J}(0) | 0 \rangle
&=&-\Delta \ &\mathcal{O}_{\Delta, J}(0)| 0 \rangle \ , \\
H_1& \mathcal{O}_{\Delta, J}(0)| 0 \rangle
&=&+J \  &\mathcal{O}_{\Delta, J}(0)| 0 \rangle \ , \\
H_0 &\mathcal{O}_{\Delta, -J}(0)| 0 \rangle
&= &-\Delta\ &  \mathcal{O}_{\Delta, -J}(0)| 0 \rangle \ , 
\\
H_1& \mathcal{O}_{\Delta, -J}(0)| 0 \rangle
&=& - J \ & \mathcal{O}_{\Delta, -J}(0)| 0 \rangle \ . \\
\end{array}
\end{equation}
We thus recognize that $\mathcal{O}_{\Delta, J}(x)$ has dimension $\Delta$ and spin $J$ while $\mathcal{O}_{\Delta, -J}(x)$ has dimension $\Delta$ and spin $-J$. The choice of the sign of $H_0$ is such that it measures $-\Delta$; this is the convention used by mathematicians to define highest weight (instead of lowest weight) representations.
Usually in $d=2$ we define holomorphic and antiholomorphic operators $\mathcal{O}$ and $\bar{\mathcal{O}}$. These are exactly the same as the ones above, namely $\mathcal{O}_{h, \bar h}\equiv \mathcal{O}_{\Delta, J}$, $\overline{\mathcal{O}_{h, \bar h}}\equiv \mathcal{O}_{\Delta, -J}$, where
the labels $h=\frac{\Delta+J}{2}$, $\bar h=\frac{\Delta-J}{2}$ are the eigenvalues with respect to the Cartan generators $L_0,\bar L_{0}$.

\subsubsection*{Simplicity of the Modules}
The concept of parabolic Verma modules plays a crucial role in the study of representation theory of CFT$_{d>2}$. Primary operators are defined to be killed by the special conformal generators, and they are labelled by their conformal dimensions and by their $SO(d)$ spin. 
The fact that a primary state transforms under $SO(d)$ means that a given primary does not define a unique state, but a finite set of states that transform as a finite dimensional representation, e.g. $|\mathcal{O}^{a}\rangle=\{|\mathcal{O}^{1}\rangle, \dots , |\mathcal{O}^{d}\rangle \} $.
The positive roots which need to annihilate $|\mathcal{O}^{a}\rangle$  are not all positive roots of $so(d+2)$, but only those of $so(d+2)$ which do not belong to $so(d)$. 
This construction makes the Verma modules `parabolic'.
In practice one defines a parabolic subalgebra $\mathfrak{p}$ which contains the Cartan. For CFT$_d$ this takes the form $\mathfrak{p}=so(2) \oplus so(d)$. Primary states are annihilated by the set of positive roots $\Psi^+=\Phi^+/\Phi^+_{\mathfrak{p}}$ (namely $K^{a}$), where $\Phi^+_{\mathfrak{p}}$ are the positive roots of $\mathfrak{p}$. The parabolic module is defined by freely acting with the negative roots $\Psi^-=\Phi^-/\Phi^-_{\mathfrak{p}}$ (namely $P^{a}$) on the highest weight, e.g. $\{|\mathcal{O}^{a}\rangle, P_{a}|\mathcal{O}^{a}\rangle , P^{(b}|\mathcal{O}^{a)}\rangle , P^{b} P_{a}  |\mathcal{O}^{a}\rangle , \dots   \}$.

The situation in $d=2$ is much simpler since the Cartan subalgebra is $so(2) \oplus so(d=2)$. This means that $\Psi^+=\Phi^+$ and $\Psi^-=\Phi^-$. In other words the modules are not parabolic. In this case the criterion of simplicity of the modules is much simpler  than for the parabolic counterpart. In particular a Verma module $M(\lambda)$ is simple if its weight $\lambda=(-\Delta,J)$ is antidominant, \emph{i.e.} $\lambda$ has to satisfy 
\begin{equation}
    2\frac{\langle \lambda +\rho, \alpha \rangle}{\langle \alpha, \alpha \rangle} \notin \mathbb{Z}_> \, , 
    \qquad 
    \forall \alpha \in \Phi^+ \, ,
\end{equation} 
where the Weyl vector $\rho$ is defined as half the sum of the positive roots,\footnote{This criterion is not sufficient for the parabolic case: the antidominant condition for $\lambda$ (here antidominance is only with respect to the roots in $\Psi^+$) need to be supplied by an extra condition on the character of the modules called the Jantzen criterion, which is reviewed in section 6 of \cite{Penedones:2015aga}. }
\be
\rho\equiv \frac{1}{2}\sum_{\alpha \in \Phi^+} \alpha =\left(1,0\right) \, .
\ee
We can exemplify this for our case. Given $\lambda$, we say that the module $M(\lambda)$ is not simple if one of the following two conditions is satisfied,
\begin{eqnarray}
  2\frac{\langle \lambda +\rho, \alpha^{++} \rangle}{\langle \alpha^{++}, \alpha^{++} \rangle} =  -\Delta +J+1 = n \, , 
  \label{cond++}
   \\
   2\frac{\langle \lambda +\rho, \alpha^{+-} \rangle}{\langle \alpha^{+-}, \alpha^{+-} \rangle} =  -\Delta -J+1 =n \, ,
    \label{cond+-}
\end{eqnarray}
where $n=1,2,\dots$. In other words we expect the presence of a primary descendant when a primary ${\cal O}$ satisfies either or both of the following two requirements $ h  =\frac{1-n}{2}$, $ \bar h  =\frac{1-\bar n}{2}$ for $n,\bar n \in \mathbb{Z}_>$, which is what we found in section \ref{sec:global_primary_desc}.
Finally we would like to find the weight $\lambda'\equiv (-\Delta',J')$ of the primary descendant when the module is not simple. $\lambda'$ can be obtained from $\lambda$ through a Weyl reflection. The Weyl reflections of a weight $\lambda$ with respect to a root $\alpha$  is defined as 
\begin{equation}
    s_{\alpha} \cdot \lambda =\lambda - 2\frac{\langle \lambda +\rho, \alpha \rangle}{\langle \alpha, \alpha \rangle} \alpha\, .
\end{equation}
In our case, when the condition \eqref{cond++} is valid the primary descendant is labelled by $\lambda'=s_{\alpha^{++}}\cdot \lambda$. 
This means that given a primary with dimension  $\Delta= J+1-n$ and spin $ J$ we get a primary descendant at $\Delta'= J+1$ and spin $J'=J-n$.
Similarly, when \eqref{cond+-} is satisfied, $\lambda'=s_{\alpha^{+-}}\cdot \lambda$, which implies that the primary with dimension $\Delta= 1-J-n$ and spin $J$ has a primary descendant at $\Delta'=1-J$ and spin $J'=J+n$. Notice that $J$ can be positive or negative. In the main text we rewrite the above conditions in terms of the absolute value of the spin $J$. In this language there are three different types of primary descendants depending on whether $|J'|$ increases, decreases or does not change with respect to $|J|$. The two formulations are of course equivalent.
In figures \ref{TipeI_primdesc}, \ref{TipeII_primdesc}, and \ref{Tipes_primdescIII} of the main text we show how the primaries and their associated primary descendants are related through the Weyl reflections defined  above.

\subsubsection*{Conformal Characters}
By looking at the weight $\lambda$ and $\lambda'$ of the primary and its primary descendant, it is easy to see that nested submodules can occur. The decomposition of the modules into irreducible ones is studied through Kazhdan-Lusztig theory. 
The main tool in this framework is the character. 
In the following we will not introduce  Kazhdan-Lusztig theory (for a review see \cite{Humphreys} and \cite{Penedones:2015aga, Bourget:2017kik}), however we will exemplify how the character contains the information about the decomposition of modules in CFT$_{d=2}$. 

In $d=2$ the global conformal character $\mbox{ch} M_{h,\bar h}(q,\bar q)$ associated to a module $M_{h,\bar h}$ is simply written as the product of the holomorphic and antiholomorphic characters $ \mbox{ch} M_{h}(q)\equiv q^h/(1-q)$,
\begin{equation}
   \mbox{ch} M_{h,\bar h}(q,\bar q)=   \mbox{ch} M_{h}(q) \,  \mbox{ch} M_{\bar h}(\bar q) \, .
\end{equation}
Let us see what happens to the character when the associated module is not simple. When $h=\frac{1-n}{2}$ for $n \in \mathbb{Z}_>$, we expect that the module becomes reducible and that it should contain a submodule with $h=\frac{1+n}{2}$. Indeed
\begin{equation}
    \mbox{ch} M_{\frac{1-n}{2},\bar h}(q,\bar q) =
    \mbox{ch}L_{\frac{1-n}{2}}(q)  \mbox{ch} M_{\bar h}(\bar q) 
    +  \mbox{ch} M_{\frac{1+n}{2},\bar h}(q,\bar q) \, ,
\end{equation}
where $\mbox{ch}L_{J}(q)\equiv \sum _{j=-|J|}^{|J|} q^j$. Notice that, for any $n\in\mathbb{Z}_>$, $\mbox{ch}L_{\frac{1-n}{2}}(q)$ contains finitely many terms and thus corresponds to a finite dimensional representation. However for generic $\bar h$ (meaning $\bar h\neq \frac{1-\bar n}{2}$) the product $\mbox{ch}L_{\frac{1-n}{2}}(q)  \mbox{ch} M_{\bar h}(\bar q)$ does not. 
When $h=\frac{1-n}{2}$ and $\bar h=\frac{1-\bar n}{2}$ we expect to see nested submodules. Indeed it is easy to check that
\begin{equation}
\badat{2}
     \mbox{ch} M_{\frac{1-n}{2},\frac{1-\bar n}{2}}(q,\bar q) &=
      \mbox{ch} M_{\frac{1-n}{2},\frac{1+\bar n}{2}}(q,\bar q)
     +  \mbox{ch} M_{\frac{1+n}{2},\frac{1-\bar n}{2}}(q,\bar q)-  \mbox{ch} M_{\frac{1+n}{2},\frac{1+\bar n}{2}}(q,\bar q)\\
     &\quad +
    \mbox{ch}L_{\frac{1-n}{2}}(q) \mbox{ch}L_{\frac{1-\bar n}{2}}(\bar q)\,.
\eadat
\end{equation}
This is exactly the expression that we would have guessed: the module contains three nested submodules which appear at the position of the primary descendants, while the remaining piece is a product of two $\mbox{ch}L$ and corresponds to a finite dimensional representation.

\section{Celestial SL(2,\texorpdfstring{$\mathbb{C}$}{C}) Transformations}
\label{sec:CelestialSL2C}

In this appendix we discuss various aspects of global conformal symmetry. We derive relations between SL(2,$\mathbb{C}$) Lorentz generators and celestial derivatives $\partial_w$ and $\partial_\bw$ satisfied by conformal primary wavefunctions at points $(w,\bw)$ on the celestial sphere.

We start by considering the generators $J_{\mathbb{R}^{1,3}}^{\mu \nu}$ of Lorentz transformations. We will consider their action on the the Hilbert space as well as on wavefunctions on the spacetime.  On the Hilbert space, we expect the representation for tensor states to be given by~\eqref{eq:actionJ}.  Meanwhile local operators in the bulk will transform via commutators with the Hilbert space generators.\footnote{The expected transformation rules only need to hold up to gauge equivalence within the quantum theory~\cite{Hamada:2017uot}.  The amplitudes themselves are gauge invariant, meaning the Mellin correlators only differ from the ones created by preparing conformal primary states by the overall normalization.}
Finite transformations can be obtained by exponentiating the generators with parameters $\omega$ as follows $U=\exp(\frac{i}{2} \omega_{\mu \nu} J_{\mathbb{R}^{1,3}}^{\mu \nu})$. On the Hilbert space, we want this representation to be unitary. Unitarity of $U$ implies that the generators $J_{\mathbb{R}^{1,3}}^{\mu \nu}$ are Hermitian. 
It can be also useful to rewrite $ J_{\mathbb{R}^{1,3}}^{\mu\nu}$ in 
 terms of the boosts $\vec{K}$ and rotations $\vec{J}$ as follows
\be\label{KJgens}
J_{\mathbb{R}^{1,3}}^{\mu\nu}=\left(\begin{array}{cccc} 0 & K_1 & K_2& K_3 \\
-K_1 &0&J_3&-J_2\\
-K_2 &-J_3&0&J_1\\
-K_3 &J_2&-J_1&0
\end{array}\right) \ .
\ee
Meanwhile, it is known that one can embed the $D-2$-dimensional conformal algebra  into the $D$-dimensional Lorentz one (e.g. see \cite{DiFrancesco:1997nk}).
We choose the following embedding\footnote{While the embedding is fairly rigid, it is not unique. 
}
\begin{equation} \label{EmbeddingConfAlg}
J_{\mathbb{C}}^{\mu\nu}=\left(\begin{array}{cccc} 
0 &-\frac{P_{1}-K^{SCT}_{1}}{2} &-\frac{P_{2}-K^{SCT}_{2}}{2} &  D \\
 \frac{P_{1}-K^{SCT}_{1}}{2}
& 0& J_{12} & \frac{P_{1}+K^{SCT}_{1}}{2} \\
 \frac{P_{2}-K^{SCT}_{2}}{2}
& -J_{12} & 0 & \frac{P_{2}+K^{SCT}_{2}}{2} \\
- D & -\frac{P_{1}+K^{SCT}_{1}}{2} &-\frac{P_{2}+K^{SCT}_{2}}{2} & 0
\end{array}\right).
\end{equation}
where $D, P_a, K^{SCT}_a, J_{ab}$ are the generators of the conformal algebra introduced in appendix \ref{App:rep_theory}.  It is thus natural to group  4D generators in a manner that reflects the conformal algebra in 2D,
\begin{equation}
\label{embedding_generators}
    \begin{array}{lll}
 L_0=\frac{J_3-i K_3}{2} \, , & L_{-1}=\frac{-J_1+i J_2+i K_1+K_2}{2} 
  \, ,
    & L_1=\frac{J_1+i
   J_2-i K_1+K_2}{2} 
    \, ,
   \\
 \bar L_0=\frac{-J_3-i K_3}{2} 
  \, ,&
   \bar L_{-1}=\frac{J_1+i J_2+i K_1-K_2}{2} 
    \, ,
   & \bar L_1 =\frac{-J_1+i J_2-i K_1-K_2}{2} 
    \, ,
   \\
\end{array}
\end{equation}
where the definition of $L_n$ in terms of the conformal algebra is given in appendix \ref{App:rep_theory}.
Since the generators $K_i$ and $J_i$ are taken to be Hermitian, we obtain the following conjugation relations for the $L_n$,
\begin{equation}
    L_n^\dagger =-\bar L_n \, ,
    \qquad 
    \bar L_n^\dagger =- L_n \, 
\end{equation}
as operators on the Hilbert space.  These appeared in our discussions of primary descendants versus null states in the main text.

We now want to relate these generators on the $4D$ Hilbert space to our descendancy relations. We can do so using the transformation properties of the conformal primary wavefunctions used to prepare the external scattering states.

Conformal primary wavefunctions $\Phi_{\Delta,J}(X^\mu;w,\bw)$ are covariant under the transformation
\be\label{deltaj2}
\Phi_{\Delta,J}\Big(\Lambda^{\mu}_{~\nu} X^\nu;\frac{a w+b}{cw+d},\frac{{\bar a} \bw+{\bar b}}{{\bar c}\bw+{\bar d}}\Big)=(cw+d)^{2h}({\bar c}\bw+{\bar d})^{2 \bar h}D(\Lambda)\Phi_{\Delta,J}(X^\mu;w,\bw)\,,
\ee
comprised of an SL(2,$\mathbb{C}$) M\"{o}bius transformation acting on the reference point $(w,\bw)$ on the celestial sphere and a Lorentz transformation with $\Lambda^{\mu}_{~\nu}$ in the corresponding vector representation of $SO(3,1)\simeq SL(2,\mathbb{C})$
and $D(\Lambda)$ the representation appropriate for the 4D indices of $\Phi_{\Delta,J}$. Here $h=\frac{1}{2}(\Delta+J)$ and $\bar h=\frac{1}{2}(\Delta-J)$.
Since the wavefunctions transform covariantly, a $4D$ transformation acts in the opposite way of its respective $2D$ transformation, namely
\begin{equation}
\label{covariance_generators}
(J_{\mathbb{R}^{1,3}}^{\mu\nu}+J_{\mathbb{C}}^{\mu\nu})\Phi_{\Delta,J}(X^\sigma;w,\bw)=0 \, 
\end{equation}
where now $J_{\mathbb{R}^{1,3}}^{\mu\nu}$ is realized via a Lie derivative on the corresponding tensor field,\footnote{The transformations $\Lambda^\mu_{~\nu} X^\nu$ and $D(\Lambda)$ appearing on the left and right sides of~\eqref{deltaj2} are both captured by a Lie derivative of the corresponding tensor field along the respective spacetime boost or rotation vector field.} and similarly for $J_{\mathbb{C}}^{\mu\nu}$ acting on a function of $(w,\bw)$ with definite weights.\footnote{See, for instance, the weighted scalars of~\cite{Barnich:2021dta}.}

It is instructive to check explicitly how this happens.  In doing so, we will be able to show that the generators in~\eqref{embedding_generators} act on the Hilbert space inherited from 4D like one would expect in a 2D CFT. To do so we start by writing $ \Lambda_{\mu \nu}$ in terms of the SL(2,$\mathbb{C}$) parameters (see ~\cite{Oblak:2015qia}),
\begin{equation}
\label{Lambda_abcd}
    \Lambda_{\mu \nu}=\frac{1}{2}
    \scalemath{0.8}{
    \left(\begin{array}{cccc}
    -(a\ba+b\bb+c\bc+d\bd) & -(a\bb+\ba b+\bc d+c\bd) & -i(a\bb-\ba b+c \bd-\bc d) & a\ba-b\bb+c\bc-d\bd\\
    a\bc+\ba c+b\bd+\bb d & a \bd+\ba d +b\bc+\bb c & i(a\bd -\ba d -b\bc+\bb c) & -a\bc -\ba c+b\bd+\bb d\\
    i(-a\bc +\ba c -b \bd +d\bb ) & i(-a\bd +\ba d-b \bc+\bb c) & a\bd+\ba d-b\bc -\bb c & i(a\bc -\ba c-b \bd+\bb d\\
    -a \ba- b\bb + c\bc +d\bd & -a \bb- \ba b+c\bd+\bc d & i(-a\bb+\ba b+c\bd -\bc d) & a\ba -b\bb -c\bc +d\bd
    \end{array}\right)\, .
    }
\end{equation}
For infinitesimal transformations $\Lambda$ takes the form $\Lambda_{\mu \nu} \approx \eta_{\mu \nu} + \omega_{\mu \nu}$ where $\omega$ are written in terms of SL(2,$\mathbb{C}$) parameters. The form of $\omega$ also defines the linear combination of the 4D generators responsible for the transformation as  $f(\Lambda^{\sigma}_{\, \rho} X^{\rho})\approx(1+\frac{i}{2} \omega_{\mu \nu} J_{\mathbb{R}^{1,3}}^{\mu \nu})f(X^{\sigma})$.
We proceed by considering infinitesimal transformations  around
\be
a=1\,,~~b=0\,,~~c=0\,,~~d=1\,,\quad \text{and} \quad \bar a=1\,,~~\bar b=0\,,~~\bar c=0\,,~~\bar d=1\,,
\ee
which obey $ad-bc=1$ and $\bar a \bar b-\bar b \bar c=1$, independently transforming the left (un-barred) and right (barred) sectors (where implicitly  we have complexified SL(2,$\mathbb{C}$) to SL(2,$\mathbb{C})\times$ SL(2,$\mathbb{C})$). 
We focus on the left sector here while the right sector is analogous. 

Let us start with the infinitesimal transformation 
\be
a=1-\frac{\epsilon}{2}\,,~~b=0\,,~~c=0\,,~~d=1+\frac{\epsilon}{2}\,,
\ee
which, at leading order in $\epsilon$, yields
\begin{equation}
    \Lambda_{\mu\nu}=\eta_{\mu \nu}-
    \scalemath{0.83}{
    \frac{i}{2  }\left(\begin{array}{cccc}
   0 & 0 & 0 & -i\epsilon\\
   0 &0& \epsilon & 0\\
   0 & -\epsilon & 0 & 0\\
   + i\epsilon & 0 & 0  & 0
    \end{array}\right)\,,
    }
    \qquad   \qquad
    w\mapsto w-\epsilon w\,.
\end{equation}
The transformation~\eqref{deltaj2} can then be expressed as
\be
\left( \frac{J_3-i K_3}{2}-w\p_w -h\right)\Phi_{\Delta,J}=0\,.
\ee
Repeating the same computation for the infinitesimal transformation
\be
a=1\,,~~b=-\epsilon\,,~~c=0\,,~~d=1\,,
\ee
for which
\begin{equation}
    \Lambda_{\mu\nu}=\eta_{\mu \nu}-
    \scalemath{0.83}{
    \frac{i}{2 }\left(\begin{array}{cccc}
    0 & i\epsilon & \epsilon & 0\\
   -i\epsilon &0& 0& -i\epsilon\\
   -\epsilon & 0 & 0 & -\epsilon\\
    0 & i\epsilon & \epsilon  & 0
    \end{array}\right)\,,
    }
    \qquad
    \qquad
    w\mapsto w-\epsilon\,,
\end{equation}
yields
\be
\left(\frac{-J_1+i J_2+i K_1+K_2}{2}-\p_w \right)\Phi_{\Delta,J}=0\,.
\ee
Finally, the infinitesimal transformation
\be
a=1\,,~~b=0\,,~~c=\epsilon\,,~~d=1\,,
\ee
gives
\begin{equation}
    \Lambda_{\mu\nu}=
   \eta_{\mu \nu}- \scalemath{0.83}{
    \frac{i}{2}
\left(
\begin{array}{cccc}
 0 & -i \epsilon  & \epsilon  & 0 \\
 i \epsilon  & 0 & 0 & -i \epsilon  \\
 -\epsilon  & 0 & 0 & \epsilon  \\
 0 & i \epsilon  & -\epsilon  & 0 \\
\end{array}
\right)    \,,
    }
    \qquad
    \qquad
    w\mapsto w-\epsilon w^2\,.
\end{equation}
This yields
\be
\left(\frac{-i K_1+ K_2+J_1+i J_2}{2}-w^2\p_w- 2h w\right)\Phi_{\Delta,J}=0\,.
\ee
Similar expressions exist for the barred quantities, taking the complex conjugates of the differential operators appearing above. 

We can now use~\eqref{covariance_generators} to show that the operators $L_i$ defined in~\eqref{embedding_generators} transform the operators~\eqref{qdelta} as expected for a local operator in a 2D CFT.  In the limit where the bulk operators are described by a quantum field theory, under consideration here, we expect local operators to transform as follows
\be
[J_{\cal H}^{\mu\nu},\mathcal{O}(X)]=-\mathcal{L}_{J_{\mathbb{R}^{1,3}}^{\mu\nu}}\mathcal{O}(X).
\ee
For gauge fields, this equality needs to hold only up to gauge transformations~\cite{Hamada:2017uot}.
With this, equation~\eqref{covariance_generators} and the definition~\eqref{qdelta}, we find
 \be\label{wlocal}
 \badat{3}
[J_{\cal H}^{\mu\nu},\O^{s,\pm}_{\Delta,J}(w,\bw)]&= i([J_{\cal H}^{\mu\nu},O^{s}(X^\mu)],\Phi^s_{\Delta^*,-J}(X_\mp^\mu;w,\bw))\\
&= -i(\mathcal{L}_{J_{\mathbb{R}^{1,3}}^{\mu\nu}}O^{s}(X^\mu),\Phi^s_{\Delta^*,-J}(X_\mp^\mu;w,\bw))\\
&= i(O^{s}(X^\mu),\mathcal{L}_{J_{\mathbb{R}^{1,3}}^{\mu\nu}}\Phi^s_{\Delta^*,-J}(X_\mp^\mu;w,\bw))\\
&= -i(O^{s}(X^\mu),\mathcal{L}_{J_{\mathbb{C}}^{\mu\nu}}\Phi^s_{\Delta^*,-J}(X_\mp^\mu;w,\bw))\\
&=-\mathcal{L}_{J_{\mathbb{C}}^{\mu\nu}}i(O^{s}(X^\mu),\Phi^s_{\Delta^*,-J}(X_\mp^\mu;w,\bw))\,.\\
\eadat\ee
The first line uses the fact that generator can be defined on the Cauchy slice where the inner product is taken, the second uses Lorentz invariance of the inner product to move the 4D Lie derivative to the wavefunction.
Equation~\eqref{wlocal} tells us that the Hilbert space generators transform the ${\cal O}_{\Delta,J}$ like one would expect for a local operator in a 2D CFT. In particular, our wavefunction computations above now tell us that $\p_w$ and $\p_\bw$ descendants indeed correspond to acting on states in the Hilbert space with $L_{-1}$ and $\bar{L}_{-1}$, as used in section~\ref{sec:global_primary_desc}.

\section{Celestial Multiplets}\label{app:modules}

With the connection between the $\p_w$ and $\p_\bw$ descendants established, we will now turn to the wavefunctions and demonstrate how to use the tetrad and spin frame to identify primary descendants of the radiative modes in section~\ref{sec:cpwdec}.  We can proceed similarly with vector fields on the spacetime and illustrate in section~\ref{sec:ConfModules} how the Poincar\'e and 4D conformal generators descend from primaries.

\subsubsection*{Useful Wavefunction Formulae}\label{UsefulIDs}
We collect here a list of relations between conformal primaries and the null tetrad and spin frame in terms of which they are constructed involving spacetime and celestial CFT derivatives.\vspace{1em}

\noindent{\bf Spacetime derivatives}~~ The members of the tetrad satisfy
\begin{equation}
    \partial_\mu l_\nu=l_\mu l_\nu\,, \quad \partial_\mu n_\nu=\eta_{\mu\nu}+n_\mu l_\nu\,, \quad \partial_\mu m_\nu = m_\mu l_\nu\,, \quad \partial_\mu \bar{m}_\nu=\bar{m}_\mu l_\nu\,,
\end{equation}
yielding $\Box l^\mu=0$, $\Box n^\mu=2l^\mu$, $\Box m^\mu=0$, $\Box \bar{m}^\mu=0$
and
\begin{equation}
    X\cdot l=-1\,, \quad X\cdot n=\frac{X^2}{2}\,, \quad X\cdot m=0\,, \quad X\cdot \bar{m}=0\,.
\end{equation}
For the elements of the spin frame we have
\begin{equation}
\partial_\mu o=\frac{1}{2} l_\mu o\,, \quad \partial_\mu \iota = \frac{1}{2}l_\mu \iota +\frac{1}{\sqrt{2}} \sigma_\mu \bar{o}\,,
\end{equation}
implying $\Box o=0$ and $\Box \iota=0$. \vspace{1em}

\noindent{\bf Celestial CFT derivatives}~~
Derivatives of the tetrad vectors obey the following relations
\begin{equation}
 \badat{4}
\p_w^n l_\mu&=2^{\frac{n}{2}}n! (\epsilon_w\cdot X)^{n-1}\varphi^n m_\mu\,,\\
\p_w^n n_\mu &=\frac{X^2}{2}\p_w^n l_\mu\,,\\
\p_w^n m_\mu&=(\epsilon_w\cdot X)\partial_w^n l_\mu\,,\\
\p_w^{n} \bar{m}_\mu&=\left\{\begin{array}{ll}
    2^{\frac{1}{2}}\varphi^1\left( v_\mu-(\epsilon_w\cdot X)\bar{m}_\mu\right)  &  n=1\\
   \frac{X^2}{2}(\epsilon_w\cdot X)^{-1}\p_w^nl_\mu\,,  & n>1
\end{array}\right.\,,
 \eadat
\end{equation}
where we defined the quantity
\be
v_\mu=\frac{X^2}{2}l_\mu+n_\mu\,.
\ee
Similar expressions as the ones above are obtained for $w\leftrightarrow \bw$ and $m \leftrightarrow \bar{m}$. The elements of the spin frame satisfy
\begin{equation}
\badat{2}
   \p_w^n o&=2^{\frac{n}{2}} \pi^{-\frac{1}{2}} {\textstyle \Gamma(n+\frac{1}{2})}(\epsilon_w\cdot X)^n o \varphi^n\,, 
   \\
   \p_\bw^n o&=2^{\frac{n}{2}} \pi^{-\frac{1}{2}} {\textstyle \Gamma(n+\frac{1}{2})}(\epsilon_\bw\cdot X)^n o \varphi^n- 2^{\frac{n}{2}}n{\textstyle(\frac{1}{2})_{n-1}}(\epsilon_\bw\cdot X)^{n-1} \nu\varphi^{n-\frac{1}{2}}\,,
\eadat
\end{equation}
where we introduced $\nu=i(1,0)$, and from which the corresponding expressions for $\iota$ are obtained due to the commutativity of $X^\mu$ with the $\{w,\bw\}$ derivatives.\vspace{1em}

\subsection{Wavefunction Descendants}
\label{sec:cpwdec}
In this section, we identify primary descendants of radiative fields starting from the explicit form of the conformal primary wavefunctions.  This complements the algebraic approach of the previous section in a manner that reflects the methods used in section~\ref{sec:prim_desc}.  We stick to the integer spin cases for brevity, while the half-integer cases follow from similar arguments.

\subsubsection*{Descendants of Radiative Primaries}
The conformal primary scalar obeys the relation
\begin{equation}
\badat{2}
    \p_w^n \varphi^\Delta
    &=2^\frac{n}{2}(\Delta)_n (\epsilon_w\cdot X)^n \varphi^{\Delta+n} \,.
\eadat
\end{equation}
To compute descendants of spinning primaries let us introduce an arbitrary reference direction $Z^\mu$ so that we automatically enforce symmetry and can simplify our notation. This yields
\be\label{holoderiv}
\p_w^n\left[(Z\cdot m)^s\varphi^\Delta\right]=2^{\frac{n}{2}}(s+\Delta)_{n}(Z\cdot m)^s\varphi^{\Delta+n}(\epsilon_w\cdot X)^n\,.
\ee
Since taking derivatives with respect to $\{w,\bw\}$ commutes with $X^2$ this relation also holds for the corresponding shadow primary.  Next, we have
\be\scalemath{0.9}{\badat{3}\label{recurse1}
\p_w^n\left[(Z\cdot \bar{m})^{s}\varphi^\Delta\right]&=2^{\frac{n}{2}}\varphi^{\Delta+n}(\epsilon_w\cdot X)^n \left[\sum_{k,\ell=0}^s C_{\Delta,s,n}(k,\ell)(Z\cdot \bar{m})^{s-k-\ell}\Big(\frac{Z\cdot v}{\epsilon_w\cdot X}\Big)^k\Big({X^2}\frac{Z\cdot m}{(\epsilon_w\cdot X)^2}\Big)^\ell\right]\,.
\eadat}\ee
We can take a derivative of both sides to determine a recursion relation for the coefficients
\be\badat{3}\label{recursion}
C_{\Delta,s,{n+1}}(k,\ell)&=(\Delta+n-s+k+2\ell)C_{\Delta,s,n}(k,\ell)\\
&+(s-k-\ell+1)C_{\Delta,s,n}(k-1,\ell)+(k+1)C_{\Delta,s,n}(k+1,\ell-1)\,.
\eadat\ee
Then, starting from 
\be\label{initial}
C_{\Delta,s,0}(k,\ell)=\delta_{k,0}\delta_{\ell,0}
\ee
 we can iterate up for any $n>0$. We find
\be\badat{3}\label{Csolved}
C_{\Delta,s,n}(k,\ell)&=(s+1-k-\ell)_{k+\ell}(\Delta-s+k+2\ell)_{n-k-2\ell}\frac{(2\ell-1)!!~n!}{k!(2\ell)!(n-k-2\ell)!}\,.
\eadat\ee
By verifying that this expression satisfies the initial conditions~\eqref{initial} and the recursion~\eqref{recursion}, we have a proof by induction.  Expressing everything in terms of Gamma functions, we have
\be\scalemath{0.9}{\badat{3}
C_{\Delta,s,n}(k,\ell)&=\frac{2^\ell}{\sqrt{\pi}}\frac{\Gamma(\Delta+n-s)\Gamma(s+1)\Gamma(n+1)\Gamma(\ell+\frac{1}{2})}{\Gamma(k+1)\Gamma(2\ell+1)\Gamma(s+1-k-\ell)\Gamma(n-k-2\ell+1)\Gamma(\Delta-s+k+2\ell)}\,.
\eadat}\ee
The right hand side only has support when $k,2\ell\le n$.  So long as $k,s,\ell$ are positive integers, we can drop the limit in the sum. 

Now, the only terms which transform with definite conformal weight are those with
\be
k+2\ell=n\,.
\ee
Because we are also only summing over $k,\ell\le s$, we can set up a finite system of equations that must be satisfied if we want a descendant at level~$n$. For instance, if we want a primary at level~1 we would need the $\ell=0,k=0$ term to vanish which requires $\Delta=s$. If we want a primary at level~2, we would need the $k=0,\ell=0$ and $k=1,\ell=0$ terms to vanish which gives
\be
(\Delta-s)(1+\Delta-s)=0,~~~s(1+\Delta-s)=0\,.
\ee
Hence, there is a level~2 primary descendant at $\Delta=s-1$ or we have one at $s=0,\Delta=0$ but this is the trivial kind.  We can summarize this nicely. For fixed $\Delta,s$  one can only have a primary descendant if 
\be
(n-k-2\ell)C_{\Delta,s,n}(k,\ell)=0
\ee
for all integer $k, n\ge0$ (where we only need to check for $k,2\ell\le n$). This is consistent with the results of section~\ref{sec:global_primary_desc}.

The complex conjugate of these expressions gives us a tower of $\p_\bw$ derivatives on all of the radiative wavefunctions. One can then work out a similar set of recursions relations for mixed derivatives. From the complex conjugate of~\eqref{holoderiv} we have 
\be\label{holoderiv2}
\p_\bw^p\left[(Z\cdot \bar{m})^s\varphi^\Delta\right]=2^{\frac{p}{2}}(s+\Delta)_{p}(Z\cdot {\bar m})^s\varphi^{\Delta+p}(\epsilon_\bw\cdot X)^p\,
\ee
so that from~\eqref{recurse1} we have\be\badat{3}
\p_w^n\p_\bw^p\left[(Z\cdot \bar{m})^s\varphi^\Delta\right]&=2^{\frac{n}{2}+\frac{p}{2}}(s+\Delta)_{p}\varphi^{\Delta+n+p}(\epsilon_\bw\cdot X)^p(\epsilon_w\cdot X)^n\\
&\times\left[\sum_{k,\ell=0}^s C_{\Delta+p,s,n}(k,\ell)(Z\cdot \bar{m})^{s-k-\ell}\Big(\frac{Z\cdot v}{\epsilon_w\cdot X}\Big)^k\Big({X^2}\frac{Z\cdot m}{(\epsilon_w\cdot X)^2}\Big)^\ell\right]\,\\
\eadat
\ee
with the same coefficients $C$ as above in~\eqref{Csolved}.  From this example, we see that there are no extra primary descendants than what we have from $\p_w^n$ or $\p_\bw^p$ acting separately.  One has factors of $(\epsilon_\bw\cdot X)$ hanging around unless $p=0$ and then that reduces to the same system of equations as above for $C$. The other way we can get a primary is if the coefficient out front vanishes and that is the same as the primary descendants we find from~\eqref{holoderiv}.

\subsection{Generators of (Conformal) Isometries}
\label{sec:ConfModules}

In this section, we move away from $w=\bw=0$ and present a set of covariant vector fields which respect the SL(2,$\mathbb{C}$) submodule structure of the celestial diamonds discussed in section~\ref{sec:diamond}.\vspace{1em}

\noindent{\bf Poincar\'e Generators as SL(2,$\mathbb{C}$) Modules}~~  We would like to find a vector field representation of the Poincar\'e algebra~\cite{Barnich:2017ubf}\footnote{We will modify our notation to avoid confusion with the Poincar\'e generators that appear elsewhere in this paper as well as the representation of this algebra in celestial amplitudes in~\cite{Stieberger:2018onx,Fotopoulos:2019vac}.}
\be\label{Poincare}
\badat{3}
[p_{i,j},p_{k,l}]&=0\,,\\
[\ell_{n},p_{k,l}]&=(\frac{1}{2}n-k)p_{n+k,l}\,,\\
[\bar{\ell}_{n},p_{k,l}]&=(\frac{1}{2}n-l)p_{k,n+l}\,,\\
[\ell_m,\ell_n]&=(m-n)\ell_{m+n}\,,\\
[{\bar \ell}_m,{\bar \ell}_n]&=(m-n){\bar \ell}_{m+n}\,,\\
[{\bar \ell}_m, \ell_n]&=0\,,\\
\eadat
\ee
via SL(2,$\mathbb{C}$) descendants of vector fields with definite conformal dimension and spin which match those of the circled nodes in figure~\ref{grav_modules}. The top node corresponds to an SL(2,$\mathbb{C}$) primary which we can construct from the tetrad~\eqref{tetrad} while the other encircled nodes correspond to $\partial_w$ and $\partial_\bw$ descendants:
\be\badat{3}
\ell_{-1}&=\alpha\varphi^{-1}\bar{m}^\mu\p_\mu\,,~~~&\bar{\ell}_{-1}&=\alpha\varphi^{-1}m^\mu\p_\mu\,,\\
\ell_0&=\frac{1}{2\sqrt{2}\alpha}\p_w \ell_{-1}\,,~~~&{\bar{\ell}}_0&=\frac{1}{2\sqrt{2}\alpha}\p_\bw {\bar \ell}_{-1}\,,\\
\ell_1&=\frac{1}{4\alpha^2}\p_w^2 \ell_{-1}\,,~~~&{\bar{\ell}}_1&=\frac{1}{4\alpha^2}\p_\bw^2 {\bar \ell}_{-1}\,,\\
\eadat
\ee
and
\be\badat{3}
p_{-\frac{1}{2},-\frac{1}{2}}&=\beta\varphi^{-1}l^\mu\p_\mu\,,\\
p_{\frac{1}{2},-\frac{1}{2}}&=\frac{1}{\sqrt{2}\alpha}\p_w p_{-\frac{1}{2},-\frac{1}{2}}\,,\\
p_{-\frac{1}{2},\frac{1}{2}}&=\frac{1}{\sqrt{2}\alpha}\p_\bw p_{-\frac{1}{2},-\frac{1}{2}}\,,\\
p_{\frac{1}{2},\frac{1}{2}}&=\frac{1}{2\alpha^2}\p_\bw\p_w p_{-\frac{1}{2},-\frac{1}{2}}\,.
\eadat
\ee
Moreover, we note that further descendants vanishes
\be
\p_w^3 \ell_{-1}=\p_\bw \ell_{-1}=0\,,~~~\p_{\bw}^3 {\bar \ell}_{-1}=\p_{w} {\bar \ell}_{-1}=0\,,~~~\p_w^2p_{-\frac{1}{2},-\frac{1}{2}}=\p_{\bw}^2p_{-\frac{1}{2},-\frac{1}{2}}=0\,.
\ee
Here $\alpha$ and $\beta$ are real constants which we can fix as follows. The choice $\alpha=\frac{1}{\sqrt{2}}$ is most natural from the point of view that the $w$ representation of SL(2,$\mathbb{C}$)
\be
[\p_w,w^2\p_w]=2w\p_w
\ee
is being used to step between Lorentz generators
\be
\p_w \ell_{-1}=2\ell_0\,.
\ee
Choosing $\beta=1$ reduces $p_{-\frac{1}{2},-\frac{1}{2}}$ to the generator for spacetime translations along $q^\mu$.  One can use the derivative relations given in appendix~\ref{UsefulIDs} to explicitly evaluate the descendant vector fields appearing here.\vspace{1em}

\noindent{\bf Bulk Conformal Generators as SL(2,$\mathbb{C}$) Modules}~~  
It is worth pointing out that we can extend the above results to reproduce the conformal algebra in 4D. We first observe that the vector fields
\be\badat{3}
k_{-\frac{1}{2},-\frac{1}{2}}&=\varphi^{-1}n^\mu\p_\mu\,,\\
k_{\frac{1}{2},-\frac{1}{2}}&=\p_w k_{-\frac{1}{2},-\frac{1}{2}}\,,\\
k_{-\frac{1}{2},\frac{1}{2}}&=\p_\bw k_{-\frac{1}{2},-\frac{1}{2}}\,,\\
k_{\frac{1}{2},\frac{1}{2}}&=\p_\bw\p_w k_{-\frac{1}{2},-\frac{1}{2}}\,.
\eadat
\ee
obey an algebra isomorphic to that of the $P_{ij}$, namely 
\be
\badat{3}
[k_{i,j},k_{k,l}]&=0\,,\\
[\ell_{n},k_{k,l}]&=(\frac{1}{2}n-k)k_{n+k,l}\,,\\
[\bar{\ell}_{n},k_{k,l}]&=(\frac{1}{2}n-l)k_{k,n+l}\,.\\
\eadat
\ee
Upon adding the generator 
\be
{\mathscr D}=X^\mu\p_\mu\,,
\ee
we complete the 4D conformal algebra
\be
\badat{3}
[{\mathscr D},p_{k,l}]&=-p_{k,l}\,,\\
[{\mathscr D},k_{k,l}]&=k_{k,l}\,,\\
[p_{i,j},k_{k,l}]&=2((i-k)(j-l){\mathscr D}+(i-k)\ell_{j+l}+(j-l)\bar{\ell}_{i+k})\,,
\eadat
\ee
so that the $k_{i,j}$ are the special conformal generators.

\section{Embedding Space Formalism}\label{app:embeddingspace}
The embedding space formalism (see e.g. \cite{Costa:2011mg}) is a convenient tool in CFTs. The idea is to make $D$-dimensional global conformal group $SO(D+1,1)$ act linearly by uplifting local insertions from $\mathbb{R}^{d}$ to $q \in \mathbb{R}^{1,d}$. In particular each point in physical space corresponds to a light ray in the null cone $q^2=0$. 
Interestingly, the embedding space is a very natural language for celestial CFTs since $\mathbb{R}^{1,d}$ is  the space where the scattering takes place.
In what follows we review how to define the wavefunctions and the shadow transforms in the embedding space and what is their relation with the physical space counterparts.
\subsubsection*{Wavefunctions in the Embedding Space}
Let us consider wavefunctions $\Phi(X;q)$ depending on a point $X^\mu \in \mathbb{R}^{1,3}$ and a generic null vector $q^\mu\in \mathbb{R}^{1,3}$.
These wavefunctions are correctly uplifted to the embedding space if  $\Phi(X;q)$ satisfies the following conditions:
\begin{enumerate}
     \item Homogeneous in $q$: $\Phi(X;\alpha q)_{\mu_1...\mu_{|J|} \nu_1...\nu_{|J|}} =\alpha^{-\Delta}\Phi(X;q)_{\mu_1...\mu_{|J|} \nu_1...\nu_{|J|}}$.
    \item Transverse: $\Phi(X;q)_{\mu_1...\mu_{|J|} \nu_1...\nu_{|J|}} q^{\nu_i} =0$.
    \item Traceless: $\Phi(X;q)_{\mu_1...\mu_{|J|} \nu_1...\nu_{|J|}} \eta^{\nu_i\nu_j} =0$.
    \item $\Phi(X;q)_{\mu_1...\mu_{|J|} \nu_1...\nu_{|J|}}\sim\Phi(X;q)_{\mu_1...\mu_{|J|} \nu_1...\nu_{|J|}}+ q^{\nu_i} \Phi'(X;q)_{\mu_1...\mu_{|J|}; \nu_1...\nu_{i-1}\nu_{i+1}...\nu_{|J|}}$.
\end{enumerate}
For our purposes we further consider wavefunctions which are symmetric in the $\mu_i$ and $\nu_i$ indices separately.  In embedding space the spin $s=0,1,2$ radial wavefunctions $\Phi(X;q)$ are given by the following bulk-to-boundary propagators
\begin{equation}\label{bulktobdy}
 \badat{3}
 \Phi(X;q)&=\frac{1}{(-q\cdot X)^\Delta}\,, \\ 
 \Phi_{\mu_1 \nu_1}(X;q)&=\frac{(-q\cdot X)\eta_{\mu_1\nu_1}+q_{\mu_1} X_{\nu_1}}{(-q\cdot X)^{\Delta+1}}\,, \\ 
 \Phi_{\mu_1 \mu_2\nu_1\nu_2}(X;q)&=\frac{[(-q\cdot X)\eta_{\mu_1\nu_1}+q_{\mu_1} X_{\nu_1}][(-q\cdot X)\eta_{\mu_2\nu_2}+q_{\mu_2} X_{\nu_2}]}{(-q\cdot X)^{\Delta+2}}\,.
\eadat
\end{equation}
To project down to the 2D physical space, we typically parametrize the null cone via 
\begin{equation}
 \label{poincare_section}
  q^\mu(\vec{w})=(1+|\vec{w}|^2,2\vec{w},1-|\vec{w}|^2)\,
\end{equation} 
the so-called `Poincar\'e section'.
In order to fully project the 4D wavefunctions we must also project the $\nu_i$ indices,
so that it transforms as a 2D conformal primary in the symmetric traceless rank-$|J|$ representation of SO(2) with conformal dimension $\Delta$.
The result is the usual wavefunction 
\begin{equation}
\Phi_{\mu_1\dots \mu_{|J|}a_1\dots a_{|J|}}(X^\mu;\vec{w}) 
\equiv  
 \Phi_{\mu_1\dots \mu_{|J|} \nu_1\dots \nu_{|J|}}(X^\mu;q(\vec{w})) \prod_{i=1}^{|J|} \partial_{a_i} q^{\mu_i}(\vec{w}) \,,
\end{equation}
where $\partial_a q^\mu=2(w_a,\delta_{ab},-w_a)$.

\subsubsection*{Shadow Transform}
\label{app:Shadow}
In 2D it is often convenient to use complex coordinates in physical space. In this section we show how to relate the shadow transform in embedding space to the one in complex coordinates.

Let us consider a local operator $\mathcal{O}_{\nu_1\dots \nu_{|J|}}(q)$ with dimension $\Delta$ and spin $J$ defined in the embedding space. 
We can define its shadow transform directly in the embedding space using~\cite{SimmonsDuffin:2012uy}
\begin{equation}\scalemath{1}{ \label{SHPhiEmbeddingDSD}
 \widetilde{\mathcal{O}}_{\nu_1\dots \nu_{|J|}}(q)=\frac{K_{\Delta,J}}{2\pi} \int \frac{d^4 q' \; \delta(q'^2)\theta(q'{}^0)}{\textrm{Vol  GL(1,$\mathbb{R}$)$^+$}}  \frac{\prod_{n=1}^{|J|}[\delta_{\nu_n}^{\rho_n} (-\frac{1}{2} q\cdot q')+\frac{1}{2} q'_{\nu_n} q^{\rho_n}]}{(-\frac{1}{2}q\cdot q')^{2-\Delta+|J|}}\mathcal{O}_{\rho_1\dots\rho_{|J|}}(q')\,,
}\end{equation}
where the integral is restricted to the positive null cone which passes through the origin in $\mathbb{R}^{1,3}$ and we quotient by the connected component of the identity GL(1,$\mathbb{R}$)$^+\subset$ GL(1,$\mathbb{R}$) to render it finite. The normalization $K_{\Delta,J}$ will be determined below to match the 2D normalization chosen in~\eqref{def:2dShadowTransform}.

We can eliminate the factor of $\textrm{Vol  GL(1,$\mathbb{R}$)$^+$}$ by choosing a section of the null cone, which we will take to be \eqref{poincare_section}. We further project the indices to 2D using
\begin{equation}
 \O_{a_1\dots a_{|J|}}(\vec{w})=\prod_{i=1}^{|J|} \partial_{a_i} q^{\mu_i}(\vec{w})\O_{\mu_1 \dots \mu_{|J|}}(q(\vec{w}))\,.
\end{equation}
We then recover the standard expression for the 2D shadow transform~\cite{Ferrara:1972xe,Ferrara:1972uq,Ferrara:1972ay,Ferrara:1973vz,Dolan:2011dv} 
\begin{equation} \label{2DrealSh}
 \widetilde{\O}_{a_1\dots a_{|J|}}(\vec{w})=\frac{K_{\Delta,J}}{2\pi} \int d^2 \vec{w}' \frac{1}{|\vec{w}-\vec{w}'|^{2(2-\Delta)}} \mathcal{I}_{a_1\dots a_{|J|},b_1,\dots b_{|J|}}(\vec{w}-\vec{w}')\mathcal{O}^{b_1 \dots b_{|J|}}(\vec{w}')\,,
\end{equation}
where $\mathcal{I}_{a_1\dots a_{|J|},b_1,\dots b_{|J|}}(\vec{w}-\vec{w}')$ is the inversion tensor for symmetric traceless tensors obtained from the symmetrized product of $|J|$ inversion tensors
\begin{equation}\label{invtens}
 \mathcal{I}_{ab}(\vec{w}-\vec{w}')=\eta_{ab}-2\frac{(w_a-w'_a)(w_b-w'_b)}{|\vec{w}-\vec{w}'|^2}\,.
\end{equation}
Note that integral in~\eqref{2DrealSh} is divergent unless $\Delta<1$ but can be extended to more general $\Delta$ by analytic continuation so that under conformal transformations~\eqref{2DrealSh} defines a conformal primary operator in the symmetric
traceless rank-$|J|$ representation of $SO(2)$ of conformal dimension $2-\Delta$~\cite{Dolan:2011dv}.

Let us now return to  complex coordinates $(w,\bar w)\in \mathbb{R}^2$ used throughout this paper which are convenient to work with in 2D CFT.  We start again from~\eqref{2DrealSh} and use 
\begin{equation}
 x^a=(w,\bw)\,, \quad x_a=\eta_{ab} x^b=\frac{1}{2}(\bw,w)\,, \quad x^2=x^a x_a=w\bw\,, \quad \eta_{ab}=\left(\begin{matrix}0&1/2\\ 1/2&0 \end{matrix} \right)\,,
\end{equation}
to express the inversion tensor $\mathcal{I}_{ab}(\vec{x})=\eta_{ab}-2\frac{x_a x_b}{x^2}$ as~\cite{Dolan:2000ut}
\begin{equation}
 \I_{ww}(\vec{x}-\vec{x}')=-\frac{1}{2} \frac{\bw-\bw'}{w-w'}\,, \quad \I_{\bw \bw}(\vec{x}-\vec{x}')=-\frac{1}{2} \frac{w-w'}{\bw-\bw'}\,,\quad \I_{w\bw}=\I_{\bw w}=0\,.
\end{equation} 
As an example we focus on spin $J=1$ (generalizations are straightforward) and we rewrite the shadow transform~\eqref{2DrealSh} as follows 
\begin{equation}\label{DOSHspin1}
\badat{4}
 \widetilde{\O}_{\bw}(w,\bw)&=\frac{K_{\Delta,J=1}}{2\pi} \int d^2w' \frac{\I_{\bw\bw}(w-w',\bw-\bw')}{[(w-w')(\bw-\bw')]^{(2-\Delta)}}  \O^\bw(w',\bw')\\
 &={-}\frac{K_{\Delta,J=1}}{2\pi} \int d^2w' \frac{\O_w(w',\bw')}{(w-w')^{2(1-h)}(\bw-\bw')^{2(1-\bh)} }\,,
 \eadat
\end{equation}
where we used $\O^\bw=\eta^{\bw w} \O_w=2\O_w$ and the fact that $\O_w$ has $J=h-\bh=+1$. The result matches \eqref{def:2dShadowTransform} up to a minus sign in the normalization.
Similarly for higher spin $J$ we have to use products of $J$ terms of the form $\I_{\bw\bw}$, which give an extra factor of $(-1)^J$. We thus conclude that \eqref{def:2dShadowTransform} matches \eqref{2DrealSh} if (recall that here we are only considering integer~$J$)
\begin{equation}
    K_{\Delta,J}=(-1)^J K_{h \bh} \, .
\end{equation}
In the main body of the paper we normalize \eqref{def:2dShadowTransform} as $ K_{h,\bar h}=2{\rm max}\{h,\bar h\}-1$. We thus get an extra sign in the 4D conformal primary wavefunctions of spin $s=1$ in~\eqref{SHCPWs} (and similarly for $s=\frac{1}{2}$) as compared to~\cite{Pasterski:2017kqt}.

\bibliographystyle{utphys}
\bibliography{CelestialDiamonds}

\end{document}